\documentclass[10pt,twocolumn,letterpaper]{article}

\usepackage{titling}
\usepackage{iccv}
\usepackage{times}
\usepackage{epsfig}
\usepackage{graphicx}
\usepackage{amsmath}
\usepackage{amssymb}
\usepackage{booktabs}
\usepackage{float}
\usepackage{adjustbox}
\usepackage{pgfplots}
\usepackage{caption}
\usepackage{subcaption}
\usepackage[accsupp]{axessibility}
\usepackage[breaklinks=true,bookmarks=false]{hyperref}

\usepackage[capitalize]{cleveref}
\crefname{section}{Sec.}{Secs.}
\Crefname{section}{Section}{Sections}
\Crefname{table}{Table}{Tables}
\crefname{table}{Tab.}{Tabs.}

\iccvfinalcopy % *** Uncomment this line for the final submission

\def\iccvPaperID{6604} % *** Enter the ICCV Paper ID here

\ificcvfinal\fi

\newcommand{\beginsupplement}{%
        \setcounter{table}{0}
        \renewcommand{\thetable}{A\arabic{table}}%
        \setcounter{figure}{0}
        \renewcommand{\thefigure}{A\arabic{figure}}%
        \setcounter{section}{0}
        \renewcommand{\thesection}{A\arabic{section}}%
}

\begin{document}

%%%%%%%%% TITLE
%\predate{}
%\postdate{}

\title{MoTIF: Learning Motion Trajectories with Local Implicit Neural \\ Functions for Continuous Space-Time Video Super-Resolution}

\newcommand*{\affaddr}[1]{#1} % No op here. Customize it for different styles.
\newcommand*{\affmark}[1][*]{\textsuperscript{#1}}
\newcommand*{\email}[1]{\texttt{#1}}
\makeatletter
\newcommand{\printfnsymbol}[1]{%
  \textsuperscript{\@fnsymbol{#1}}%
}
\author{Yi-Hsin Chen\thanks{Both authors contributed equally to this work.}
\qquad Si-Cun Chen\printfnsymbol{1} 
\qquad Yi-Hsin Chen
\qquad Yen-Yu Lin
\qquad Wen-Hsiao Peng \\
National Yang Ming Chiao Tung University, Taiwan \\ 
\tt\small \{yhchen12101, sicun.mapl, karta6120\}.cs09@nycu.edu.tw \\ \tt\small lin@cs.nycu.edu.tw \tt\small wpeng@cs.nctu.edu.tw
}

%\date{}

\maketitle
% Remove page # from the first page of camera-ready.
\ificcvfinal\thispagestyle{empty}\fi

\begin{abstract}
This work addresses continuous space-time video super-resolution (C-STVSR) that aims to up-scale an input video both spatially and temporally by any scaling factors. One key challenge of C-STVSR is to propagate information temporally among the input video frames. To this end, we introduce a space-time local implicit neural function. It has the striking feature of learning forward motion for a continuum of pixels. We motivate the use of forward motion from the perspective of learning individual motion trajectories, as opposed to learning a mixture of motion trajectories with backward motion. To ease motion interpolation, we encode sparsely sampled forward motion extracted from the input video as the contextual input. Along with a reliability-aware splatting and decoding scheme, our framework, termed MoTIF, achieves the state-of-the-art performance on C-STVSR. The source code of MoTIF is available at \href{https://github.com/sichun233746/MoTIF}{https://github.com/sichun233746/MoTIF}.
%The source code of MoTIF will be available upon acceptance.
%The source code of MoTIF is available at [link to become available upon acceptance]. 

%Our Fourier analysis shows that the forward motion of a pixel, when viewed as a function of time characterizing its motion trajectory, tends to be smoother and easier to approximate than the backward motion.
%\vspace{-1.0mm}
\end{abstract}

\section{Introduction}
\label{sec:intro}
\begin{figure}[t!]
  \centering
  \begin{subfigure}[b]{1.0\linewidth}
  % \includegraphics[trim={0 2120 0 0}
  %       ,clip,width=1.0\linewidth]
  %       {Figures/teaser_1111.png}
  %   \caption{VideoINR~\cite{VideoINR}}
  %\includegraphics[trim={0 3050 0 0}
  %\includegraphics[trim={0 770 0 0}
  \includegraphics[trim={0 565 0 10}
        ,clip,width=0.92\linewidth]
        {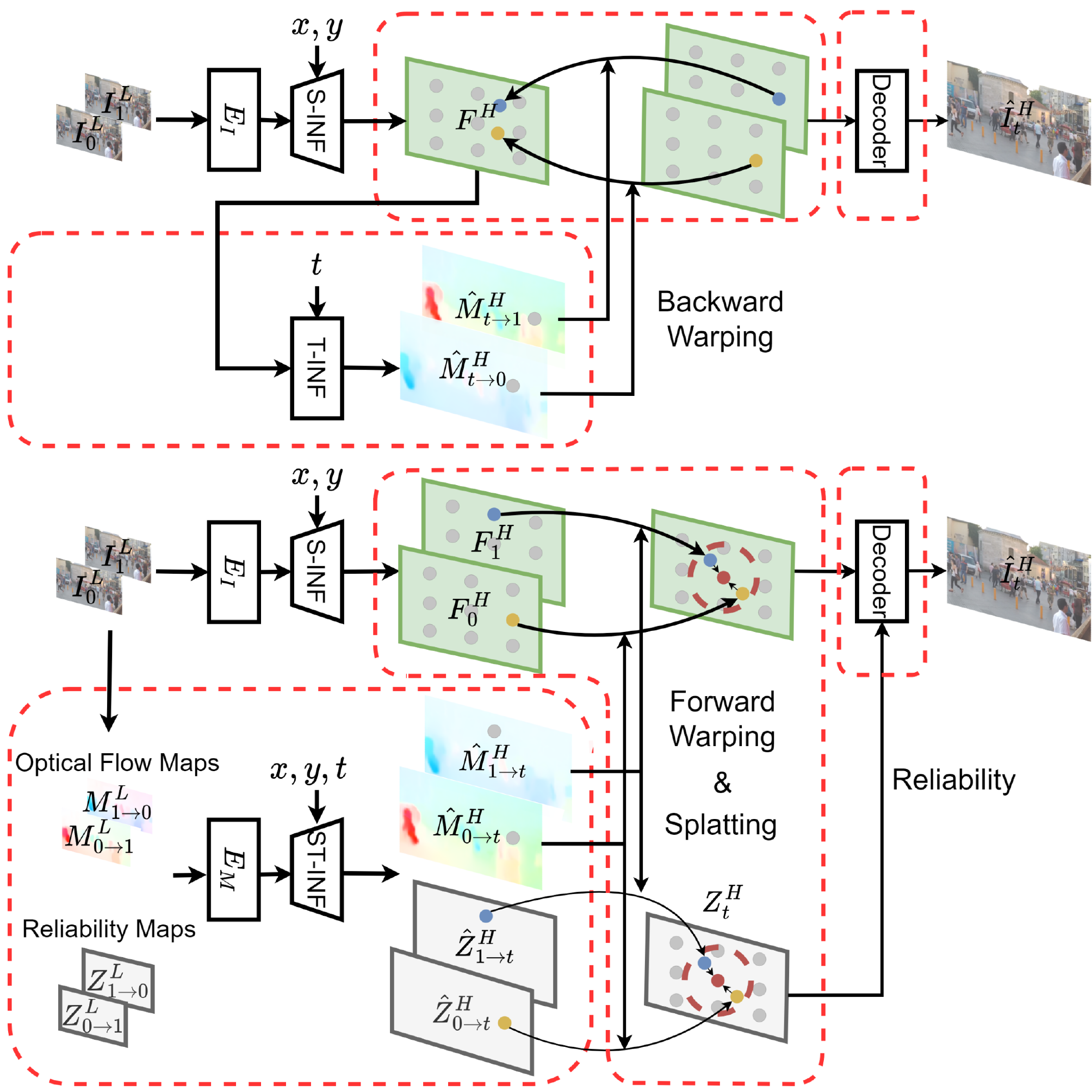}  
    \caption{VideoINR~\cite{VideoINR}}
    \label{fig:teaser_a}
  \end{subfigure}
   
  \begin{subfigure}[b]{1.0\linewidth}
  % \includegraphics[trim={0 0 0 2300}
  %       ,clip,width=1.0\linewidth]
  %       {Figures/teaser_1111.png}
  %\includegraphics[trim={0 0 0 2300}
  %\includegraphics[trim={0 0 0 580}
  \includegraphics[trim={0 0 0 410}
        ,clip,width=0.92\linewidth]
        {Figures/teaser_final.pdf}
    \caption{MoTIF}
    \label{fig:teaser_b}
  \end{subfigure}
  \vspace{-5mm}
  \caption{
    Illustrations of (a) VideoINR~\cite{VideoINR} and (b) MoTIF. The red dash lines highlight their major differences. }%We use an spatialtemporal implicit neural function taking motion latents to generate forward motions while VideoINR~\cite{VideoINR} use an temporal implicit neural function taking reference features to generate backward motions.
  %\vspace{-4mm}
  \vspace{-1.3em}
  \label{fig:teaser}
\end{figure}

%\vspace{-1.5mm}
This work addresses continuous space-time video super-resolution (C-STVSR). The task of C-STVSR is to increase simultaneously the spatial resolution and temporal frame-rate of an input video by any scaling factors with only one single model. It is to be distinguished from fixed-scale space-time video super-resolution (F-STVSR), for which a model is learned to perform space-time super-resolution for only one specific spatiotemporal scale. As compared to F-STVSR, C-STVSR is more flexible and practical in real-world scenarios, which often call for up-scaling low-resolution and low-frame-rate videos of varied spatiotemporal resolutions on heterogeneous video-enabled devices.

C-STVSR remains largely under-explored. One trivial solution to C-STVSR is to perform continuous video frame interpolation~\cite{DAIN, SuperSloMo, softmax, Context_Aware, QVI}, followed by interpolating individual video frames with continuous image super-resolution~\cite{LIIF, UltraSR, LTE}, or the other way around. However, their divide-and-conquer nature of treating C-STVSR as two independent sub-tasks--i.e.~temporal interpolation and spatial super-resolution--misses the opportunity to attain the best achievable performance. By leveraging the spatiotemporal information in an end-to-end optimized fashion, some recent works~\cite{StarNet, ZSM, TMNet, RSTT} for F-STVSR adopt a one-stage approach, combining the extraction of individual frame features and the temporal aggregation of these features as a unified task. Nonetheless, these F-STVSR methods can hardly be extended straightforwardly to C-STVSR. 

% \begin{figure}[t!]
%   \centering
%   \includegraphics[width=1.0\linewidth]{Figures/forward_backward.png}

%   \caption{Example of caption.}
%   \label{fig:forward_backward}
% \end{figure}
\begin{figure}[t!]
  \centering
  
  \begin{subfigure}[b]{0.49\linewidth}
  \includegraphics[trim={350 0 0 0}
            ,clip,width=1.0\linewidth]{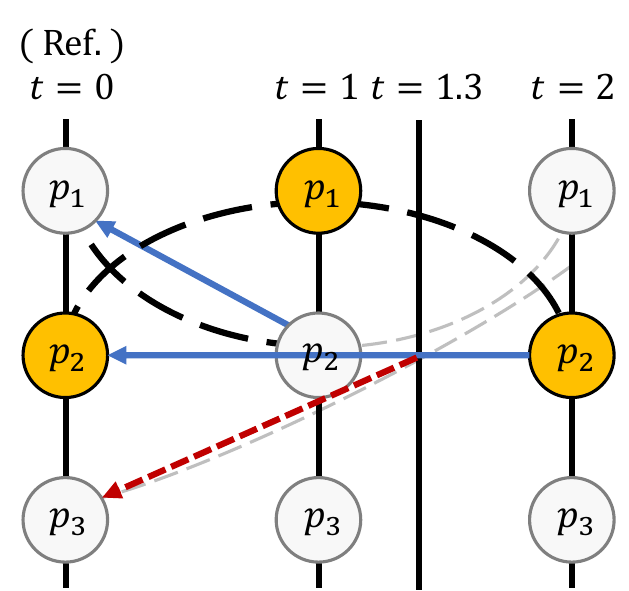}
    \caption{Backward Motion}
    \label{fig:forward_backward_a}
  \end{subfigure}
  \begin{subfigure}[b]{0.49\linewidth}
  \includegraphics[trim={0 0 350 0}
            ,clip,width=1.0\linewidth]{Figures/motion_v3.pdf}
    \caption{Forward Motion}
    \label{fig:forward_backward_b}
  \end{subfigure}
  
  \caption{
  Illustration of backward and forward motion. \textcolor{black}{The circles denote pixels accessible in the input video. The dashed lines display the motion trajectories of pixels in the reference frame at $t=0$. The blue arrows are backward/forward motion in the form of displacement vectors. The red arrows show the displacement vectors for an arbitrary time instance that are to be predicted from blue arrows.}
  }
  \vspace{-1.em}
  %\caption{
  %\textbf{An illustration of modeling forward motion and backward motion.}
  %The dash curves represent motion trajectories. The blue arrows represent the motions we can access to while the red arrow represents the predicted motion.
  %}
  \label{fig:forward_backward}
\end{figure}

Inspired by continuous image super-resolution~\cite{LIIF, UltraSR, LTE}, VideoINR~\cite{VideoINR} presents an early attempt at C-STVSR. Given any query coordinates $(x,y,t)$ in the continuous spatiotemporal space, it takes the latent representation of the input video as the contextual information to decode the corresponding RGB value. The process involves learning a spatial implicit neural function (S-INF in Fig.~\ref{fig:teaser} (a)) for super-resoluting the frame features, followed by learning another temporal implicit neural function (T-INF in Fig.~\ref{fig:teaser} (a)) to generate motion estimates at time $t$ to backward warp the super-resoluted frame features. However, learning implicitly \textit{backward motion} (indicating displacement vectors that identify matching pixels/features in the reference frame) as a function of time is challenging. Essentially, the backward motion at the same spatial coordinates $(x,y)$ yet at different time instances $t$ may capture the motion trajectories of different pixels/features in the reference frame. \textcolor{black}{For example, in Fig.~\ref{fig:forward_backward}~(a), the backward motion vectors of $p_2$ at $t=1$ and $t=2$ are governed by the two distinct motion trajectories that originate from pixels $p_1$ and $p_2$ in the reference frame at $t=0$, respectively. In other words, the backward motion vectors at $p_2$, when viewed as a function of time, are a mixture of multiple motion trajectories. This could potentially introduce undesirable randomness and discontinuities in the resulting time function, which must be learned by T-INF in Fig.~\ref{fig:teaser} (a).} Furthermore, learning implicitly such a time function based solely on frame features complicates the task.  

%\textcolor{black}{Taking yellow dot's trajectory in Fig.~\ref{fig:forward_backward_a} as example,} \textcolor{black}{we conjecture that typical motion trajectories of pixels vary smoothly with time (Sections~\ref{subsec:estimate_explicit_motion}) and are relatively easier to model with an implicit neural function.}

To circumvent the aforementioned issues, we propose learning \textit{forward motion} of pixels in the form of motion trajectories with a space-time implicit neural function (ST-INF in Fig.~\ref{fig:teaser} (b)). Considering each reference frame in the input video as sitting at the origin in time, our ST-INF takes $(x,y,t)$ as input and outputs a displacement vector that specifies where the pixel at the coordinates $(x,y)$ of the reference frame will appear in a synthesized frame at time $t$. \textcolor{black}{That is, it encodes the motion trajectory of the pixel at $(x,y)$, e.g. the highlighted motion trajectory of $p_2$ in the reference frame at $t=0$ in Fig.~\ref{fig:forward_backward}~(b).} 
%Because $(x,y)$ can take any continuous values, the implicit neural function encodes motion trajectories (i.e.~forward motion) for a contiguous spectrum of pixels. 
Moreover, to facilitate the learning of such a neural function in an explicit way, we supply \textit{forward optical flow maps} estimated between reference frames as the contextual information (i.e.~$M^L_{0 \rightarrow 1}, M^L_{1 \rightarrow 0}$ in Fig.~\ref{fig:teaser} (b)). \textcolor{black}{Our space-time neural function is also learned to predict the reliability of every motion trajectory (i.e.~$\hat{Z}^H_{0 \rightarrow t},\hat{Z}^H_{1 \rightarrow t}$ in Fig.~\ref{fig:teaser} (b)), which is essential to ensure the quality of forward warping. Explicit motion modeling allows us to extract rough reliability estimates from the input video for better prediction.}
%since they may be disrupted by occlusion in the scene. 

%This is in direct contrast to learning implicitly backward motion from frame features as with VideoINR~\cite{VideoINR}. Last but not least, 
%\textcolor{purple}{We argue that typical motion trajectories of pixels vary smoothly with time (Sections~\ref{subsec:estimate_explicit_motion} and \ref{subsec:forward_backward}) and are relatively easier to model with an implicit neural function.}

Fig.~\ref{fig:teaser} (b) depicts our end-to-end trainable C-STVSR framework, MoTIF. The main contributions of our work include: (1) we propose a space-time local implicit neural function that predicts \textit{forward} motion and its reliability in a continuous manner; (2) we propose a reliability-aware splatting and decoding scheme that fuses simultaneously information from multiple reference frames; and (3) \textcolor{black}{our MoTIF achieves the state-of-the-art performance on C-STVSR and provides out-of-distribution generalization.} %is applicable to any number of reference frames.
% \textcolor{purple}{(2) we provide a Fourier analysis of forward and backward motion, justifying the use of forward motion as a better representation;}

\begin{figure}[t!]
  \centering
  \includegraphics[width=0.50\linewidth]{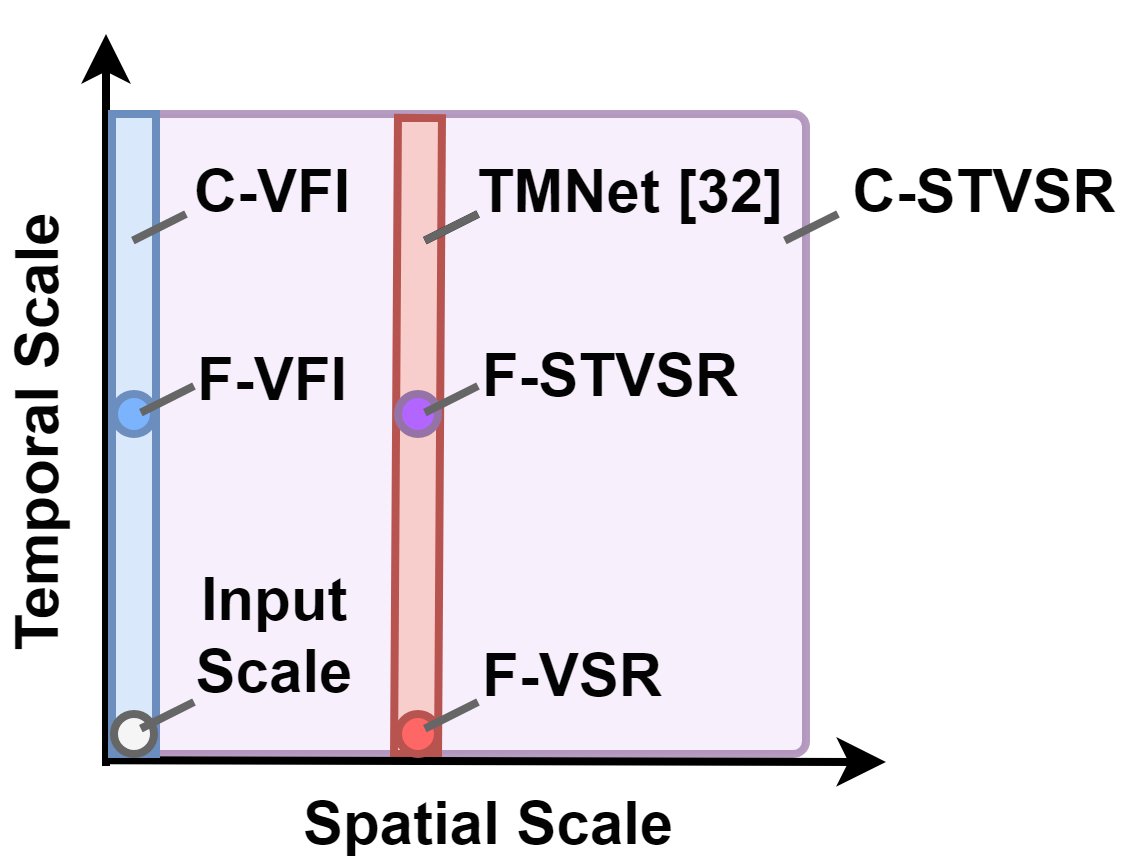} %venn_0306_v4
  \caption{Illustration of fixed-scale video frame interpolation (F-VFI), continuous video frame interpolation (C-VFI), fixed-scale video super-resolution (F-VSR), fixed-scale space-time video super-resolution (F-STVSR), TMNet~\cite{TMNet}, and continuous space-time video super-resolution (C-STVSR) in terms of their supported space-time scales.}
  %\caption{A Venn diagram categorizes prior works into fixed video frame interpolation (F-VFI), continuous video frame interpolation (C-VFI), fixed video super-resolution (F-VSR), fixed space-time video super-resolution (F-STVSR), TMNet~\cite{TMNet}, and continuous space-time video super-resolution (C-STVSR)}
  \vspace{-4mm}
  \label{fig:releated}
\end{figure}
\section{Related Work}
\label{sec:related}
This section surveys methods for video frame interpolation, video super-resolution, and space-time video super-resolution. Fig.~\ref{fig:releated} presents a Venn diagram to illustrate how C-STVSR, the focus of our work, is related to these methods in terms of their supported space-time scales. As shown, the fixed-scale methods--e.g.~fixed-scale video super-resolution~\cite{BasicVSR}, fixed-scale video frame interpolation~\cite{MEMC, AdaCoF} and F-STVSR~\cite{RSTT, ZSM, StarNet}--perform only one specific type of space-time interpolation. Their supported space-time scales are visualized as singletons in Fig.~\ref{fig:releated}. In comparison, the continuous-scale methods--such as continuous video frame interpolation~\cite{DAIN, SuperSloMo, softmax, Context_Aware, QVI} and TMNet\cite{TMNet}--are able to cover a continuum of temporal scales. Of these approaches, C-STVSR~\cite{VideoINR} is the most flexible and challenging one, with its supported space-time scales covering the entire space-time space.

%In this section, we introduce works related to our paper. We conclude these works into space-time interpolation methods and implicit neural representation methods. As shown in Fig.~\ref{fig:releated}, space-time interpolation methods can be grouped into seven types. Fixed VSR\cite{}, fixed VFI\cite{} and fixed STVSR\cite{} can only perform a unique type of space-time interpolation, forming a dot in the space-time interpolation space. Continuous VFI\cite{}, continuous VSR\cite{} and TMNet\cite{TMNet} can adjust their space-time interpolation ratio along one axis, forming a line in the space-time interpolation space. As for continuous STVSR\cite{VideoINR}, it can adapt its space-time interpolation ratio along space and time, forms an area that covers the entire space-time interpolation space.

%\input{Chapter/2_related_work/mix_STVSR}

\subsection{Video Frame Interpolation} 
Video frame interpolation~\cite{DAIN, SuperSloMo, softmax, QVI, PhaseNet, AdaCoF, MEMC} aims to increase the frame rate of a video by interpolating between existing reference frames. The key to successful frame interpolation is to predict how the pixels/features of the reference frames progress temporally to the interpolated frame. The flow-based methods~\cite{Context_Aware, DAIN, softmax, Splatting_based, voxel, SuperSloMo} rely on optical flow maps to propagate features/pixels from the neighboring reference frames, whereas the kernel-based methods~\cite{AdaCoF, Deformable_Separable, CDFI} estimate motion implicitly as kernels for motion compensation with deformable convolution. Most flow-based approaches adopt backward warping~\cite{SuperSloMo, voxel, ABME, BMBC}, but more recently, forward warping~\cite{DAIN, Context_Aware, softmax, Splatting_based} emerges as an attractive alternative. Forward warping, however, is faced with the challenge that multiple features/pixels in the reference frame may be mapped to the same location in the target frame. To tackle this issue, Niklaus~\emph{et al.}~\cite{softmax} introduce softmax splatting, weighting the conflicting features/pixels according to the reliability of their forward motion. 

\begin{figure*}[t!]
  \centering
  \includegraphics[width=1.0\linewidth]{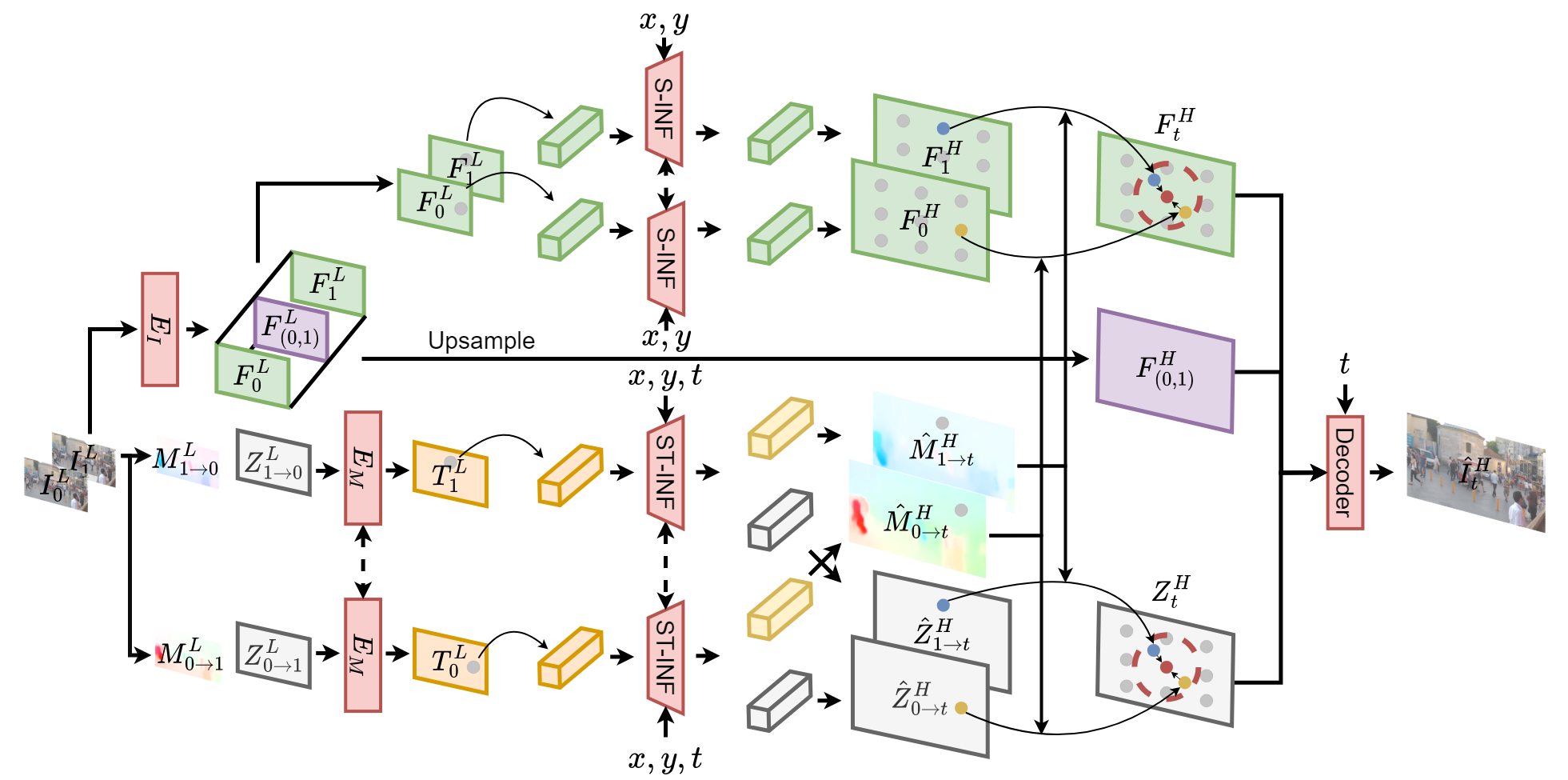}
  \caption{
    The proposed MoTIF for C-STVSR, where the dash double arrows represent the shared-weight networks. 
  }
  \vspace{-.6em}
%   \caption{
%     An overview of our \textcolor{red}{Motion Trajectory Implicit Function (MoTIF)}. The dash double arrows represent shared-weight networks.
%   }
  \label{fig:overview}
\end{figure*}
\subsection{Video Super-Resolution} 
Video super-resolution is to increase the spatial resolution of a video. Its central theme is to exploit temporal information from neighboring frames in order to complement a low-resolution video frame in recovering its missing high-frequency details. Early deep learning-based methods~\cite{VSR_Detail_Revealing, VSR_real_time, VSR_recurrent, TOFlow, BasicVSR} rely on optical flows to align the features/pixels of the neighboring frames. However, optical flow estimation can be expensive. As such, Tian~\emph{et al.}~\cite{TDAN} adopt deformable convolution for temporal alignment. Wang~\emph{et al.}\cite{EDVR} extend the idea to perform temporal alignment in a coarse-to-fine manner. These works target fixed-scale video super-resolution.

%propose the Pyramid, Cascading and Deformable (PCD) module, which extends the implicit alignment to multi-scale scheme for more precise alignment. 

%Tian \emph{et al.}\cite{} adopts deformable convolutions to implicitly align features of input frames. In addition to this, Wang \emph{et al.}\cite{EDVR} propose the Pyramid, Cascading and Deformable (PCD) module, which extends the implicit alignment to multi-scale scheme for more precise alignment. However, these methods can only perform fixed video super-resolution and can not straightforward extend to continuous video super-resolution.

%Video super-resolution (VSR) aims to increase the spatial resolution of the input video. Early deep learning based methods\cite{} rely on explicit optical flow estimation to align the contexts or features of the input video. Suffer from large parameter numbers and long throughput of optical flow estimation, Tian \emph{et al.}\cite{} adopts deformable convolutions to implicitly align features of input frames. In addition to this, Wang \emph{et al.}\cite{EDVR} propose the Pyramid, Cascading and Deformable (PCD) module, which extends the implicit alignment to multi-scale scheme for more precise alignment.
\subsection{Space-Time Video Super-Resolution}
Recognizing that both video frame interpolation and video super-resolution involve aggregating temporal information from neighboring frames, Haris~\emph{et al.}~\cite{StarNet} adopt a unified network to address space-time video super-resolution (STVSR). STVSR is much more challenging than the previous two tasks, as the low-resolution neighboring frames are the only source of information to interpolate a high-resolution video frame. Along this line of research, Xiang~\emph{et al.}~\cite{ZSM} propose using bidirectional deformable ConvLSTM to mine useful space-time information from the input video in an end-to-end fashion. Based on~\cite{ZSM}, Xu~\emph{et al.}~\cite{TMNet} introduce a temporal modulation block, which allows STVSR to be continuous in the temporal scale. By contrast, both~\cite{StarNet} and~\cite{ZSM} support only F-STVSR. 

More recently, Chen~\emph{et al.}~\cite{VideoINR} present the first work on end-to-end learned C-STVSR, allowing both the spatial and temporal scales to be continuous. Inspired by~\cite{LIIF}, which learns local implicit neural functions for continuous image super-resolution, their C-STVSR scheme includes a spatial and a temporal implicit neural function. The former generates the pixel features at any given spatial coordinates $(x,y)$ for super-resolution, while the latter predicts the \textit{backward} motion for any spatiotemporal coordinates $(x,y,t)$ to propagate temporally the resulting features to time $t$. Both neural functions are local; they refer to neighboring latents extracted from the input video as additional contextual information. 

%\textcolor{black}{Unlike~\cite{VideoINR}, which implicitly uses backward motion, our space-time local implicit neural function models \textit{forward} motion trajectories, while explicitly referring to the forward motion estimated between reference frames as the contextual input. Sections~\ref{subsec:estimate_explicit_motion} and \ref{sec:ablation} analyze both motion types, showing that forward motion typically forms a smoother time function.} %Moreover, we additionally introduce the reliability-aware splatting and decoding schemes to incorporate the reliability information in fusing information from reference frames.}

\section{Proposed Method}
\label{sec:method}
%In this section, we first sketch our C-STVSR framework in Sec. \ref{subsec:method_overview}. Then we introduce our explicit motion trajectory implicit function in Sec. \ref{subsec:estimate_explicit_motion}; multi-frame forward warping in Sec. \ref{subsec:multi_frame_forward_warping}; frame reconstruction module in Sec. \ref{subsec:frame_reconstruction}; Finally, the training details are obtained in Sec. \ref{subsec:implementation_details}. 

Given two low-resolution RGB video frames $I_{0}^L, I_{1}^L \in \mathbb{R}^{3 \times H \times W}$ of size $H \times W$, our task is to interpolate a high-resolution video frame $I_t^H \in \mathbb{R}^{3 \times H' \times W'}$ with an arbitrary scale $s=W'/W=H'/H \geq 1$ and at any time $t \in [0,1]$. 

\subsection{System Overview}
\label{subsec:method_overview}
Fig.~\ref{fig:overview} depicts our proposed MoTIF, which comprises four major components and operates as follows. First, given $I_{0}^L$ and $I_{1}^L$, (1) the encoder $E_I$ converts them into their latent representations $F_{0}^L,F_{1}^L,F_{(0,1)}^L \in \mathbb{R}^{C \times H \times W}$, where $F_{(0,1)}^L$ serves as a rough estimate of the feature of the target frame $I_t^H$. Similar to recent STVSR works~\cite{TMNet, VideoINR}, we adopt the off-the-shelf video-based encoder from~\cite{ZSM}, which fuses information from both $I_{0}^L$ and $I_{1}^L$ in generating $F_{0}^L,F_{1}^L$ and $F_{(0,1)}^L$. Second, (2) the spatial local implicit neural function (S-INF) is queried to super-resolute $F_{0}^L,F_{1}^L$ as $F_{0}^H,F_{1}^H \in \mathbb{R}^{C \times H' \times W'}$, respectively. Our S-INF follows the design of LIIF~\cite{LIIF}. Third, considering $I_{0}^L$ as sitting at the origin in time, (3) the motion encoder $E_M$ encodes $M_{0 \rightarrow 1}^L \in \mathbb{R}^{2 \times H \times W}$--namely, the forward optical flow map capturing the forward motion from $I_{0}^L$ to $I_{1}^L$--together with its reliability map $Z_{0 \rightarrow 1}^L \in \mathbb{R}^{3 \times H \times W}$ into $T_0^{L} \in \mathbb{R}^{C \times H \times W}$. The optical flow estimation is not always perfect; $Z_{0 \rightarrow 1}^L$ indicates how reliable $M_{0 \rightarrow 1}^L$ is across spatial locations $(x,y)$ (Section~\ref{subsec:estimate_explicit_motion}). Forth, using $T_0^{L}$ as the motion latent, (4) our space-time local implicit neural function (ST-INF) renders a high-resolution, forward motion map $\hat{M}_{0 \rightarrow t}^H \in \mathbb{R}^{2 \times H' \times W'}$ and its reliability map $\hat{Z}_{0 \rightarrow t}^H \in \mathbb{R}^{H' \times W'}$ according to the query space-time coordinates $(x,y,t)$. $\hat{M}_{0 \rightarrow t}^H$ specifies the forward motion of the features in $F_{0}^H$ and is utilized to forward warp $F_{0}^H$ to $F_t^H$ (Section~\ref{subsec:estimate_explicit_motion}). The same motion encoding, rendering and warping processes are repeated for $I_{1}^L$, in aggregating temporally the information from all the reference frames. Lastly, we follow~\cite{softmax} to perform softmax splatting to create $F_t^H$ and $Z_t^H$, which are further combined with $F_{(0,1)}^H$ to decode the high-resolution video frame $\hat{I}_t^H$ at time $t$ (Section~\ref{subsec:multi_frame_forward_warping}). $Z_t^H$ indicates how good $F_t^H$ is across spatial locations. It is used to condition the pixel-based decoding of the RGB values from $F_t^H$ and $F_{(0,1)}^H$.

\begin{figure}[t!]
  \centering
  \includegraphics[width=0.6\linewidth]{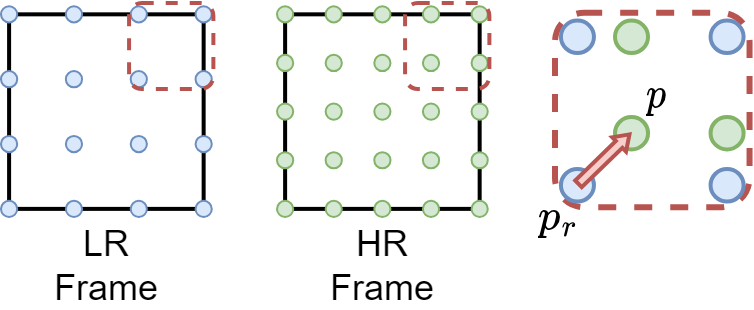}
  \vspace{-0.1mm}
  \caption{
      Illustration of low-resolution coordinates (blue dots) and high-resolution coordinates (green dots).
  }
  \vspace{-.6em}
  \label{fig:coordinate}
\end{figure}
\subsection{Space-time Local Implicit Neural Functions}
\label{subsec:estimate_explicit_motion}
The very core of our C-STVSR scheme is the space-time local implicit neural function (ST-INF) in Fig.~\ref{fig:overview}. Our ST-INF has the striking feature of predicting forward motion rather than backward motion. That is, it specifies how the feature at coordinates $p=(x,y)$ in $F_0^H$ or $F_1^H$ are propagated temporally to any designated time $t$. The forward motion is represented in the form of displacement vectors along with their reliability values. For example, to get the forward motion $\hat{M}^H_{0 \rightarrow t}(p)$ and its reliability value $\hat{Z}^H_{0 \rightarrow t}(p)$ for propagating the feature $F_0^H(p)$ of $F_0^H$ at $p=(x,y)$, it is queried as follows:
\begin{equation}
\{\hat{Z}^H_{t_r \rightarrow t}(p), \hat{M}^H_{t_r \rightarrow t}(p)\} = f_\theta(v_r,p-p_r,t-t_r),
\label{eq:ST-INF}
\end{equation}
where $v_r=T_0^L(p_r)$ is the motion latent at $p_r=(x_r,y_r)$ that is nearest to the query coordinates $p=(x,y)$, $t_r=0$ is the temporal location where the reference frame $I_{0}^L$ sits, and $\theta$ is the network parameters. Fig.~\ref{fig:coordinate} depicts an example of the geometrical relationship between $p$ and $p_r$. The sum $p+\hat{M}^H_{0 \rightarrow t}(p)$ gives the landing location of the query feature $F_0^H(p)$ at time $t$. %In addition to $\hat{M}^H_{0 \rightarrow t}(p)$, $f_\theta$ also returns its reliability value $\hat{Z}^H_{0 \rightarrow t}(p)$. 
In much the same way, $\hat{M}^H_{1 \rightarrow t}(p)$ and $\hat{Z}^H_{1 \rightarrow t}(p)$ for propagating the feature $F_1^H(p)$ can be obtained by having in Eq.~\eqref{eq:ST-INF} $v_r=T_1^L(p_r)$ and $t_r=1$, i.e. the temporal location of $I_{1}^L$.

In Eq.~\eqref{eq:ST-INF}, both $p=(x,y)$ and $t$ can take any values. Together they can refer to any space-time coordinates. Therefore, $f_\theta$ is able to generate forward motion in a continuous manner to warp $F_0^H,F_1^H$ of any spatial resolution to any time instance $t \in [0,1]$. However, in essence, $f_\theta$ is a local function that predicts forward motion in the vicinity of the reference space-time coordinates $p_r,t_r$ by referring to the local motion latent $v_r$.

\vspace{-1em}
\paragraph{Learning motion trajectories.}
Learning forward motion can be interpreted as learning motion trajectories along the temporal axis. To see this, in Eq.~\eqref{eq:ST-INF}, we fix $p=(x,y)$ at some coordinates, e.g.~$p_2$ in Fig.~\ref{fig:forward_backward} (b), take $t_r=0$, and view $f_\theta$ as a function of time $t$. With this setting, the forward motion predicted by $f_\theta$ specifies a displacement vector indicating where $p_2$ should appear at time $t$. Collectively, the displacement vectors evaluated at different time instances $t$'s define the motion trajectory of $p_2$. Generally, this motion trajectory is a smooth function of time and is relatively easier to approximate. While it is completely feasible to change the output semantics of $f_\theta$ to learn backward motion, \textcolor{black}{the resulting time function can be discontinuous.} The reason is illustrated in Fig.~\ref{fig:forward_backward} (a), where fixing the query coordinates $p$ at $p_2$, $f_\theta$ returns at every time instance $t$ a backward displacement vector identifying the location of the matching pixel/feature in the reference frame at $t_r=0$. In this case, the displacement vectors evaluated for the same $p_2$ yet at different time instances $t$'s may correspond to the distinct motion trajectories of different matching pixels. This suggests that $f_\theta$ has to model a less smooth function of time. \textcolor{black}{Section~\ref{sec:ablation} presents an ablation study to justify the use of forward motion.}
%we argue (and analyze in Section~\ref{subsec:forward_backward}) that the resulting time function is less smooth

\vspace{-1em}
\paragraph{Learning motion latents.}
Predicting the forward motion of a pixel (or a feature vector) at any given $p=(x,y)$ and for any $t$ is a non-trivial task. We formulate the problem as learning a $f_\theta$ that interpolates between forward motion sampled sparsely in both the spatial and temporal dimensions. This is achieved by providing $f_\theta$ with the motion latent that encodes the sparsely sampled forward motion as the contextual input. Take Eq.~\eqref{eq:ST-INF} as an example, where $f_\theta$ is queried to predict the forward motion of $F_0^H(p)$ for time $t$. The prediction is conditioned on the nearest motion latent $T_0^L(p_r)$, which captures the forward motion $M_{0 \rightarrow 1}^L$ estimated from $I_{0}^L$ to $I_{1}^L$ in the vicinity of $p_r$. In this work, we adopt Raft-lite~\cite{Raft} to estimate the forward optical flow map $M_{0 \rightarrow 1}^L$. Recognizing that the flow estimation is often not perfect, we follow~\cite{Splatting_based} to quantify the reliability of the resulting flow map $M_{0 \rightarrow 1}^L$ based on three metrics, including (1) the intensity warping error, (2) the flow warping error, and (3) the local variances of the flow map. Further details of these metrics are provided in the supplementary document. The reliability evaluation with each of these metrics yields a real-valued map of size the same as $M_{0 \rightarrow 1}^L$, reflecting the reliability of $M_{0 \rightarrow 1}^L$ across spatial locations. These maps are concatenated channel-wisely to form $Z_{0 \rightarrow 1}^L$, which is encoded jointly with $M_{0 \rightarrow 1}^L$ by the motion encoder $E_M$ as $T_0^L$. Section~\ref{sec:ablation} shows that $Z_{0 \rightarrow 1}^L$ benefits $f_\theta$ considerably in interpolating forward motion.

\subsection{Multi-Frame Forward Warping}
\label{subsec:multi_frame_forward_warping}
To come up with a prediction of $F_t^H$ for decoding a high-resolution video frame $\hat{I}_t^H$ at time $t$, we aggregate temporally $F_0^H,F_1^H$, each of which represents the high-resolution feature of a reference frame (Fig.~\ref{fig:overview}). Inspired by ~\cite{softmax}, we adopt softmax splatting to resolve the potential issue that multiple features from $F_0^H$, $F_1^H$ or both may be forward warped to the same location in $F_t^H$. Considering that our task is to interpolate and super-resolute a new frame from the ground up, we perform softmax splatting after $F_0^H$, $F_1^H$ have both been forward warped to time $t$. Our approach differs from~\cite{softmax}, which targets video frame interpolation and applies softmax splatting separately to individual reference frames for late fusion. In symbols, we have
\begin{equation}
\label{eq:warp_F}
{F}_t^H(p)=\sum\limits_{i=0}^{1}\sum\limits_{q}\frac{b(u) \cdot \exp{(\alpha\cdot\hat{Z}_{i \rightarrow t}^H(q))} \cdot F_{i}^H(q)}
{\sum\limits_{i=0}^{1}\sum\limits_{q}b(u) \cdot \exp{(\alpha\cdot\hat{Z}_{i \rightarrow t}^H(q))}},
\end{equation}
where the feature ${F}_t^H(p)$ of ${F}_t^H$ at $p$ is formulated as a weighted sum of all the reference features ${F}_0^H(q),{F}_1^H(q)$, with the weighting determined by the distance $u=p-(q+\hat{M}^H_{i \rightarrow t}(q))$, the bilinear kernel $b(u)=\max(0,1-|u_x|) \cdot \max(0,1-|u_y|)$, as well as the reliability $\hat{Z}_{i \rightarrow t}^H(q)$ of the forward motion at $q$. $\alpha=-20$ is the temperature of the softmax operation. Since the bilinear kernel has a finite support, only those ${F}_0^H(q),{F}_1^H(q)$ warped to the neighborhood of $p$ will actually contribute to the evaluation of ${F}_t^H(p)$. 

Additionally, we generate a map $Z_t^H$ to indicate how good ${F}_t^H$ is across spatial locations. Intuitively, if ${F}_t^H(p)$ is synthesized from those ${F}_0^H(q),{F}_1^H(q)$ whose forward motion is unreliable, the quality of ${F}_t^H(p)$ should be downgraded. $Z_t^H(p)$ serves as a conditioning factor for decoding the RGB values at $p$, and is obtained by 
\begin{equation}
\label{eq:warp_Z}
Z_t^H(p)=\max\limits_{i=0,1}\max\limits_{q} b(u) \cdot \exp{(\alpha\cdot\hat{Z}_{i\xrightarrow[]{}t}^H(q))},
\end{equation} which takes the maximum value among the (unnormalized) contributing weights from ${F}_0^H(q),{F}_1^H(q)$. When none of these contributing ${F}_0^H(q),{F}_1^H(q)$ has reliable forward motion, the quality of ${F}_t^H(p)$ is regarded as poor.

%\textcolor{blue}{Our approach is motivated partly by the principle of video super-resolution that the missing high-resolution sub-pixel information of a video frame could be recovered from observing simultaneously its neighboring frames.}
%Our approach and differs from~\cite{softmax}, which targets frame interpolation and  

%all the reference feature maps to obtain a single feature map $F_t^H$, which is more suitable for the STVSR scenario \textcolor{red}{that each pixel of the reference frames contributes to a sub-pixel of the target frame}. 

%Motion trajectories of different pixels from each reference frame could map to the same location at time instance $t$. To address this conflict issue, we adopt softmax splatting ~\cite{softmax} to aggregate the conflicting mappings by the reliability of each forward motion. Different from ~\cite{softmax}, we forward warp all the reference feature maps to obtain a single feature map $F_t^H$, which is more suitable for the STVSR scenario \textcolor{red}{that each pixel of the reference frames contributes to a sub-pixel of the target frame}. 

To synthesize a high-resolution video frame $\hat{I}_t^H$, we implement a pixel-wise decoder that incorporates a multi-layer perceptron. It decodes the RGB values at $p$ by taking as inputs $F_{t}^H(p)$, $F_{(0,1)}^H(p)$, $Z_{t}^H(p)$, and $t$ (the rightmost part of Fig.~\ref{fig:overview}). 

\subsection{Training Objective}
\label{subsec:training_objectives}
%We train our model end-to-end with the following objective:
We train our MoTIF end-to-end with the following objective:
\begin{equation}
\mathcal{L}= \mathcal{L}_{char}(\hat{I}_t^H,I_t^H) + \beta \sum_{i=0}^1 \mathcal{L}_{char}(\hat{M}_{i \rightarrow t}^H,{M}_{i \rightarrow t}^H),
\label{eq:loss}
\end{equation}
where $\mathcal{L}_{char}(\hat{x},x)=\sqrt{\|{\hat{x}-x}\|^2+\epsilon^2}$ is the Charbonnier loss~\cite{Charbonnier} and $\beta$ is a hyper-parameter. $\epsilon,\beta$ are set empirically to $10^{-3}$ and $0.01$, respectively. Our objective requires both the decoded frame $\hat{I}_t^H$ and the predicted forward motion $\hat{M}_{i \rightarrow t}^H$ to approximate their respective ground-truths.

\subsection{Comparison with Prior Works}
\label{subsec:novelty}
\textcolor{black}{Both our MoTIF and VideoINR~\cite{VideoINR} use implicit neural functions to tackle space-time video super-resolution (STVSR). As illustrated in Fig.~\ref{fig:teaser}, our MoTIF differs from VideoINR~\cite{VideoINR} in three significant ways:}

\textcolor{black}{First, for the C-STVSR task, our MoTIF uses \textit{forward} motion rather than \textit{backward} motion. This aspect in its own right has a significant impact on the quality of the generated videos (Section~\ref{sec:ablation} and Table~\ref{tab:forward v.s. backward}).}

\textcolor{black}{Second, our MoTIF models motion \textit{explicitly} rather than \textit{implicitly}. This allows our ST-INF to directly learn to interpolate between motion trajectories derived from a pre-trained optical flow estimation model. The supplementary document provides additional results, showing that our MoTIF can work well with well-behaved, off-the-shelf optical flow estimation networks. Using explicit motion also allows us to evaluate the reliability information $Z_{0 \rightarrow 1}^L,Z_{1 \rightarrow 0}^L$ based on the input video for better predicting $\hat{Z}_{0 \rightarrow t}^H,\hat{Z}_{1 \rightarrow t}^H$ (see Fig.~\ref{fig:overview}). %Section~\ref{sec:ablation} and Table~\ref{tab:explicit v.s. implicit} show that explicit motion modeling is an essential and integral part of our design. 
%we show that implicit motion modeling indeed causes performance drop and we also
%Section~\ref{sec:ablation} and Table~\ref{tab:explicit v.s. implicit} justify the benefit of this change.
}

\textcolor{black}{Third, our MoTIF introduces the reliability-aware splatting and decoding schemes, which are not seen in VideoINR~\cite{VideoINR}. Their benefits are studied in Section~\ref{sec:ablation} and Table~\ref{tab:ablation_components}.}

\textcolor{black}{Different from~\cite{softmax}, our reliability-aware splatting adopts early fusion of reference frames by forward warping all the reference features to the target frame according to Eq.~\eqref{eq:warp_F}. In contrast,~\cite{softmax} applies softmax splatting to each individual reference frame, followed by late fusing the results with a synthesis network. Our reliability-aware decoding scheme, which incorporates the reliability information for decoding (Eq.~\eqref{eq:warp_Z}) is not seen in~\cite{softmax}}.

\section{Experiments}

\definecolor{ForestGreen}{RGB}{34,139,34}
\begin{table*}
  \caption{Performance comparison on the F-STVSR task. \textcolor{red}{Red}, \textcolor{blue}{blue}, and \textcolor{ForestGreen}{green} indicate the best, the second best, and the third best performance, respectively. Quality metrics: PSNR/SSIM.}
  %\caption{Comparison with state-of-the-art methods at fixed scale.The best three results are highlighted in \textcolor{red}{red}, \textcolor{blue}{blue}, and \textcolor{ForestGreen}{bold}.}
  \fontsize{8}{8}\selectfont
  \centering
  \begin{tabular}{@{}cc|ccccccccccc@{}}
    \toprule
        VFI &
        VSR &
        \multicolumn{1}{c}{Vid4} &
        \multicolumn{1}{c}{GoPro-\emph{Center}} &
        \multicolumn{1}{c}{GoPro-\emph{Average}} &
        \multicolumn{1}{c}{Adobe-\emph{Center}} &
        \multicolumn{1}{c}{Adobe-\emph{Average}} &
        Parameters \\
        Method &
        Method &
        &%PSNR / SSIM &
        &%PSNR / SSIM &
        &%PSNR / SSIM &
        &%PSNR / SSIM &
        &%PSNR / SSIM &
        (Millions) \\
    \midrule
        SuperSloMo~\cite{SuperSloMo} &
        Bicubic &
        22.42 / 0.5645 &
        27.04 / 0.7937 &
        26.06 / 0.7720 &
        26.09 / 0.7435 &
        25.29 / 0.7279 &
        19.8 \\
        SuperSloMo~\cite{SuperSloMo} &
        EDVR\cite{EDVR} &
        23.01 / 0.6136 &
        28.24 / 0.8322 &
        26.30 / 0.7960 &
        27.25 / 0.7972 &
        25.95 / 0.7682 &
        19.8+20.7 \\
        SuperSloMo~\cite{SuperSloMo} &
        BasicVSR~\cite{BasicVSR} &
        23.17 / 0.6159 &
        28.23 / 0.8308 &
        26.36 / 0.7977 &
        27.28 / 0.7961 &
        25.94 / 0.7679 &
        19.8+6.3 \\
    \midrule
        QVI~\cite{QVI} &
        Bicubic~\cite{EDVR} &
        22.11 / 0.5498 &
        26.50 / 0.7791 &
        25.41 / 0.7554 &
        25.57 / 0.7324 &
        24.72 / 0.7114 &
        29.2 \\
        QVI~\cite{QVI} &
        EDVR~\cite{EDVR} &
        23.60 / 0.6471 &
        27.43 / 0.8081 &
        25.55 / 0.7739 &
        26.40 / 0.7692 &
        25.09 / 0.7406 &
        29.2+20.7 \\
        QVI~\cite{QVI} &
        BasicVSR~\cite{BasicVSR} &
        23.15 / 0.6428 &
        27.44 / 0.8070 &
        26.27 / 0.7955 &
        26.43 / 0.7682 &
        25.20 / 0.7421 &
        29.2+6.3 \\
    \midrule
        DAIN~\cite{DAIN} &
        Bicubic &
        22.57 / 0.5732 &
        26.92 / 0.7911 &
        26.11 / 0.7740 &
        26.01 / 0.7461 &
        25.40 / 0.7321 &
        24.0 \\
        DAIN~\cite{DAIN} &
        EDVR~\cite{EDVR} &
        23.48 / 0.6547 &
        28.01 / 0.8239 &
        26.37 / 0.7964 &
        27.06 / 0.7895 &
        26.01 / 0.7703 &
        24.0+20.7 \\
        DAIN~\cite{DAIN} &
        BasicVSR~\cite{BasicVSR} &
        23.43 / 0.6514 &
        28.00 / 0.8227 &
        26.46 / 0.7966 &
        27.07 / 0.7890 &
        26.23 / 0.7725 &
        24.0+6.3 \\
    \midrule
        \multicolumn{2}{c|}{Zooming SlowMo~\cite{ZSM}} &
        25.72 / 0.7717  &
        \textcolor{ForestGreen}{30.69} / \textcolor{ForestGreen}{0.8847}  &
        - / - &
        \textcolor{blue}{30.26} / \textcolor{blue}{0.8821}  &
        - / -  &
        11.10 \\
        \multicolumn{2}{c|}{TMNet~\cite{TMNet}} &
        \textcolor{red}{25.96} / \textcolor{red}{0.7803} &
        30.14 / 0.8692 &
        \textcolor{ForestGreen}{28.83} / \textcolor{ForestGreen}{0.8514} &
        29.41 / 0.8524 &
        \textcolor{ForestGreen}{28.30} / \textcolor{ForestGreen}{0.8354} &
        12.26 \\
    \midrule
        \multicolumn{2}{c|}{VideoINR-\emph{fixed}~\cite{VideoINR}} &
        \textcolor{ForestGreen}{25.78} / \textcolor{ForestGreen}{0.7730} &
        \textcolor{blue}{30.73} / \textcolor{blue}{0.8850} &
        - / - &
        \textcolor{ForestGreen}{30.21} / \textcolor{ForestGreen}{0.8805} &
        - / - &
        11.31 \\
        \multicolumn{2}{c|}{VideoINR~\cite{VideoINR}} &
        25.61 / 0.7709 &
        30.26 / 0.8792 &
        \textcolor{blue}{29.41} / \textcolor{blue}{0.8669} &
        29.92 / 0.8746 &
        \textcolor{blue}{29.27} / \textcolor{blue}{0.8651} &
        11.31 \\
    \midrule
        \multicolumn{2}{c|}{Ours} &
        \textcolor{blue}{25.79} / \textcolor{blue}{0.7745} &
        \textcolor{red}{31.04} / \textcolor{red}{0.8877} &
        \textcolor{red}{30.04} / \textcolor{red}{0.8773} &
        \textcolor{red}{30.63} / \textcolor{red}{0.8839} &
        \textcolor{red}{29.82} / \textcolor{red}{0.8750} &
        12.55 \\
        %\multicolumn{2}{c|}{Ours$^*$} &
        %25.78 & 0.7737 &
        %31.44 & 0.9003 &
        %30.77 & 0.8948 &
        %31.03 & 0.8919 &
        %30.37 & 0.8849 &
        %12.55 \\
    \bottomrule
  \end{tabular}
  \label{tab:comparison_fixed}
\end{table*}
\begin{table*}
  \caption{\textcolor{black}{PSNR/SSIM performance} comparison on the C-STVSR task (on Gopro). \textbf{Bold} indicates the best performance.}% Quality metrics: PSNR/SSIM.}
  %\caption{Comparison with state-of-the-art methods at continuous scales.}
  \fontsize{8}{8}\selectfont
  \centering
  \begin{tabular}{@{}cc|ccccc@{}}
    \toprule
        Temporal Scale
        & 
        Spatial Scale
        & 
        %VFI &
        %VSR &
        SuperSloMo~\cite{SuperSloMo} +  LIIF~\cite{LIIF} 
        & 
        DAIN~\cite{DAIN} + LIIF~\cite{LIIF} 
        & 
        TMNet~\cite{TMNet} &
        VideoINR~\cite{VideoINR} &
        Ours \\
    \midrule
    \midrule
        $\times6$ & 
        $\times4$ &
        %- &
        %- &
        26.70 / 0.7988 &
        26.71 / 0.7998 &
        30.49 / 0.8861 &
        30.78 / 0.8954 &
        \textbf{31.56} / \textbf{0.9064}
        \\
        $\times6$ & 
        $\times6$ &
        %- &
        %- &
        23.47 / 0.6931 &
        23.36 / 0.6902 &
        - &
        25.56 / 0.7671 &
        \textbf{29.36} / \textbf{0.8505}
        \\
        $\times6$ & 
        $\times12$ &
        %- &
        %- &
        21.92 / 0.6495 &
        22.01 / 0.6499 &
        - &
        24.02 / 0.6900 &
        \textbf{25.81} / \textbf{0.7330}
        \\
    \midrule
        $\times12$ & 
        $\times4$ &
        %- &
        %- &
        25.07 / 0.7491 &
        25.14 / 0.7497 &
        26.38 / 0.7931 &
        27.32 / 0.8141 &
        \textbf{27.77} / \textbf{0.8230}
        \\
        $\times12$ & 
        $\times6$ &
        %- &
        %- &
        22.91 / 0.6783 &
        22.92 / 0.6785 &
        - &
        24.68 / 0.7358 &
        \textbf{26.78} / \textbf{0.7908}
        \\
        $\times12$ & 
        $\times12$ &
        %- &
        %- &
        21.61 / 0.6457 &
        21.78 / 0.6473 &
        - &
        23.70 / 0.6830 &
        \textbf{24.72} / \textbf{0.7108}
        \\
    \midrule
        $\times16$ & 
        $\times4$ &
        %- &
        %- &
        24.42 / 0.7296 &
        24.20 / 0.7244 &
        24.72 / 0.7526 &
        25.81 / 0.7739 &
        \textbf{25.98} / \textbf{0.7758}
        \\
        $\times16$ & 
        $\times6$ &
        %- &
        %- &
        23.28 / 0.6883 &
        22.80 / 0.6722 &
        - &
        23.86 / 0.7123 &
        \textbf{25.34} / \textbf{0.7527}
        \\
        $\times16$ & 
        $\times12$ &
        %- &
        %- &
        21.80 / 0.6481 &
        22.22 / 0.6420 &
        - &
        22.88 / 0.6659 &
        \textbf{23.88} / \textbf{0.6923}
        \\
    \midrule
    \midrule
        $\times6$ & 
        $\times1$ &
        %xx.xx / 0.xxxx &
        %- &
        - &
        - &
        - &
        32.34 / 0.9545 &
        \textbf{34.77} / \textbf{0.9696}
        \\
    \midrule
        $\times1$ & 
        $\times4$ &
        %- &
        %xx.xx / 0.xxxx &
        - &
        - &
        33.02 / 0.9206 &
        32.26 / 0.9198 &
        \textbf{33.84} / \textbf{0.9328}
        \\
    \bottomrule
  \end{tabular}
  \label{tab:comparison_continuous}
  
  %\vspace{-1.em}
\end{table*}
%\subsection{Experimental Setup}
%\label{subsec:exprimental_setup}
\textcolor{black}{To our best knowledge, VideoINR~\cite{VideoINR} is the only prior work that addresses specifically C-STVSR. We thus follow its training and test protocols, unless otherwise specified. VideoINR~\cite{VideoINR} is also included as the major baseline method.}
%For experiments, we follow the training and test protocol specified in VideoINR~\cite{VideoINR}, unless otherwise specified.  

\vspace{-1em}
%\noindent\textbf{Training Datasets.}
\paragraph{Training Datasets.}
We train our model on Adobe240 dataset\cite{Adobe240}, which contains 133 720P hand-held videos. Of these videos, 100 are used for training, 16 for validation, and 17 for test. In each video, we take 9 consecutive frames to form a training sample, where the 1\textsuperscript{st} and 9\textsuperscript{th} frames are bicubic down-sampled and used as the low-resolution, low-frame-rate input. 

\vspace{-1em}
%\noindent\textbf{Evaluation.} 
\paragraph{Evaluation.} 
We compare the competing methods on Vid4~\cite{Vid4}, Adobe240~\cite{Adobe240}, and Gopro~\cite{Gopro} datasets. Unless otherwise specified, the spatial scaling factor defaults to 4. On Vid4, the temporal scaling factor is fixed at 2 to test single-frame interpolation. On Adobe240-\textit{average} and Gopro-\textit{average}, the temporal scaling factor is set to 8 for multi-frame interpolation. Under the same setting, we also report results on Adobe240-\textit{center} and Gopro-\textit{center} for only the 1\textsuperscript{st}, 4\textsuperscript{th} and 9\textsuperscript{th} frames (namely, single-frame interpolation).

\vspace{-1em}
%\noindent\textbf{Baselines and Quality Metrics.}
\paragraph{Baselines and Quality Metrics.}
The baseline methods include (1) two-stage F-STVSR methods, namely video frame interpolation (SuperSloMo~\cite{SuperSloMo}, QVI~\cite{QVI}, DAIN~\cite{DAIN}) plus video super-resolution (Bicubic Interpolation, EDVR~\cite{EDVR}, BasicVSR~\cite{BasicVSR}); (2) one-stage F-STVSR methods (Zooming SloMo~\cite{ZSM}); (3) two-stage C-STVSR methods, namely continuous video frame interpolation (SuperSloMo~\cite{SuperSloMo}, DAIN~\cite{DAIN}) plus continuous image super-resolution (LIIF~\cite{LIIF}); (4) one-stage C-STVSR methods (VideoINR~\cite{VideoINR}); and (5) TMNet~\cite{TMNet}. The quality metrics are Peak Signal-to-Noise Ratio (PSNR) and Structural Similarity Index (SSIM) on the Y channel.

\vspace{-1em}
%\noindent\textbf{Implementation and Training Details.}
\paragraph{Implementation and Training Details.}
We adopt the same two-stage training strategy as VideoINR~\cite{VideoINR}. The spatial scaling factor is set to 4 for the first \textcolor{black}{450,000} iterations, and is chosen uniformly from $[1,4]$ in the following \textcolor{black}{150,000} iterations. The training batch size is 24; within each batch, every input frame is down-sampled spatially by the same factor and cropped to $32 \times 32$. For training stability, we use the ground-truth forward motion in place of the predicted forward motion with a certain probability, the value of which is attenuated from 1 to 0 in the first 150,000 iterations. %Our framework adopts Zooming SloMo\cite{ZSM} as encoder and Raft-lite\cite{Raft} as motion estimator.
We adopt Adam optimizer with $\beta_1=0.9,\beta_2=0.999$ and cosine annealing to decay the learning rate from $10^{-4}$ to $10^{-7}$ for every 150,000 iterations. For data augmentation, we perform random rotation and horizontal-flipping. Both our S-INF and ST-INF (Fig.~\ref{fig:overview}) are implemented with 3-layer SIRENs~\cite{SIREN}, the hidden dimensions of which are 64, 64, and 256. More network details are in the supplementary document.

\subsection{Comparison with State-of-the-art Methods}
%\paragraph{Quantitative Results.} 
Table~\ref{tab:comparison_fixed} presents qualitative results, comparing the  competing methods on the F-STVSR task. Both VideoINR~\cite{VideoINR} and our MoTIF are trained for C-STVSR, whereas the other methods are trained for F-STVSR \textcolor{black}{and their results are excerpted from~\cite{VideoINR}}. Notably, VideoINR-\textit{fixed} is trained specifically for single-frame interpolation. %(see Evaluation in Section~\ref{subsec:exprimental_setup}). 
%\textcolor{black}{The results of the other well-established prior works are excerpted from~\cite{VideoINR}.}
From Table~\ref{tab:comparison_fixed}, several observations can be made. (1) Our MoTIF outperforms VideoINR~\cite{VideoINR} in all the test cases. It also outperforms VideoINR-\textit{fixed}, although not trained for F-STVSR. (2) While both VideoINR~\cite{VideoINR} and our MoTIF adopt the same $E_I$ encoder from Zooming SloMo~\cite{ZSM}, VideoINR~\cite{VideoINR} performs worse than Zooming SloMo~\cite{ZSM} under the single-frame interpolation on Vid4, GoPro-\textit{Center}, and Adobe-\textit{Center}; on the contrary, our MoTIF is superior to Zooming SloMo~\cite{ZSM}. (3) On Vid4, our MoTIF performs slightly worse than TMNet~\cite{TMNet}. This may be because TMNet~\cite{TMNet} is trained on Vimeo-90K dataset~\cite{TOFlow}, which shares similar characteristics to Vid4. (4) All the one-stage methods (our MoTIF, \cite{TMNet,VideoINR,ZSM}) performs better than the two-stage methods (video frame interpolation plus video super-resolution) due to end-to-end optimization. (5) Our MoTIF (with Raft-lite~\cite{Raft} included) has a similar model size to VideoINR~\cite{VideoINR}. \textcolor{black}{Section~\ref{sec:complex_analysis} further shows that MoTIF (including Raft) has comparable or even lower GMACs than VideoINR, and thus similar or higher FPS}.%\textcolor{black}{In supplementary, we further compare our MoTIF with the state-of-the-art F-STVSR methods following the common test protocol~\cite{ZSM, TMNet} of the F-STVSR task to perform training with 4 reference frames.}

Table~\ref{tab:comparison_continuous} further presents results on the C-STVSR task, with most of the spatiotemproal scaling factors not seen during training. Except TMNet~\cite{TMNet}, which supports  continuous temporal scaling but only 4x spatial scaling, all the methods are able to achieve C-STVSR. Again, our MoTIF achieves the best performance in all the test cases, confirming its better generalization to unseen scaling factors.  Notably, on the video frame interpolation task (i.e.~temporal scale = 6 and spatial scale = 1), MoTIF outperforms VideoINR~\cite{VideoINR} by 2.5dB in terms of PSNR. This underlines the merit of using forward motion for better modeling. 

%and on video super-resolution (temporal scale = 1), to show our MoTIF is general to cover whole space-time space as we illustrate in Fig.~\ref{fig:releated}. For the video frame interpolation task our 

\begin{figure*}[ht!]
    \centering
    \begin{tabular}{cc}
         VideoINR~\cite{VideoINR}  &
        \begin{tabular}{cccccccc}
            \includegraphics[width=0.105\linewidth]{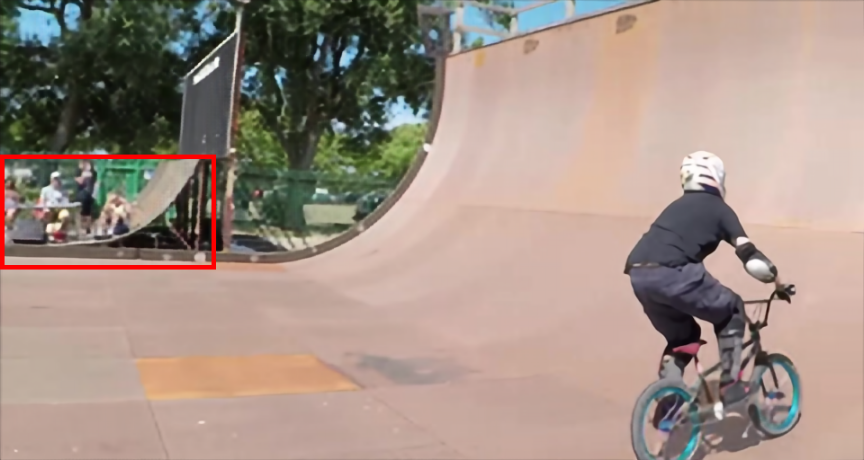}&
            \includegraphics[width=0.105\linewidth]{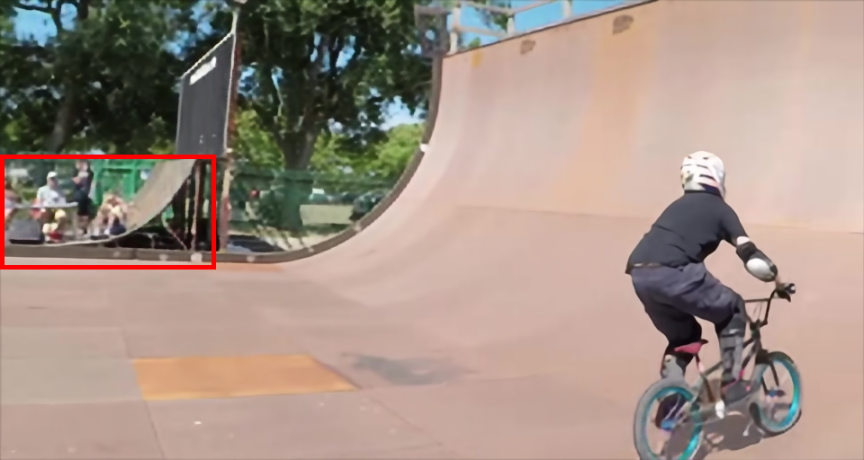}&
            \includegraphics[width=0.105\linewidth]{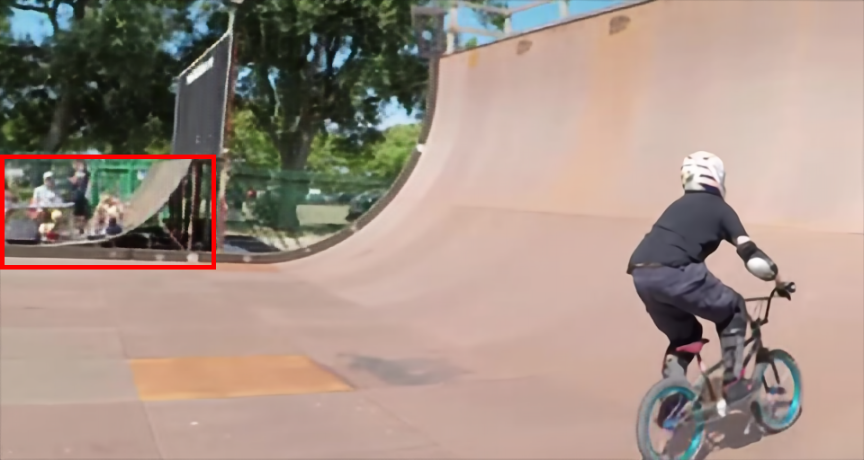}&
            \includegraphics[width=0.105\linewidth]{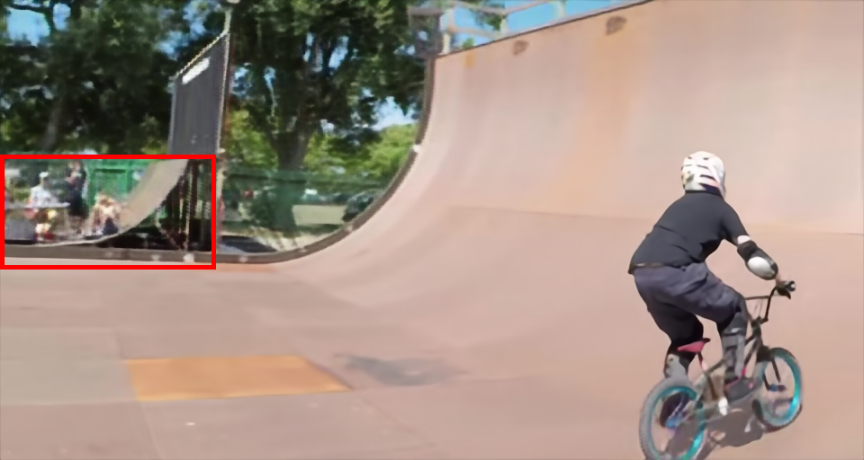}&
            \includegraphics[width=0.105\linewidth]{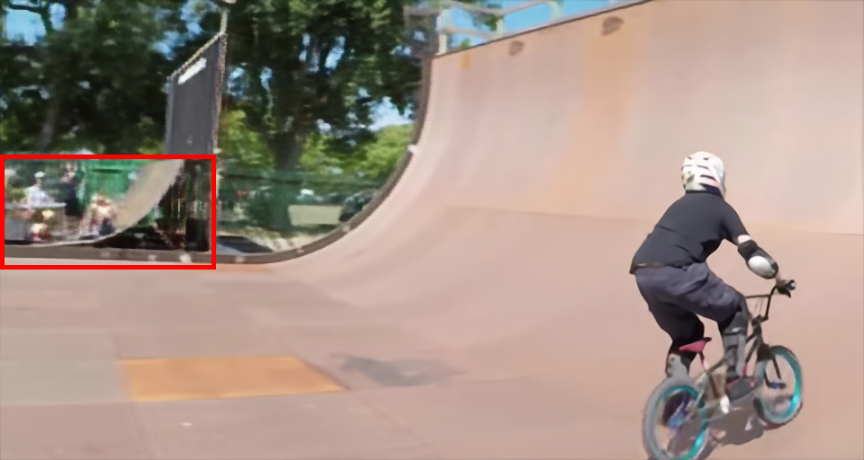}&
            \includegraphics[width=0.105\linewidth]{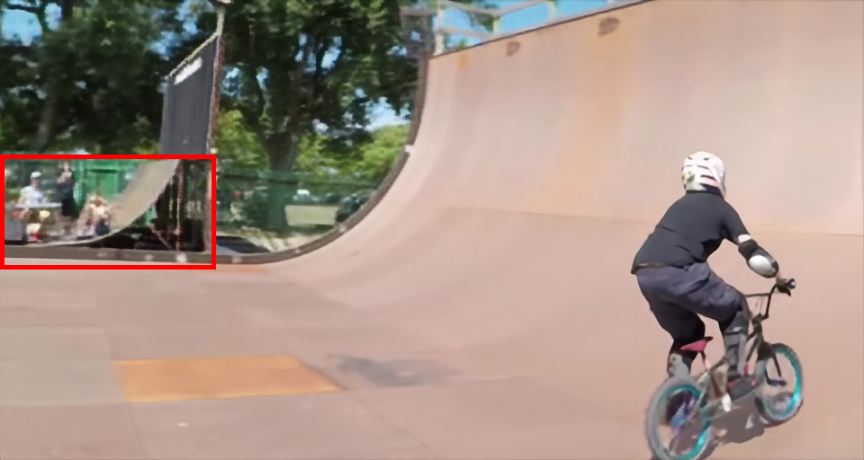}&
            \includegraphics[width=0.105\linewidth]{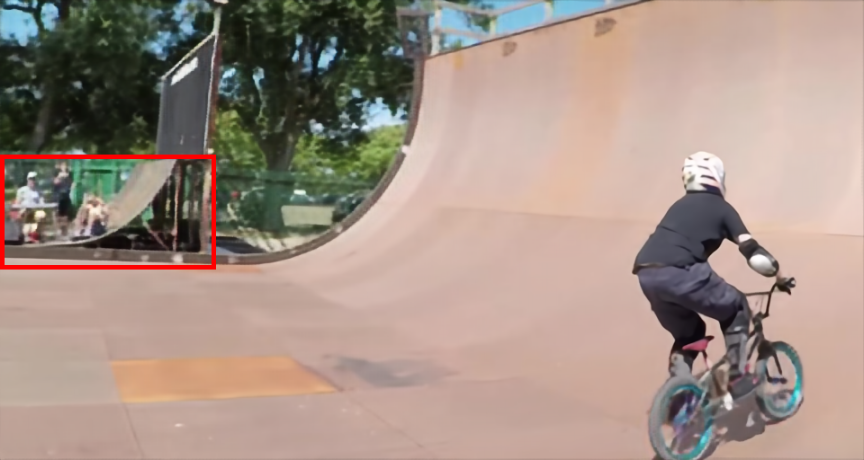}&
            \includegraphics[width=0.105\linewidth]{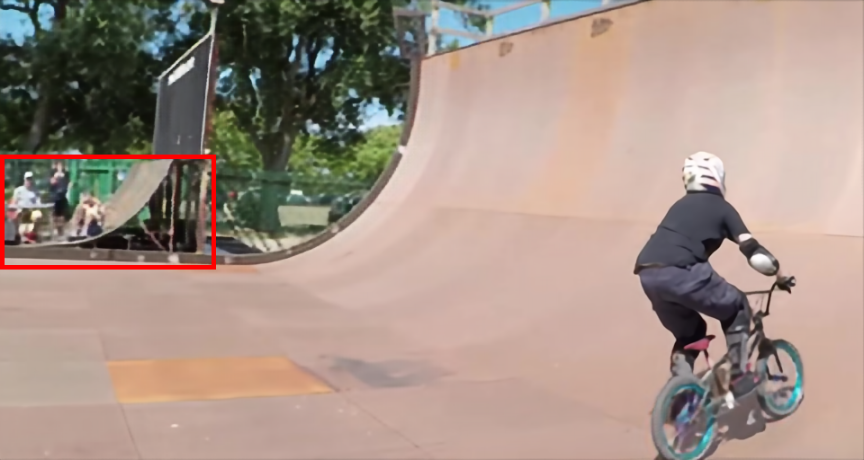}
            \\
            \includegraphics[trim={2 193 650 157}
            ,clip,width=0.105\linewidth]{Figures/quantitative_down/bmx/VideoINR/0_box.png}&
            \includegraphics[trim={2 193 650 157}
            ,clip,width=0.105\linewidth]{Figures/quantitative_down/bmx/VideoINR/1_box.png}&
            \includegraphics[trim={2 193 650 157}
            ,clip,width=0.105\linewidth]{Figures/quantitative_down/bmx/VideoINR/2_box.png}&
            \includegraphics[trim={2 193 650 157}
            ,clip,width=0.105\linewidth]{Figures/quantitative_down/bmx/VideoINR/3_box.png}&
            \includegraphics[trim={2 193 650 157}
            ,clip,width=0.105\linewidth]{Figures/quantitative_down/bmx/VideoINR/4_box.png}&
            \includegraphics[trim={2 193 650 157}
            ,clip,width=0.105\linewidth]{Figures/quantitative_down/bmx/VideoINR/5_box.png}&
            \includegraphics[trim={2 193 650 157}
            ,clip,width=0.105\linewidth]{Figures/quantitative_down/bmx/VideoINR/6_box.png}&
            \includegraphics[trim={2 193 650 157}
            ,clip,width=0.105\linewidth]{Figures/quantitative_down/bmx/VideoINR/7_box.png}
        \end{tabular}
        \\ \\ \\
        \begin{tabular}{c}
         Ours 
        \\ \\ \\ \\ \\ \\ \\ \\ \\ \\
        \end{tabular}
        &
        \begin{tabular}{cccccccc}
            \includegraphics[width=0.105\linewidth]{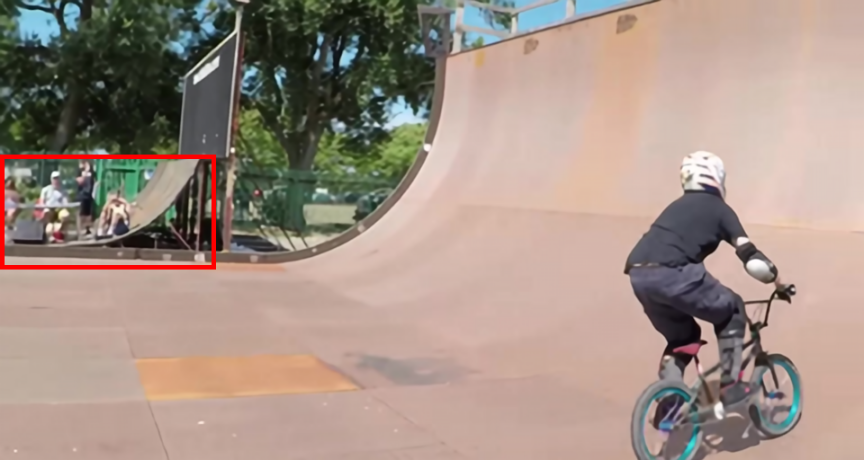}&
            \includegraphics[width=0.105\linewidth]{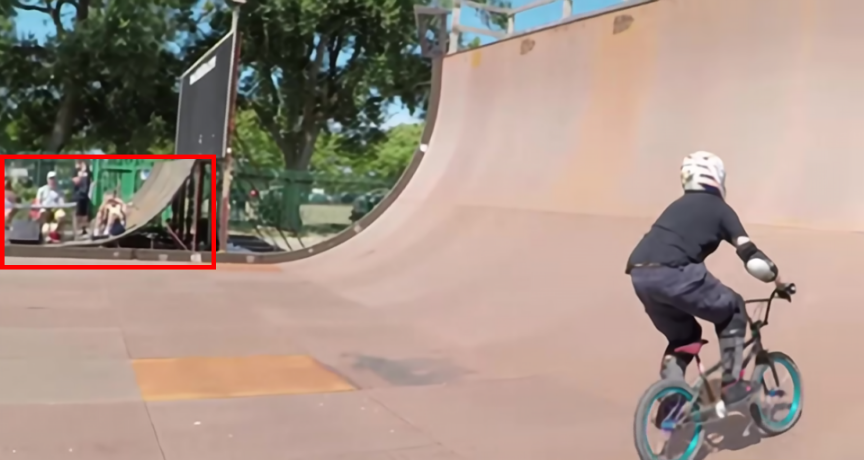}&
            \includegraphics[width=0.105\linewidth]{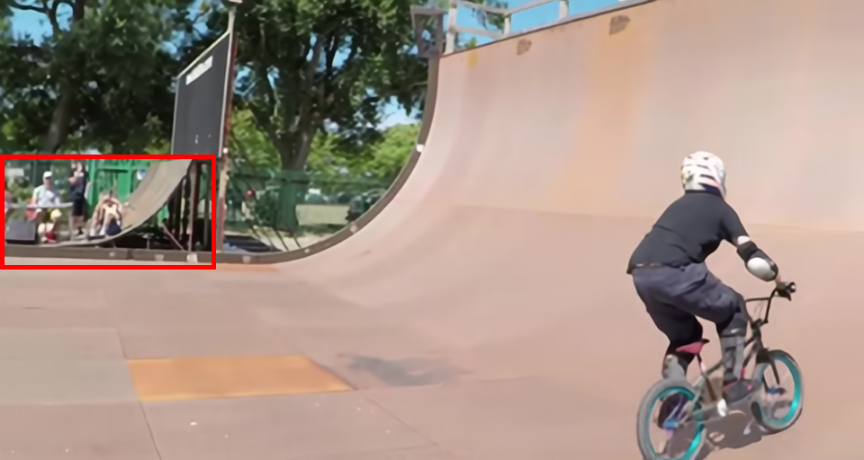}&
            \includegraphics[width=0.105\linewidth]{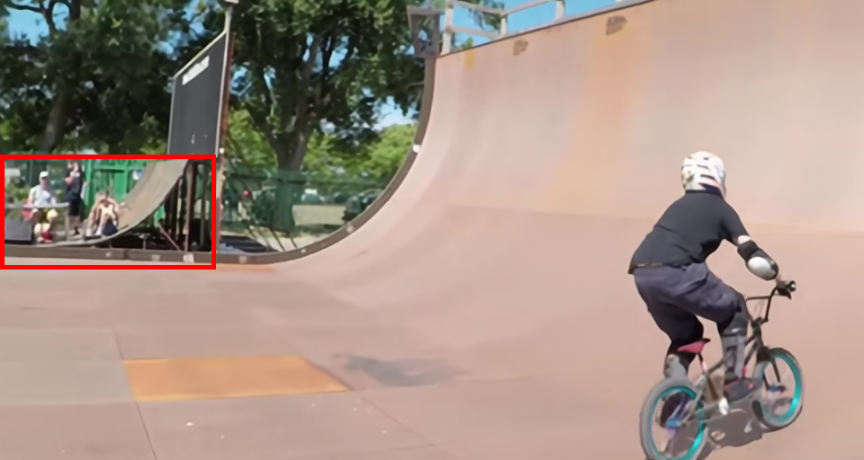}&
            \includegraphics[width=0.105\linewidth]{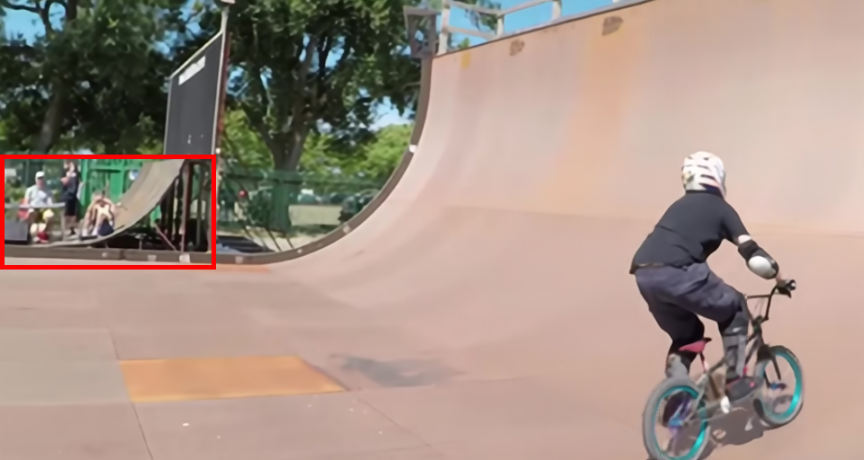}&
            \includegraphics[width=0.105\linewidth]{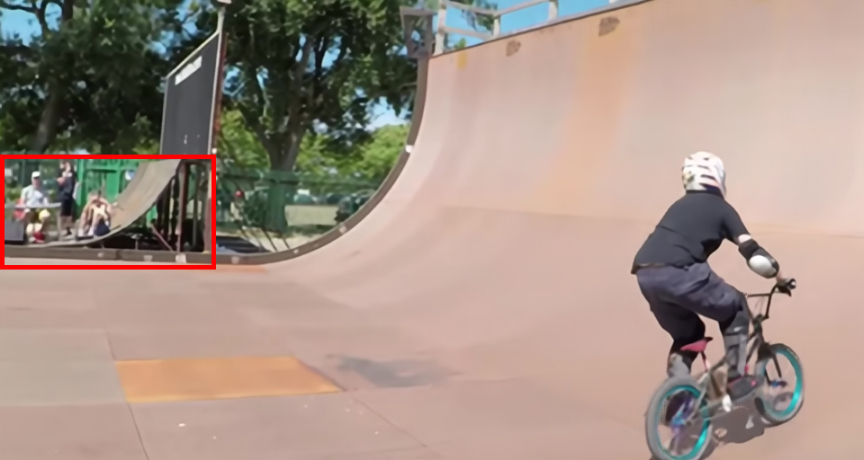}&
            \includegraphics[width=0.105\linewidth]{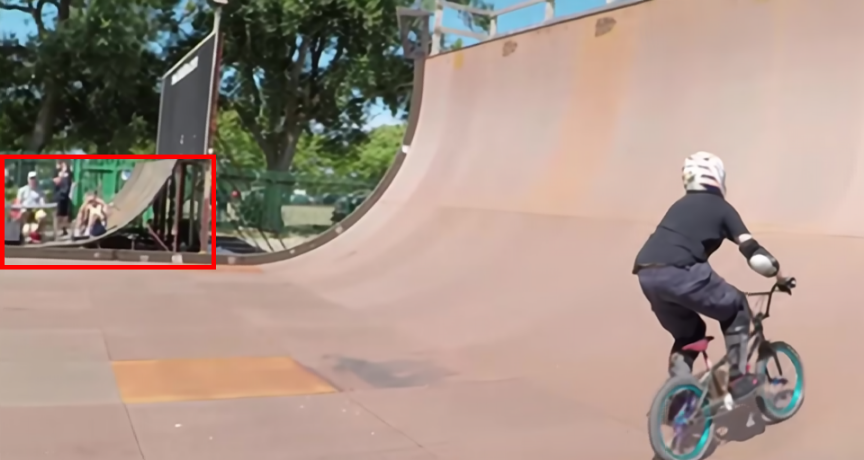}&
            \includegraphics[width=0.105\linewidth]{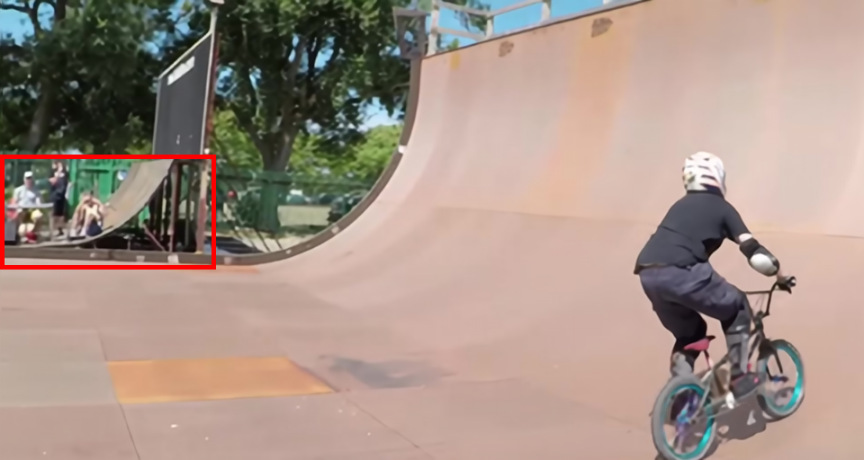}
            \\
            \includegraphics[trim={2 193 650 157}
            ,clip,width=0.105\linewidth]{Figures/quantitative_down/bmx/Ours/0_box.png}&
            \includegraphics[trim={2 193 650 157}
            ,clip,width=0.105\linewidth]{Figures/quantitative_down/bmx/Ours/1_box.png}&
            \includegraphics[trim={2 193 650 157}
            ,clip,width=0.105\linewidth]{Figures/quantitative_down/bmx/Ours/2_box.png}&
            \includegraphics[trim={2 193 650 157}
            ,clip,width=0.105\linewidth]{Figures/quantitative_down/bmx/Ours/3_box.png}&
            \includegraphics[trim={2 193 650 157}
            ,clip,width=0.105\linewidth]{Figures/quantitative_down/bmx/Ours/4_box.png}&
            \includegraphics[trim={2 193 650 157}
            ,clip,width=0.105\linewidth]{Figures/quantitative_down/bmx/Ours/5_box.png}&
            \includegraphics[trim={2 193 650 157}
            ,clip,width=0.105\linewidth]{Figures/quantitative_down/bmx/Ours/6_box.png}&
            \includegraphics[trim={2 193 650 157}
            ,clip,width=0.105\linewidth]{Figures/quantitative_down/bmx/Ours/7_box.png}
        %\end{tabular}

        \\
        \\
        \\
        \\
        
        %&
        %\begin{tabular}{cccccccc}
        T=0&T=0.125&T=0.25&T=0.375&T=0.5&T=0.625&T=0.75&T=0.875
        \end{tabular}
        
        \\
        \\
        \\
        \\
        \\
        \\
        
        % TMNet~\cite{TMNet}  &
        % \begin{tabular}{cccccc}
        %     \includegraphics[width=0.14\linewidth]{Figures/quantitative/motor/TMNet/0_box.png}&
        %     \includegraphics[width=0.14\linewidth]{Figures/quantitative/motor/TMNet/1_box.png}&
        %     \includegraphics[width=0.14\linewidth]{Figures/quantitative/motor/TMNet/2_box.png}&
        %     \includegraphics[width=0.14\linewidth]{Figures/quantitative/motor/TMNet/3_box.png}&
        %     \includegraphics[width=0.14\linewidth]{Figures/quantitative/motor/TMNet/4_box.png}&
        %     \includegraphics[width=0.14\linewidth]{Figures/quantitative/motor/TMNet/5_box.png}
        %     \\
        %     \includegraphics[trim={302 302 152 2}
        %     ,clip,width=0.14\linewidth]{Figures/quantitative/motor/TMNet/0_box.png}&
        %     \includegraphics[trim={302 302 152 2}
        %     ,clip,width=0.14\linewidth]{Figures/quantitative/motor/TMNet/1_box.png}&
        %     \includegraphics[trim={302 302 152 2}
        %     ,clip,width=0.14\linewidth]{Figures/quantitative/motor/TMNet/2_box.png}&
        %     \includegraphics[trim={302 302 152 2}
        %     ,clip,width=0.14\linewidth]{Figures/quantitative/motor/TMNet/3_box.png}&
        %     \includegraphics[trim={302 302 152 2}
        %     ,clip,width=0.14\linewidth]{Figures/quantitative/motor/TMNet/4_box.png}&
        %     \includegraphics[trim={302 302 152 2}
        %     ,clip,width=0.14\linewidth]{Figures/quantitative/motor/TMNet/5_box.png}
        % \end{tabular}
        %\\ \\ \\
         VideoINR~\cite{VideoINR} &
        \begin{tabular}{cccccc}
            \includegraphics[width=0.14\linewidth]{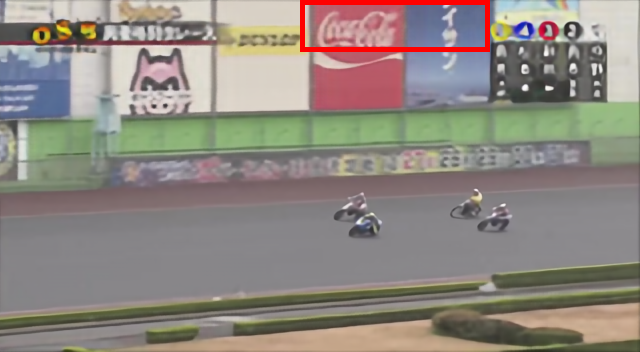}&
            \includegraphics[width=0.14\linewidth]{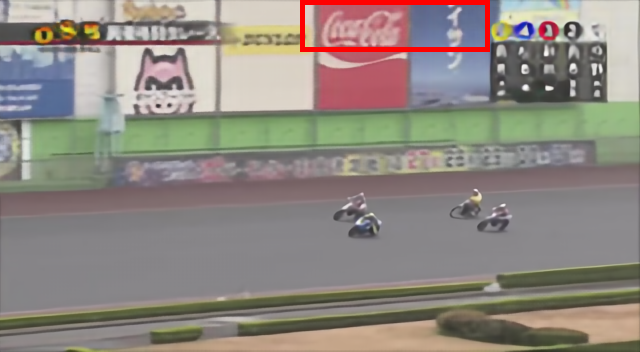}&
            \includegraphics[width=0.14\linewidth]{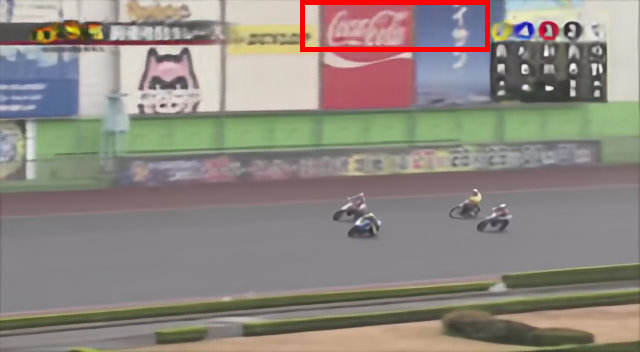}&
            \includegraphics[width=0.14\linewidth]{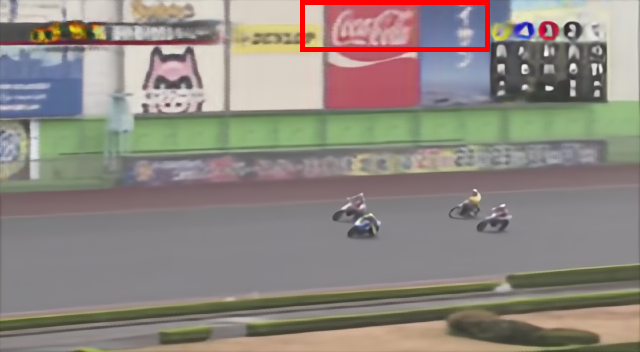}&
            \includegraphics[width=0.14\linewidth]{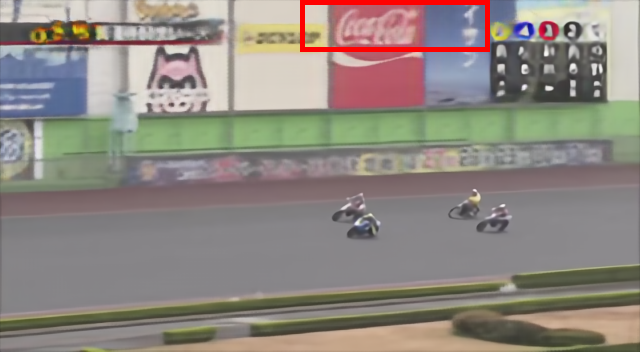}&
            \includegraphics[width=0.14\linewidth]{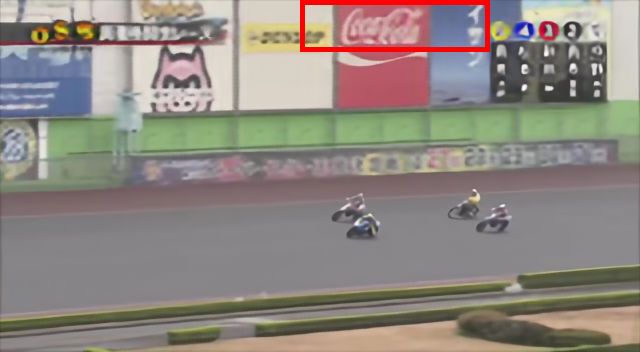}
            \\
            \includegraphics[trim={302 302 152 2}
            ,clip,width=0.14\linewidth]{Figures/quantitative/motor/VideoINR/0_box.png}&
            \includegraphics[trim={302 302 152 2}
            ,clip,width=0.14\linewidth]{Figures/quantitative/motor/VideoINR/1_box.png}&
            \includegraphics[trim={302 302 152 2}
            ,clip,width=0.14\linewidth]{Figures/quantitative/motor/VideoINR/2_box.png}&
            \includegraphics[trim={302 302 152 2}
            ,clip,width=0.14\linewidth]{Figures/quantitative/motor/VideoINR/3_box.png}&
            \includegraphics[trim={302 302 152 2}
            ,clip,width=0.14\linewidth]{Figures/quantitative/motor/VideoINR/4_box.png}&
            \includegraphics[trim={302 302 152 2}
            ,clip,width=0.14\linewidth]{Figures/quantitative/motor/VideoINR/5_box.png}
        \end{tabular}
        \\ \\ \\
        \begin{tabular}{c}
         Ours 
        \\ \\ \\ \\ \\ \\ \\ \\ \\ \\
        \end{tabular}
        &
        \begin{tabular}{cccccc}
            \includegraphics[width=0.14\linewidth]{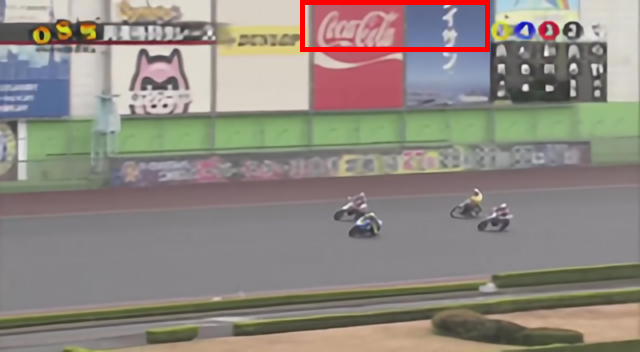}&
            \includegraphics[width=0.14\linewidth]{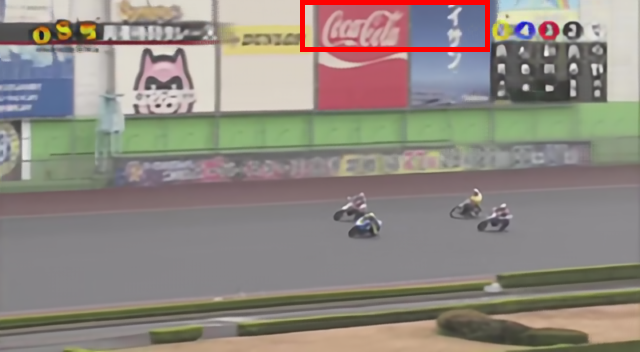}&
            \includegraphics[width=0.14\linewidth]{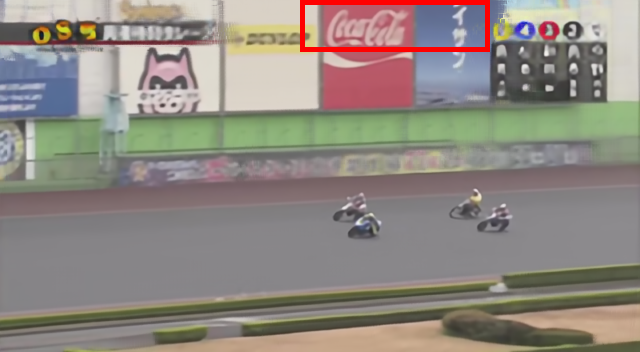}&
            \includegraphics[width=0.14\linewidth]{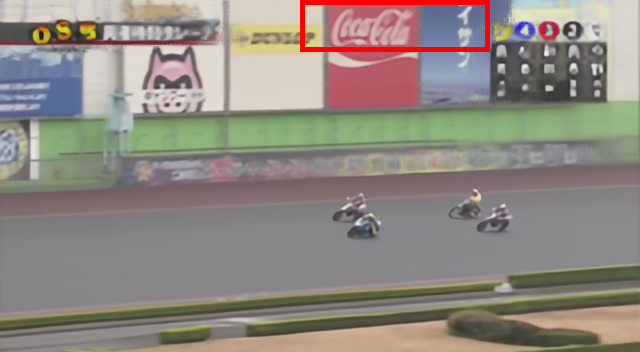}&
            \includegraphics[width=0.14\linewidth]{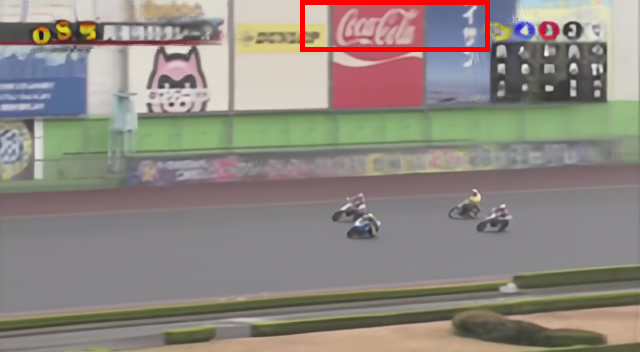}&
            \includegraphics[width=0.14\linewidth]{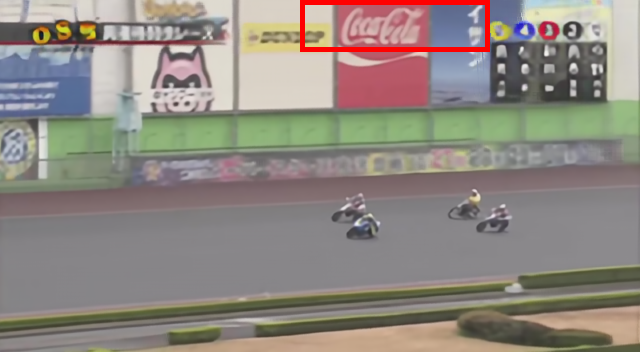}
            \\
            \includegraphics[trim={302 302 152 2}
            ,clip,width=0.14\linewidth]{Figures/quantitative/motor/Ours/0_box.png}&
            \includegraphics[trim={302 302 152 2}
            ,clip,width=0.14\linewidth]{Figures/quantitative/motor/Ours/1_box.png}&
            \includegraphics[trim={302 302 152 2}
            ,clip,width=0.14\linewidth]{Figures/quantitative/motor/Ours/2_box.png}&
            \includegraphics[trim={302 302 152 2}
            ,clip,width=0.14\linewidth]{Figures/quantitative/motor/Ours/3_box.png}&
            \includegraphics[trim={302 302 152 2}
            ,clip,width=0.14\linewidth]{Figures/quantitative/motor/Ours/4_box.png}&
            \includegraphics[trim={302 302 152 2}
            ,clip,width=0.14\linewidth]{Figures/quantitative/motor/Ours/5_box.png}
        %\end{tabular}
        \\
        \\
        \\
        \\
        %&
        %\begin{tabular}{cccccc}
        T=0 &T=0.17&T=0.33&T=0.50&T=0.67&T=0.83
        \end{tabular}
        \\
    \end{tabular}
    \caption{
        Subjective quality comparison. The temporal scaling factor of the upper example is 8 (in-distribution), whereas that of the lower example is 6 (out-of-distribution). Zoom in for better visualization.
    }
    %\vspace{-0.5em}
    \label{tbl:qualitative_results}
\end{figure*}

%\vspace{-1em}
%\paragraph{Qualitative Results.}
In Fig.~\ref{tbl:qualitative_results}, our MoTIF shows consistently better subjective quality than VideoINR~\cite{VideoINR}. More results are provided in the supplementary document.

\setlength{\tabcolsep}{0.1pt}
\renewcommand{\arraystretch}{0.1}

\renewcommand{\arraystretch}{1.}
\section{Complexity Comparison}
\label{sec:complex_analysis}
Table~\ref{tab:comparison_continuous} characterizes the complexity of the competing methods~\cite{ZSM,TMNet,VideoINR}. We follow~\cite{ZSM} to report frames per second (FPS) on Vid4~\cite{Vid4} dataset; that is, FPS is evaluated to be the ratio of the total number of output frames to the total runtime for processing the entire dataset. We also report the corresponding multiply-accumulate (MAC) operations per frame. These numbers are evaluated on one Tesla V100. From Table~\ref{tab:comparison_continuous}, our MoTIF has comparable FPS and GMACs to the other baseline methods on the lower-scale tasks (i.e. temporal scale = 2 and spatial scale = 4) while showing higher FPS and lower GMACs than the baseline methods on the tasks with higher temporal and spatial scales.

Note that when we increase the temporal scale while fixing the spatial scale, the FPS increases and the MAC per frame decreases. The same observation holds true for all the competing methods. This is because higher temporal scales invoke less frequent feature extraction to generate $F_0^L,F_1^L,F_{(0,1)}^L$. For example, a temporal scale of 16 (respectively, 2) implies that the feature extraction process is invoked only once every 16 frames (respectively, 2 frames). Given that the total number of frames to be processed is fixed, more frequent feature extraction leads to lower FPS and higher GMACs per frame. It is also seen that the complexity advantage of our MoTIF over VideoINR becomes more obvious when the temporal scale becomes higher and the spatial scale remains fixed. This is mainly because VideoINR backward warps the latent representations $F_{0}^H,F_{1}^H,F_{(0,1)}^H \in \mathbb{R}^{C \times H' \times W'}$ simultaneously and this operation is done twice. In comparison, our MoTIF forward warps the latent $F_{0}^H,F_{1}^H\in \mathbb{R}^{C \times H' \times W'}$ individually and this operation is done only once.

\begin{table*}
  \caption{Complexity comparison on the C-STVSR task. \textcolor{red}{Red} indicates the best performance. Complexity metrics: FPS ($\uparrow)$ / GMACs ($\downarrow$) per frame. The FPS and GMACs are evaluated based on processing the entire Vid4~\cite{Vid4} dataset on one Tesla V100.}
  %\caption{Comparison with state-of-the-art methods at continuous scales.}
  \setlength{\tabcolsep}{7.0pt}
  \fontsize{8}{8}\selectfont
  \centering
  \begin{tabular}{@{}cc|ccccc@{}}
    \toprule
        Temporal Scale & 
        Spatial Scale & 
        ZSM~\cite{ZSM} &
        TMNet\cite{TMNet} &
        VideoINR\cite{VideoINR} &
        Ours \\
    \midrule
    \midrule
        $\times2$ & 
        $\times4$ &
        \textcolor{red}{25.52} / \textcolor{red}{42.27} &
        24.75 / 45.36 &
        16.05 / 51.18 &
        15.15 / 52.11
        \\
        $\times4$ & 
        $\times4$ &
        - &
        \textcolor{red}{23.68} / 46.30  &
        21.69 / 36.59 &
        20.70 / \textcolor{red}{33.90}
        \\
        $\times8$ & 
        $\times4$ &
        - &
        22.84 / 46.57 &
        26.17 / 26.65 &
        \textcolor{red}{27.71} / \textcolor{red}{21.49}
        \\
        $\times16$ & 
        $\times4$ &
        - &
        21.54 / 48.15 &
        29.06 / 21.50 &
        \textcolor{red}{42.50} / \textcolor{red}{14.69}
        \\
    \midrule
        $\times2$ & 
        $\times6$ &
        - &
        - &
        11.76 / \textcolor{red}{69.51} &
        \textcolor{red}{13.59} / 70.82
        \\
        $\times4$ & 
        $\times6$ &
        - &
        - &
        15.13 / 54.93 &
        \textcolor{red}{17.51} / \textcolor{red}{47.91}
        \\
        $\times8$ & 
        $\times6$ &
        - &
        - &
        17.63 / 44.85&
        \textcolor{red}{21.20} / \textcolor{red}{32.54}
        \\
        $\times16$ & 
        $\times6$ &
        - &
        - &
        18.14 / 40.18&
        \textcolor{red}{25.20} / \textcolor{red}{24.46}
        \\
    \midrule
        $\times2$ & 
        $\times8$ &
        - &
        - &
        9.24 / \textcolor{red}{95.19} &
        \textcolor{red}{10.72} / 97.44
        \\
        $\times4$ & 
        $\times8$ &
        - &
        - &
        10.40 / 80.61&
        \textcolor{red}{13.84} / \textcolor{red}{68.95}
        \\
        $\times8$ & 
        $\times8$ &
        - &
        - &
        11.93 / 70.34&
        \textcolor{red}{16.30} / \textcolor{red}{49.22}
        \\
        $\times16$ & 
        $\times8$ &
        - &
        - &
        14.09 / 66.32&
        \textcolor{red}{23.93} / \textcolor{red}{35.49}
        \\
    \bottomrule
  \end{tabular}
  \label{tab:comparison_continuous}
  
  %\vspace{-1.em}
\end{table*}

\subsection{Ablation Experiments} 
\label{sec:ablation}

\vspace{-2.5em}
\textcolor{black}{\paragraph{Backward~vs.~Forward Motion.}
Table~\ref{tab:forward v.s. backward} presents results for an ablation experiment that replaces forward motion with backward motion in our MoTIF. This replacement includes the following changes: (1) learning ST-INF to predicting backward motion $\hat{M}_{t \rightarrow 0}^H, \hat{M}_{t \rightarrow 1}^H$ and their reliability maps $\hat{Z}_{t \rightarrow 0}^H, \hat{Z}_{t \rightarrow 1}^H$, (2) applying backward warping with $\hat{M}_{t \rightarrow 0}^H, \hat{M}_{t \rightarrow 1}^H$ to each reference feature $F^H_0, F^H_1$, and (3) synthesizing a high-resolution video frame $\hat{I}_t^H$ by taking as inputs the two backward warped reference features, their warped reliability maps, $F^H_{(0,1)}$, and $t$. From Table~\ref{tab:forward v.s. backward}, using backward motion instead of forward motion in MoTIF results in a considerable PSNR drop (0.4-1dB) across the test cases. Our supplementary document provides additional Fourier analyses to compare forward and backward motion.
}

\begin{table}
    \centering
    %\vspace{-0.15cm}
    %\caption{Forward vs. backward motion in our MoTIF following the x8 temporal scaling setting in Table 1. Metrics: PSNR/SSIM.}
    \caption{\textcolor{black}{Backward~vs.~forward motion in MoTIF. Quality metrics: PSNR/SSIM.} \textbf{Bold} indicates the best performance.}% under the difficult setting in Table 1: 8x temporal and 4x spatial scaling.}}
    %\vspace{-0.25cm}
    %\setlength{\tabcolsep}{2.0pt}
    \setlength{\tabcolsep}{4.0pt}
    \fontsize{8}{8}\selectfont
    \centering
    \begin{tabular}{c|c|c}
    \toprule

                         & Backward & Forward \\
    \midrule
    Vid4                 & 25.35 / 0.7696 & \textbf{25.79} / \textbf{0.7745} \\
    GoPro-\emph{Center}  & 29.98 / 0.8765 & \textbf{31.04} / \textbf{0.8877} \\
    GoPro-\emph{Average} & 29.38 / 0.8693 & \textbf{30.04} / \textbf{0.8773} \\
    Adobe-\emph{Center}  & 29.73 / 0.8723 & \textbf{30.63} / \textbf{0.8839} \\
    Adobe-\emph{Average} & 29.14 / 0.8658 & \textbf{29.82} / \textbf{0.8750} \\
    \bottomrule
\end{tabular}
    \label{tab:forward v.s. backward}
    \vspace{-0.15cm}
\end{table}
\begin{table}[]
    \centering
    %\vspace{-0.15cm}
    \caption{\textcolor{black}{Explicit~vs.~implicit motion modeling in MoTIF. Quality metrics: PSNR/SSIM.} \textbf{Bold} indicates the best performance.}% under the difficult setting in Table 1: 8x temporal and 4x spatial scaling.}}
    %\caption{Comparison of explicit and implicit motion modeling in our MoTIF following the setting of Table 1. Quality metrics: PSNR/SSIM.}
    %\vspace{-0.25cm}
    \setlength{\tabcolsep}{3.0pt}
    \fontsize{8}{8}\selectfont
    \centering
    \begin{tabular}{c|c|c|c}
    \toprule
    
    %                    & GoPro-Average    & Adobe-Average    \\
    % \midrule
    % VideoINR           & 29.41  / 0.8669 & 29.27 / 0.8651 \\
    % MoTIF   (Implicit) & 29.81  / 0.8744  & 29.59 / 0.8719 \\
    % MoTIF   (Explicit) & 30.04  / 0.8773  & 29.82 / 0.8750 \\  

                          & VideoINR     & MoTIF (Implicit)  & MoTIF (Explicit)   \\
    \midrule
    Vid4                  & 25.61 / 0.7709 & 25.71 / 0.7721  & \textbf{25.79} / \textbf{0.7745}       \\
    GoPro-\emph{Center}  & 30.26 / 0.8792 & 30.58 / 0.8856  & \textbf{31.04} / \textbf{0.8877}     \\
    GoPro-\emph{Average} & 29.41 / 0.8669 & 29.81 / 0.8744  & \textbf{30.04} / \textbf{0.8773}     \\
    Adobe-\emph{Center}  & 29.92 / 0.8746 & 30.24 / 0.8796  & \textbf{30.63} / \textbf{0.8839}     \\
    Adobe-\emph{Average} & 29.27 / 0.8651 & 29.59 / 0.8719  & \textbf{29.82} / \textbf{0.8750}     \\
    \bottomrule
    %%\hline
    \end{tabular}
    \label{tab:explicit v.s. implicit}
    \vspace{-0.5cm}
\end{table}

\vspace{-4em}
\textcolor{black}{\paragraph{Implicit~vs.~Explicit Motion Modeling.} Table~\ref{tab:explicit v.s. implicit} presents ablation results based on predicting the high-resolution forward motion without using a pre-trained optical flow estimation network,~i.e.~the implicit motion modeling. In this case, $F_0^L,F_1^L$ (see Fig.~\ref{fig:overview}) are used as inputs to our ST-INF. Compared with the explicit method, the implicit method has 0.1-0.4dB PSNR loss. Notably, even with the implicit method, MoTIF outperforms VideoINR by 0.1-0.4dB in PSNR, suggesting that the other components of MoTIF are essential.}

\vspace{-1em}
\paragraph{Feature Warping and Reliability Maps.}
Table~\ref{tab:ablation_components} presents ablation results to understand the contributions of different components in MoTIF. Four variants of MoTIF are investigated, including (a) using only $F_{(0,1)}^H$ for decoding, (b) using both $F_{(0,1)}^H$ and $F_t^H$ for decoding, (c) using $F_{(0,1)}^H,F_t^H$ for decoding while encoding the reliability maps $Z_{0 \rightarrow 1}^L,Z_{1 \rightarrow 0}^L$ of forward motion into the motion latents, and (d) the full model (i.e.~variant (c) plus $Z_t^H$). From Table~\ref{tab:ablation_components}, the considerable PSNR gain of (b) over (a) indicates that our ST-INF is effective in interpolating forward motion for propagating the reference features. The incremental improvement from (b) to (c) suggests that the reliability maps $Z_{0 \rightarrow 1}^L,Z_{1 \rightarrow 0}^L$ help to improve the quality of the interpolated forward motion. Last but not least, the additional use of $Z_t^H$ in the decoding process ((d) vs. (c)) does allow the decoder to better combine $F_{(0,1)}^H$ and $F_t^H$.

\vspace{-1em}
\paragraph{Tri-linear Motion and More Reference Frames.}
Table~\ref{tab:ablation_motion} investigates the benefits of our space-time implicit neural function (ST-INF) by comparing its performance with tri-linear motion interpolation and by showing its applicability to more reference frames. From Table~\ref{tab:ablation_motion}, we observe that when ST-INF is replaced with tri-linear motion interpolation--i.e.,~$\hat{M}_{0 \rightarrow t}^H,\hat{M}_{1 \rightarrow t}^H$ are interpolated tri-linearly from $M_{0 \rightarrow 1}^L,M_{1 \rightarrow 0}^L$, respectively--the performance drops by 0.1-0.2dB in PSNR. Although ST-INF provides seemingly moderate gain, its advantage becomes obvious when the number of reference frames goes beyond two. In this case, ST-INF can benefit from encoding more forward motion into the motion latents. Conceptually, this amounts to taking more forward motion samples, which help to construct accurate motion trajectories. In the ablation experiment, two more reference frames $I_{-1}^L,I_{2}^L$ are made available; we thus encode jointly information from $\{M_{0 \rightarrow i}^L,Z_{0 \rightarrow i}^L\}_{i=-1,1,2}$ as $T_0^L$, and information from $\{M_{1 \rightarrow i}^L,Z_{1 \rightarrow i}^L\}_{i=-1,0,2}$ as $T_1^L$. As a consequence, the PSNR improves by 0.5-0.7dB. We expect the gain to be even higher if we generate more motion latents to propagate information from these extra reference frames (instead of only $I_0^L,I_1^L$). The simple tri-linear motion interpolation cannot benefit similarly from having more reference frames.

\begin{table}
  \caption{Ablation experiment on individual components. (d) is the proposed MoTIF. Quality metric: PSNR.}
  \setlength{\tabcolsep}{8.0pt}
  \fontsize{8}{8}\selectfont
  \centering
  \begin{tabular}{@{}c|ccccc@{}}
    \toprule
        Settings &
        (a) &
        (b) &
        (c) &
        %(d) &
        (d)
        \\
    \midrule
        $F_{(0,1)}^H$ &
        V &
        V &
        V &
        %V &
        V 
        \\
        $F_t^H$ &
        &
        V &
        V &
        %V &
        V 
        \\
        $Z^L_0, Z^L_1$ &
        &
        &
        V &
        %V &
        V 
        \\
        %predict Z &
        %&
        %&
        %&
        %V &
        %V 
        %\\
        $Z^H_t$ &
        &
        &
        &
        %&
        V 
        \\
    \midrule
    \midrule
        Vid4 &
        22.38& %/ 0.5663 &
        25.26& %/ 0.7312 &
        25.61& %/ 0.7694 &
        %25.58 / 0.7688 &
        25.79& %/ 0.7745
        \\
        Gopro-\emph{Center} &
        26.68& %/ 0.7792 &
        30.54& %/ 0.8806 &
        30.97& %/ 0.8869 &
        %30.96 / 0.8869 &
        31.04& %/ 0.8877
        \\
        Gopro-\emph{Average} &
        26.44& %/ 0.7726 &
        29.72& %/ 0.8714 &
        29.97& %/ 0.8765 &
        %29.94 / 0.8764 &
        30.08& %/ 0.8780
        \\
        Adobe-\emph{Center} &
        25.82& %/ 0.7549 &
        30.04& %/ 0.8750 &
        30.49& %/ 0.8827 &
        %30.51 / 0.8823 &
        30.63& %/ 0.8839
        \\
        Adobe-\emph{Average} &
        25.61& %/ 0.7491 &
        29.40& %/ 0.8669 &
        29.68& %/ 0.8735 &
        %29.65 / 0.8730 &
        29.82& %/ 0.8750
        \\
    \bottomrule
  \end{tabular}
  \label{tab:ablation_components}
  %\vspace{-0.2cm}
\end{table}

\begin{table}
  \caption{Ablation experiment on tri-linear motion interpolation and multiple reference frames. Quality metrics: PSNR/SSIM.}
  %\caption{Ablation studies on flow. Forward Motions: Ours w/o reliability map}
  \setlength{\tabcolsep}{3pt}
  \fontsize{8}{8}\selectfont
  \centering
  \begin{tabular}{c|c|c|c}
    \toprule
                          & Tri-linear Motion     & Ours (2 ref.)  & Ours (4 ref.)   \\
    \midrule
    Vid4                 & 25.57 / 0.7728 & 25.79 / 0.7745  & 26.32 / 0.7864       \\
    GoPro-\emph{Center}  & 30.89 / 0.8860 & 30.96 / 0.8868  & 31.44 / 0.9003     \\
    GoPro-\emph{Average} & 29.93 / 0.8759 & 30.08 / 0.8780  & 30.77 / 0.8948     \\
    Adobe-\emph{Center}  & 30.42 / 0.8818 & 30.63 / 0.8839  & 31.03 / 0.8919     \\
    Adobe-\emph{Average} & 29.64 / 0.8727 & 29.82 / 0.8750  & 30.37 / 0.8849  
        \\
    \bottomrule
  \end{tabular}
  \label{tab:ablation_motion}
  \vspace{-1.em}
\end{table}

\section{Conclusion}
This paper introduces a C-STVSR framework, known as MoTIF. It features a space-time implicit neural function for encoding forward motion, and a reliability-aware splatting and decoding scheme for fusing spatiotemporal information from multiple reference frames. \textcolor{black}{We show that learning forward motion amounts to learning individual motion trajectories rather than a mixture of motion trajectories as with learning backward motion. In addition, for better aggregating temporal information via forward warping, performing splatting and decoding based on the reliability of forward motion is crucial. With all these techniques combined,} MoTIF demonstrates superior quantitative and qualitative performance to the state-of-the-art methods for C-STVSR.

%forward motion rather than backward motion has a significant impact on the quality of the generated videos; (2) modeling motion explicitly rather than implicitly help to improve the video quality; (3) introducing reliability-aware splatting and decoding further benefit the video quality. With all the techniques combined,} MoTIF demonstrates superior quantitative and qualitative performance to the state-of-the-art methods for C-STVSR and provides out-of-distribution generalization.

% (1) using forward motion rather than backward motion has a significant impact on the quality of the generated videos; (2) modeling motion explicitly rather than implicitly help to improve the video quality; (3) introducing reliability-aware splatting and decoding further benefit the video quality. With all the techniques combined,

%\textcolor{purple}{We justify the use of forward motion through a Fourier analysis, which shows that forward motion is smoother than backward motion in both spatial and temporal dimensions.}

\section*{Acknowledgement}
This work is supported by National Science and Technology Council, Taiwan under Grants NSTC 111-2634-F-A49-010- and MOST 110-2221-E-A49-
065-MY3, and National Center for High-performance Computing.

{\small
\bibliographystyle{ieee_fullname}
\bibliography{egbib}
}

\beginsupplement

% \predate{}
% \postdate{}

%%%%%%%%% TITLE

\title{MoTIF: Learning Motion Trajectories with Local Implicit Neural \\ Functions for Continuous Space-Time Video Super-Resolution \\ \textit{Supplementary Materials}}

%\author{}
\author{
Yi-Hsin Chen\printfnsymbol{1} 
\qquad Si-Cun Chen\printfnsymbol{1} 
\qquad Yi-Hsin Chen
\qquad Yen-Yu Lin
\qquad Wen-Hsiao Peng \\
National Yang Ming Chiao Tung University, Taiwan \\ 
\tt\small \{yhchen12101, sicun.mapl, karta6120\}.cs09@nycu.edu.tw \\ \tt\small lin@cs.nycu.edu.tw \tt\small wpeng@cs.nctu.edu.tw
}

%\date{}

\maketitle
% Remove page # from the first page of camera-ready.
\ificcvfinal\thispagestyle{empty}\fi

%%%%%%%%% BODY TEXT
%\vspace{-10cm}
This document provides additional results for
\begin{itemize}
     \setlength{\itemsep}{0pt}
    % \setlength{\parsep}{0pt}
    % \setlength{\parskip}{0pt}
    %\item Complexity comparison in terms of runtime and GMACs in Section~\ref{sec:complex_analysis};
    %\item Comparison with using tri-linear motion and more reference forward motions by MoTIF in Section~\ref{sec:trilinear};
    \item More comparison with the F-STVSR methods in Section~\ref{sec:vimeo};
    \item Replacing Raft-lite in MoTIF with other pre-trained flow estimation network in Section~\ref{sec:model_dependent};
    \item Fourier analysis of forward and backward motion in Section~\ref{sec:fourier};
    \item Subjective comparison in Section~\ref{sec:qualitative};
    \item Implementation details in Section~\ref{sec:implementation_detials}.
    %\item Visualizing the motion trajectories generated by our MoTIF in Section~\ref{sec:trajectory};    
    % fixed STVSR are trained in Vimeo
\end{itemize}

\begin{table*}
  \caption{Performance comparison on the F-STVSR task. \textcolor{red}{Red}, \textcolor{blue}{blue}, and \textbf{bold} indicate the best, the second best, and the third best performance, respectively. Quality metrics: PSNR/SSIM. Our MoTIF, although trained for the C-STVSR task, shows comparable performance to RSTT-L and TMNet on the F-STVSR task. Both RSTT-L and TMNet are the state-of-the-art one-stage F-STVSR methods. They are not able to support the C-STVSR task. See Section~\ref{sec:vimeo}.}
  \setlength{\tabcolsep}{8.0pt}
  \fontsize{8}{8}\selectfont
  \centering
  \begin{tabular}{@{}cc|ccccccccc@{}}
    \toprule
        \begin{tabular}{c}
            VFI
            \\
            Method
        \end{tabular}&
        \begin{tabular}{c}
            VSR
            \\
            Method
        \end{tabular}
        &
        \multicolumn{1}{c}{Vid4~\cite{Vid4}} &
        \multicolumn{1}{c}{Vimeo-Fast~\cite{TOFlow}} &
        \multicolumn{1}{c}{Vimeo-Medium~\cite{TOFlow}} &
        \multicolumn{1}{c}{Vimeo-Slow~\cite{TOFlow}} 
        \\
    \midrule
        SuperSloMo\cite{SuperSloMo} &
        Bicubic &
        22.84 / 0.5772 &
        31.88 / 0.8793 &
        29.94 / 0.8477 &
        28.37 / 0.8102 &
        \\
        SuperSloMo\cite{SuperSloMo} &
        RCAN\cite{RCAN} &
        23.80 / 0.6397 &
        34.52 / 0.9076 &
        32.50 / 0.8884 &
        30.69 / 0.8624 &
        \\
        SuperSloMo\cite{SuperSloMo} &
        RBPN\cite{RBPN} &
        23.76 / 0.6362 &
        34.73 / 0.9108 &
        32.79 / 0.8930 &
        30.48 / 0.8584 &
        \\
        SuperSloMo\cite{SuperSloMo} &
        EDVR\cite{EDVR} &
        24.40 / 0.6706 &
        35.05 / 0.9136 &
        33.85 / 0.8967 &
        30.99 / 0.8673 &
        \\
    \midrule
        SepConv\cite{SepConv} &
        Bicubic &
        23.51 / 0.6273 &
        32.27 / 0.8890 &
        30.61 / 0.8633 &
        29.04 / 0.8290 &
        \\
        SepConv\cite{SepConv} &
        RCAN\cite{RCAN} &
        24.92 / 0.7236 &
        34.97 / 0.9195 &
        33.59 / 0.9125 &
        32.13 / 0.8967 &
        \\
        SepConv\cite{SepConv} &
        RBPN\cite{RBPN} &
        26.08 / 0.7751 &
        35.07 / 0.9238 &
        34.09 / 0.9229 &
        32.77 / 0.9090 &
        \\
        SepConv\cite{SepConv} &
        EDVR\cite{EDVR} &
        25.93 / 0.7792 &
        35.23 / 0.9252 &
        34.22 / 0.9240 &
        32.96 / 0.9112 &
        \\
    \midrule
        DAIN\cite{DAIN} &
        Bicubic &
        23.55 / 0.6268 &
        32.41 / 0.8910 &
        30.67 / 0.8636 &
        29.06 / 0.8289 &
        \\
        DAIN\cite{DAIN} &
        RCAN\cite{RCAN} &
        25.03 / 0.7261 &
        35.27 / 0.9242 &
        33.82 / 0.9146 &
        32.26 / 0.8974 &
        \\
        DAIN\cite{DAIN} &
        RBPN\cite{RBPN} &
        25.96 / 0.7784 &
        35.55 / 0.9300 &
        34.45 / 0.9262 &
        32.92 / 0.9097 &
        \\
        DAIN\cite{DAIN} &
        EDVR\cite{EDVR} &
        26.12 / 0.7836 &
        35.81 / 0.9323 &
        34.66 / 0.9281 &
        33.11 / 0.9119 &
        \\
    \midrule
        \multicolumn{2}{c|}{STARnet\cite{StarNet}} &
        26.06 / \textcolor{red}{0.8046}  &
        36.19 / 0.9368  &
        34.86 / 0.9356  &
        33.10 / 0.9164  &
        \\
        \multicolumn{2}{c|}{Zooming SlowMo\cite{ZSM}} &
        26.31 / 0.7976  &
        \textbf{36.81} / \textbf{0.9415}  &
        35.41 / 0.9361  &
        33.36 / 0.9138  &
        \\
        \multicolumn{2}{c|}{TMNet\cite{TMNet}} &
        \textcolor{red}{26.43} / \textcolor{blue}{0.8016}  &
        \textcolor{red}{37.04} / \textcolor{red}{0.9435}  &
        \textcolor{blue}{35.60} / \textcolor{blue}{0.9380}  &
        \textcolor{red}{33.51} / \textcolor{red}{0.9159}  &
        \\
        \multicolumn{2}{c|}{RSTT-L\cite{RSTT}} &
        \textcolor{red}{26.43} / 0.7994  &
        36.80 / 0.9403  &
        \textcolor{red}{35.66} / \textcolor{red}{0.9381}  &
        \textcolor{blue}{33.50} / \textbf{0.9147}  &
        \\
    \midrule
        \multicolumn{2}{c|}{Ours} &
        \textcolor{red}{26.43} / \textbf{0.8013}  &
        \textcolor{blue}{36.88} / \textcolor{blue}{0.9427}  &
        \textbf{35.53} / \textbf{0.9372}  &
        \textbf{33.46} / \textcolor{blue}{0.9148}  &
        \\
        %\multicolumn{2}{c|}{Ours$^*$} &
        %25.78 & 0.7737 &
        %31.44 & 0.9003 &
        %30.77 & 0.8948 &
        %31.03 & 0.8919 &
        %30.37 & 0.8849 &
        %12.55 \\
    \bottomrule
  \end{tabular}
  \label{tab:comparison_fixed_supp}
\end{table*}
\section{More Comparisons with F-STVSR Methods}
\label{sec:vimeo}
This experiment compares our MoTIF with the state-of-the-art F-STVSR methods. Similar comparison is provided in the main paper, following the setting of VideoINR~\cite{VideoINR}, in which the training is done on Adobe240fps~\cite{Adobe240} dataset and with 2 reference frames. Here, we follow the common test protocol~\cite{ZSM, TMNet} of the F-STVSR task to perform training with 4 reference frames. 

In the present case, we have access to $I_{-1}^L,I_0^L,I_1^L,I_2^L$, in generating a high-resolution video frame $I_t^H, t \in [-1,2]$. To extend our scheme to 4 reference frames, (1) we follow ZSM~\cite{ZSM} to generate the reference features $F_{-1}^L, F_0^L, F_1^L, F_2^L$ and the intermediate features $F_{(-1,0)}^L, F_{(0,1)}^L, F_{(1,2)}^L$. (2) We then have the motion latent $T_0^L$ encode jointly information from multiple pairs $\{M_{0 \rightarrow i}^L,Z_{0 \rightarrow i}^L\},i=-1,1,2$ of the forward flow map $M_{0 \rightarrow i}$ and its reliability map $Z_{0 \rightarrow i}^L$, with $i$ referring to the reference frames except $I_{0}^L$. The same process is repeated to generate the other motion latents $T_i^L, i=-1,1,2$. (3) Based on these motion latents, we aggregate features $F_i^H$ from the 4 reference frames to synthesize $F_t^H,Z_t^H$. (4) During the decoding of the RGB values, the intermediate feature is chosen from $F_{(-1,0)}^L, F_{(0,1)}^L, F_{(1,2)}^L$ depending on which interval $t$ 
sits in. For example, if $t=-0.3$, the intermediate feature is $F_{(-1,0)}^L$, and  if $t=1.8$, the intermediate feature is $F_{(1,2)}^L$.

From Table~\ref{tab:comparison_fixed_supp}, we see that our MoTIF, although trained for the C-STVSR task, shows comparable performance to RSTT-L~\cite{RSTT} and TMNet~\cite{TMNet} on the F-STVSR task. Both RSTT-L~\cite{RSTT} and TMNet~\cite{TMNet} are the state-of-the-art one-stage F-STVSR methods. They, however, are not able to support the C-STVSR task. VideoINR~\cite{VideoINR} is not included in this comparison since it accepts only 2 reference frames.
%It is to be noted that although TMNet~\cite{TMNet} performs slightly better than our MoTIF, Tables 1 and 2 of the main paper show that \cite{TMNet}, which can only supports continuous temporal scaling and fixed 4x spatial scaling, can not perform well in 
%\textcolor{red}{Because most F-STVSR methods with 4 reference frames adopt Vimeo-90K~\cite{TOFlow} dataset for training, we conduct a experiment that also train our MoTIF on Vimeo-90K~\cite{TOFlow}

%Table~\ref{tab:comparison_fixed} provides comparison results on the F-STVSR task with 4 reference frames. 

 %Note that following VideoINR~\cite{VideoINR}, the performance comparison in \textcolor{red}{Tables 1 and 2 of the main paper} is based on using 2 reference frames, i.e.~$I_0^L,I_1^L$.
\section{Raft-lite vs. PWC-Net in MoTIF}
\label{sec:model_dependent}
Following the same experimental setup in Section~\ref{sec:vimeo}, Table~\ref{tab:model dependent} provides additional results by replacing Raft-lite~\cite{Raft} in MoTIF with the pre-trained PWC-Net~\cite{PWC-net}. As shown, the change in PSNR/SSIM is minor. This indicates that our MoTIF can work well with well-behaved, off-the-shelf flow estimation networks. %Moreover, the PWC-Net leads to the same conclusion as Raft in terms of their Fourier analyses (Fig. B1 vs. 2nd and 4th rows of Fig. A1 in our supplementary). This indicates that our MoTIF can work well with well-behaved, off-the-shelf flow estimation networks.}

\begin{table}[h]
    \centering
    \vspace{-0.15cm}
    %\caption{Pre-trained Raft vs. PWC-Net in our MoTIF following the setting of Table A1. Metrics: PSNR/SSIM.}
    \caption{PSNR/SSIM comparison of the pre-trained Raft~\cite{Raft} and PWC-Net~\cite{PWC-net} in MoTIF.}% following the setting in Table A3.}}
    %\vspace{-0.25cm}
    \setlength{\tabcolsep}{2.0pt}
    \fontsize{7.4}{7.4}\selectfont
    \centering
    \begin{tabular}{@{}c|cccc@{}}
    \toprule
    %\hline
    Flow Estimator&
    Vid4 &
    Vimeo-Fast &
    Vimeo-Medium &
    Vimeo-Slow \\
    \midrule
    %\hline
    Raft-lite~\cite{Raft} &
    26.43 / 0.8013 &
    36.88 / 0.9427 &
    35.53 / 0.9372 &
    33.46 / 0.9148 \\
    PWC-Net~\cite{PWC-net} &
    26.40 / 0.8001 &
    36.89 / 0.9432 &
    35.52 / 0.9366 &
    33.48 / 0.9161 \\
    \bottomrule
    %%\hline
    \end{tabular}
    \label{tab:model dependent}
    \vspace{-1em}
\end{table}
\section{Fourier Analysis Results}
\label{sec:fourier}
\begin{figure*}[t!]
  \centering

  \begin{subfigure}[b]{0.115\linewidth}
  \includegraphics[width=1.0\linewidth]{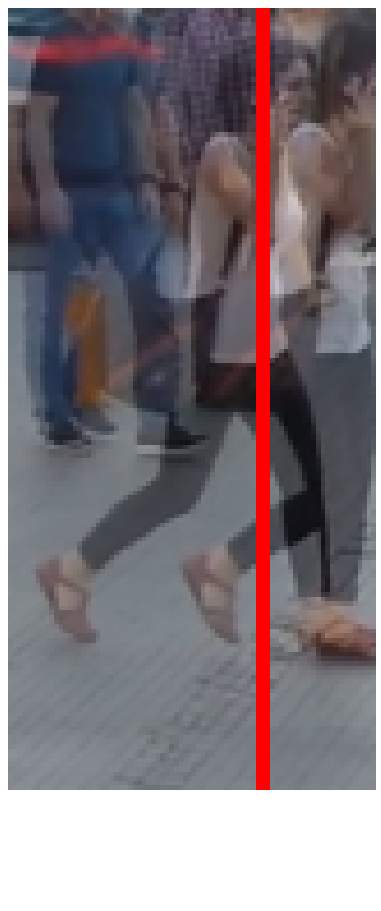}
    \label{fig:forward_backward_analysis_a}
  \end{subfigure}
  \begin{subfigure}[b]{0.115\linewidth}
  \includegraphics[width=1.0\linewidth]{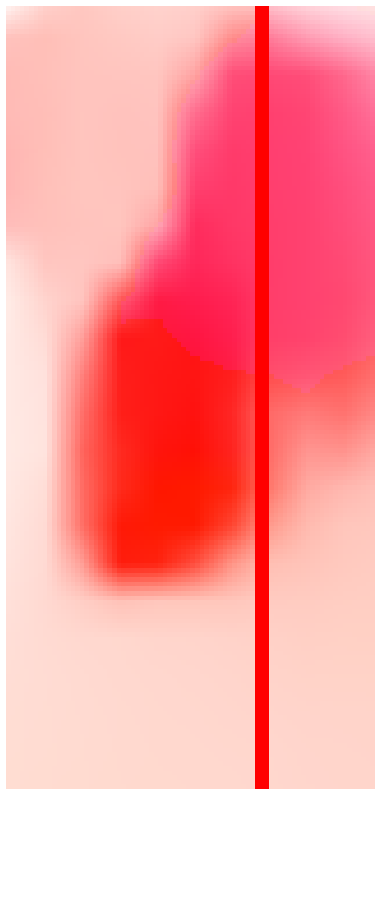}
    \label{fig:forward_backward_analysis_b}
  \end{subfigure}
  \begin{subfigure}[b]{0.340\linewidth}
  \includegraphics[width=1.0\linewidth]{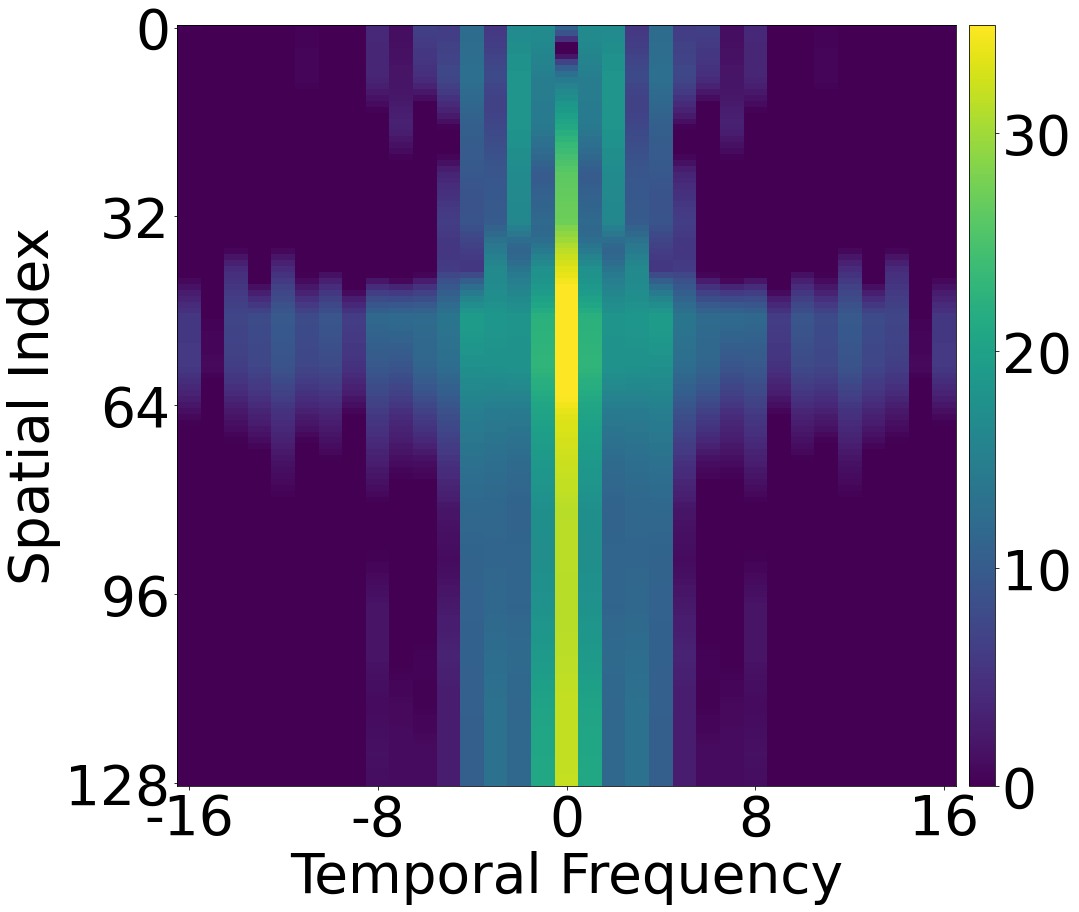}
    \label{fig:forward_backward_analysis_c}
  \end{subfigure}
  \begin{subfigure}[b]{0.340\linewidth}
  \includegraphics[width=1.0\linewidth]{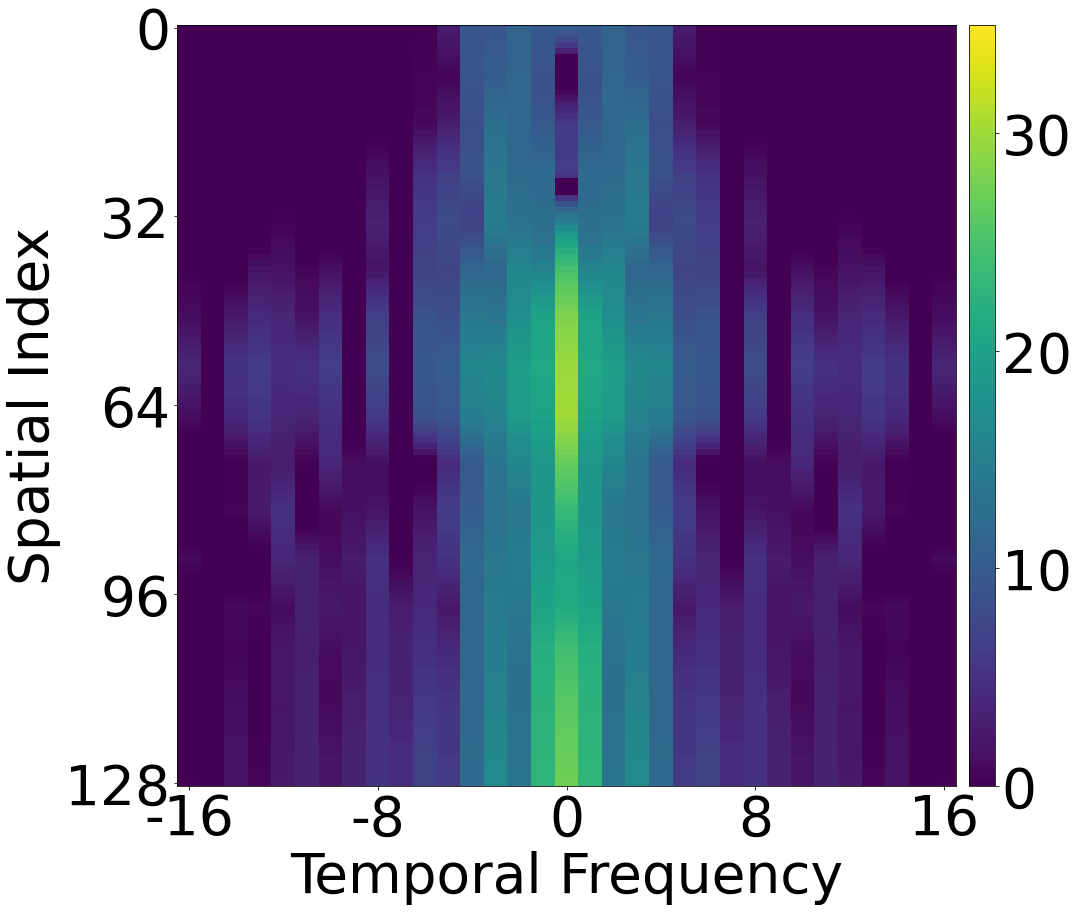}
    \label{fig:forward_backward_analysis_d}
  \end{subfigure}  
  
  \begin{subfigure}[b]{0.115\linewidth}
  \includegraphics[width=1.0\linewidth]{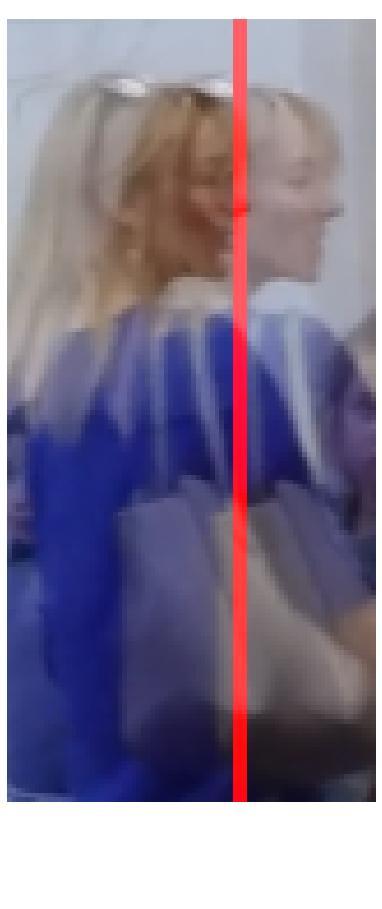}
    \label{fig:forward_backward_analysis_a}
  \end{subfigure}
  \begin{subfigure}[b]{0.115\linewidth}
  \includegraphics[width=1.0\linewidth]{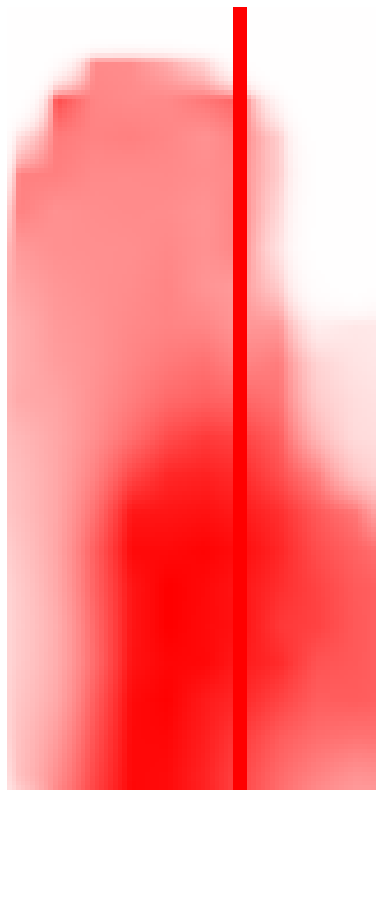}
    \label{fig:forward_backward_analysis_b}
  \end{subfigure}
  \begin{subfigure}[b]{0.340\linewidth}
  \includegraphics[width=1.0\linewidth]{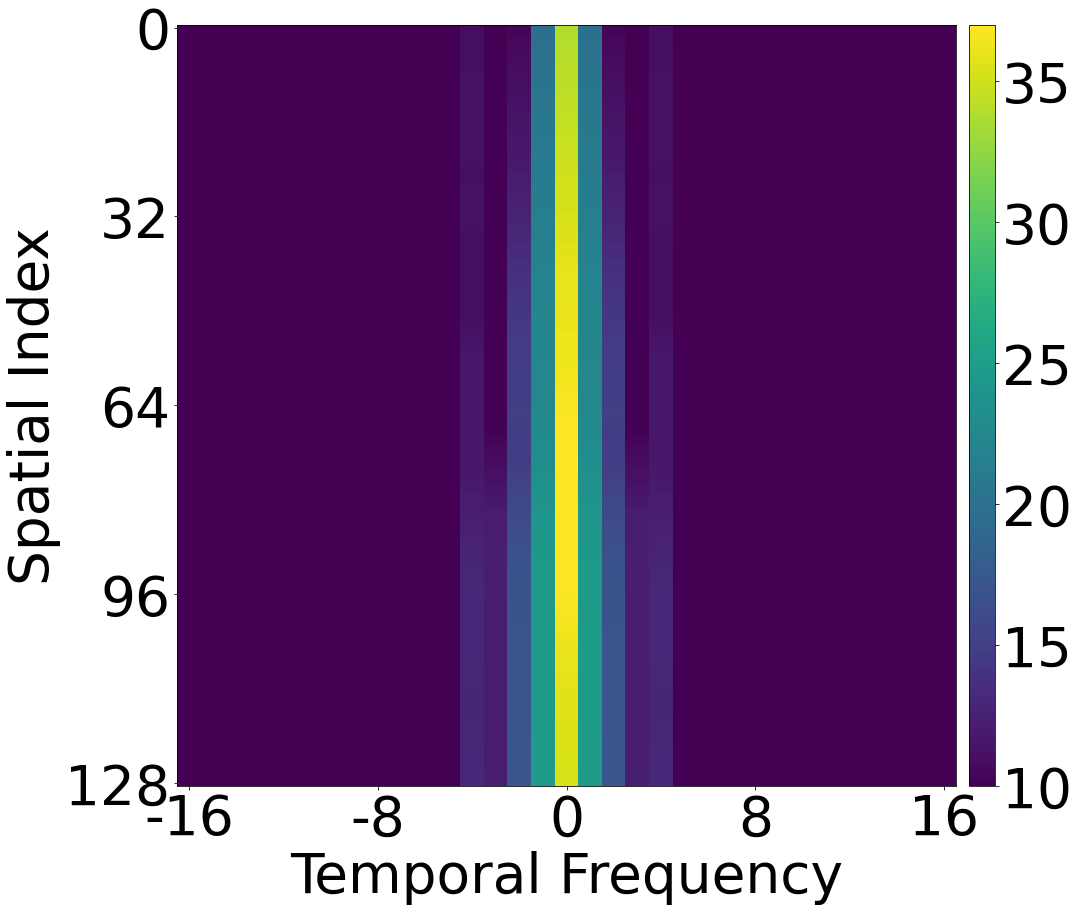}
    \label{fig:forward_backward_analysis_c}
  \end{subfigure}
  \begin{subfigure}[b]{0.340\linewidth}
  \includegraphics[width=1.0\linewidth]{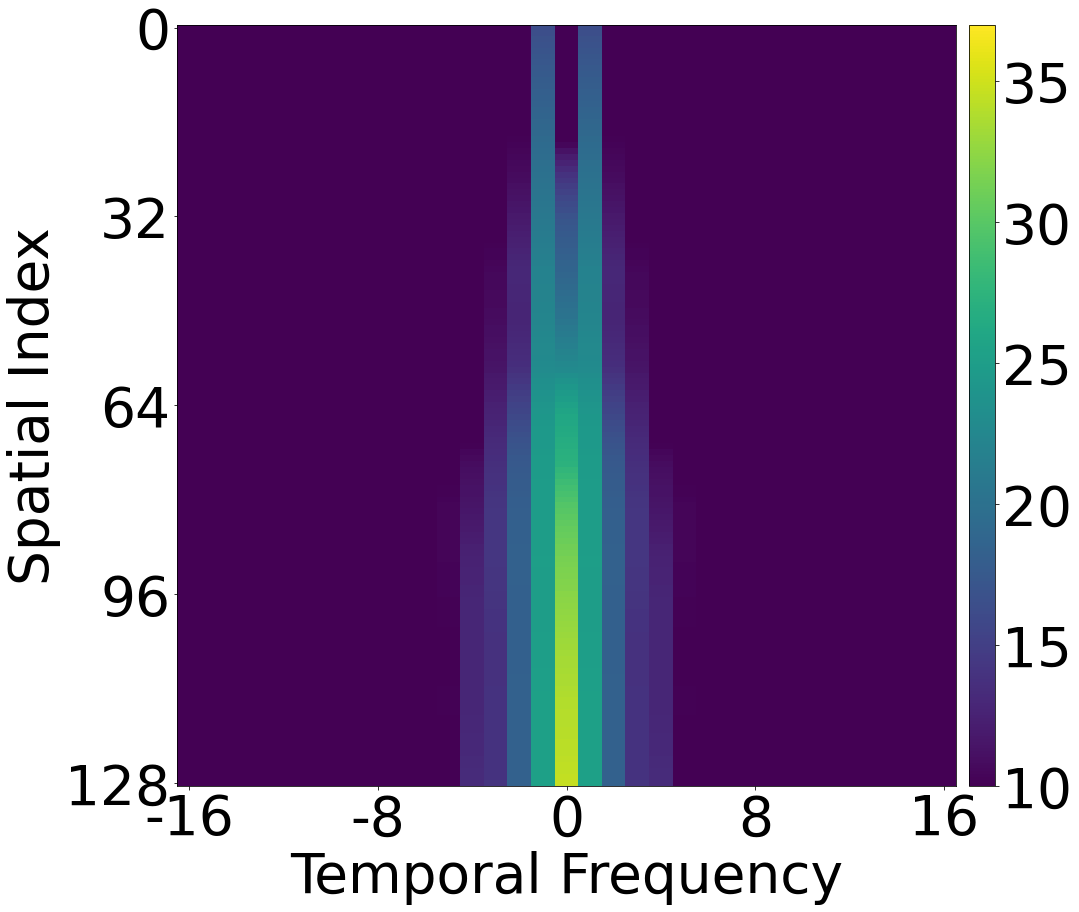}
    \label{fig:forward_backward_analysis_d}
  \end{subfigure}

  \begin{subfigure}[b]{0.115\linewidth}
  \includegraphics[width=1.0\linewidth]{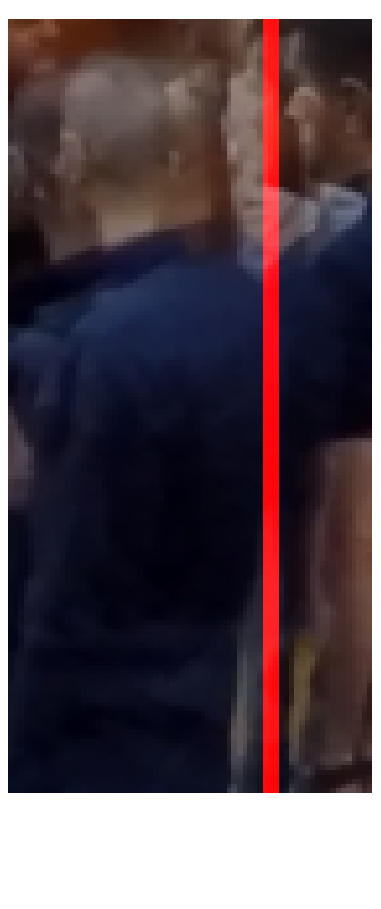}
    \label{fig:forward_backward_analysis_a}
  \end{subfigure}
  \begin{subfigure}[b]{0.115\linewidth}
  \includegraphics[width=1.0\linewidth]{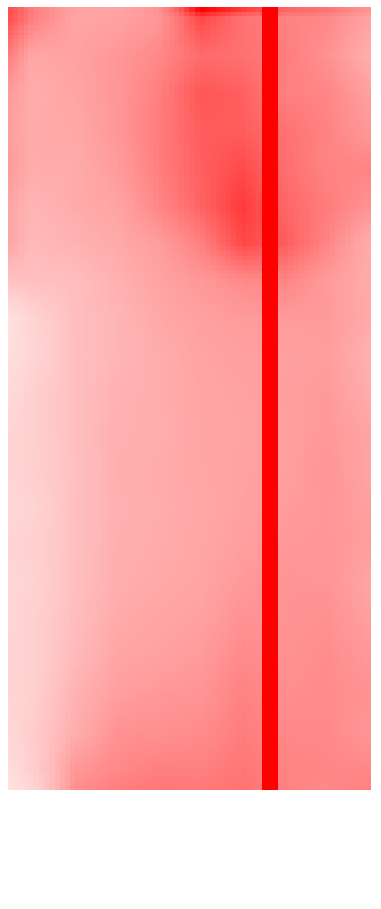}
    \label{fig:forward_backward_analysis_b}
  \end{subfigure}
  \begin{subfigure}[b]{0.340\linewidth}
  \includegraphics[width=1.0\linewidth]{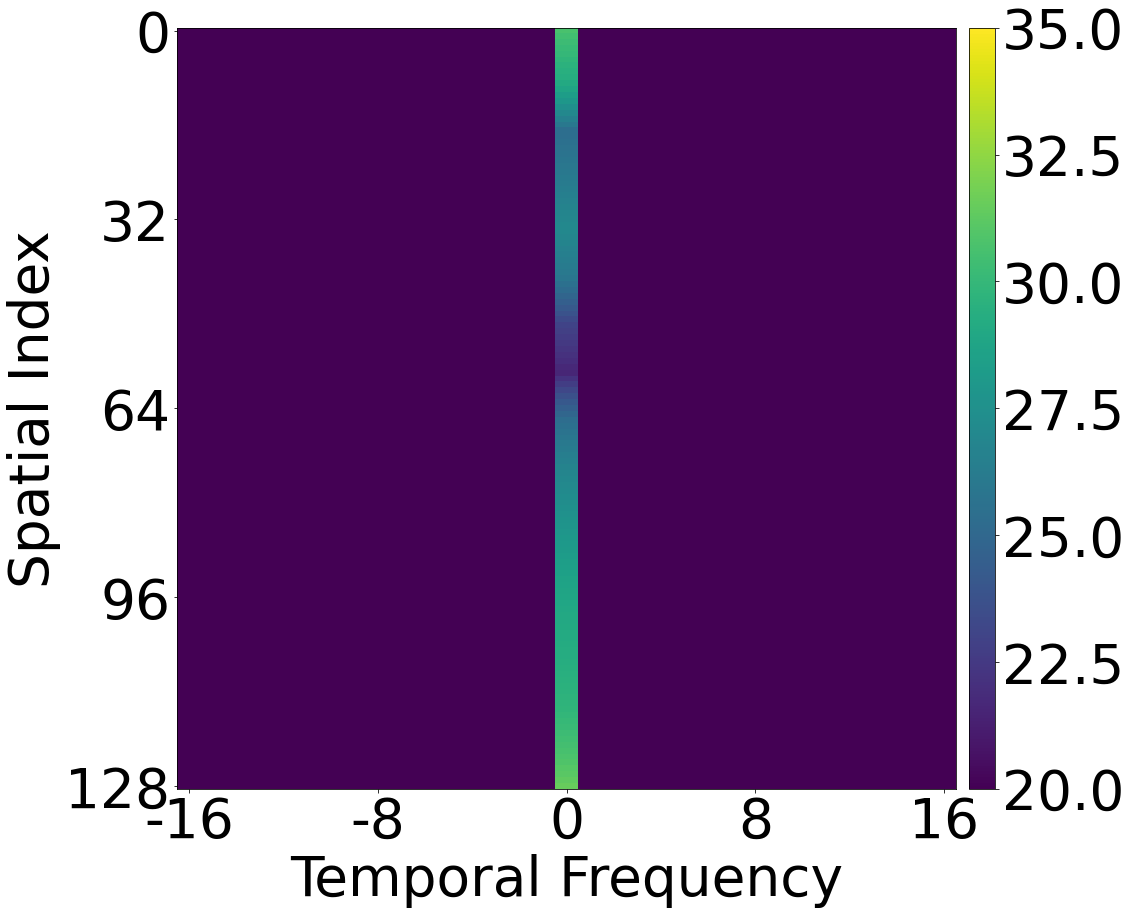}
    \label{fig:forward_backward_analysis_c}
  \end{subfigure}
  \begin{subfigure}[b]{0.340\linewidth}
  \includegraphics[width=1.0\linewidth]{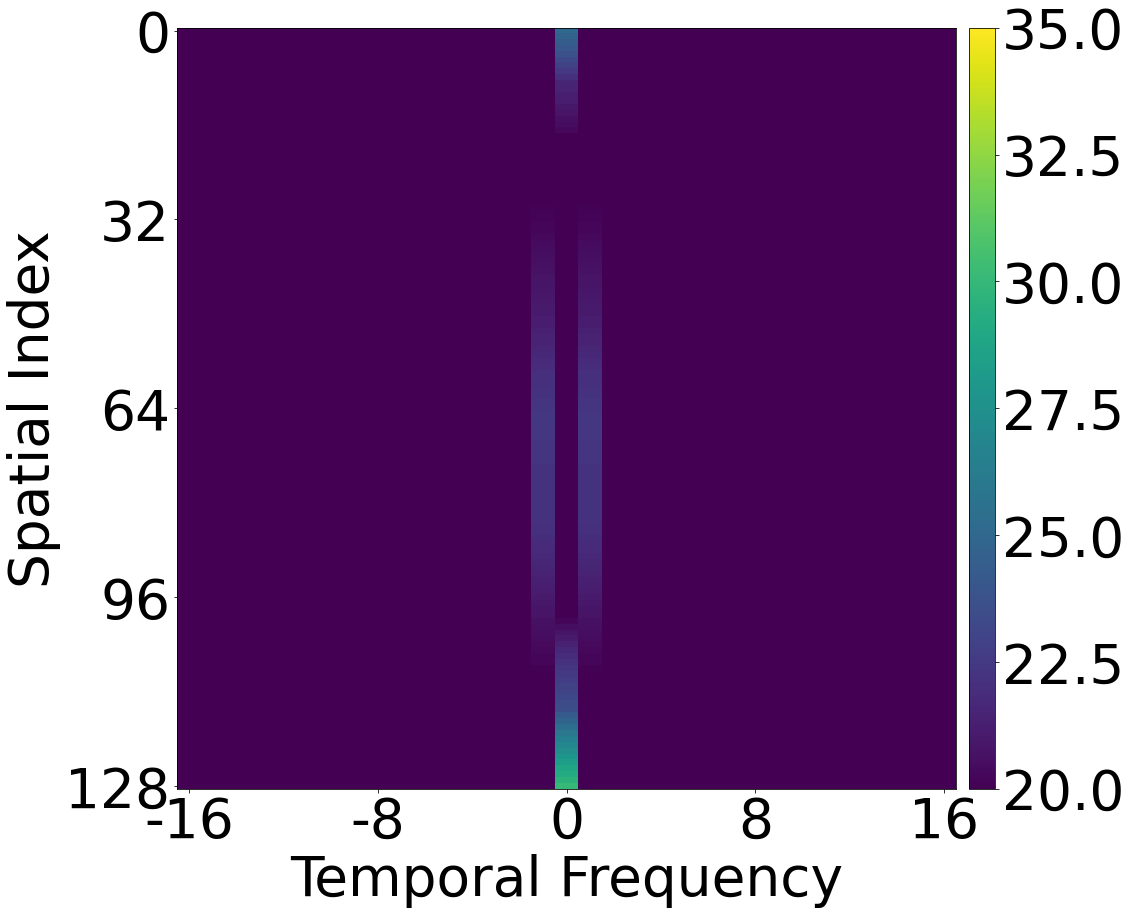}
    \label{fig:forward_backward_analysis_d}
  \end{subfigure}  
  
  \begin{subfigure}[b]{0.115\linewidth}
  \includegraphics[width=1.0\linewidth]{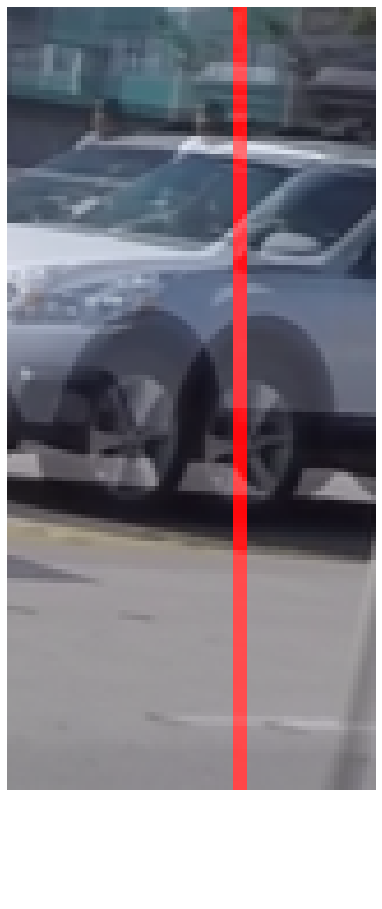}
    \label{fig:forward_backward_analysis_a}
  \end{subfigure}
  \begin{subfigure}[b]{0.115\linewidth}
  \includegraphics[width=1.0\linewidth]{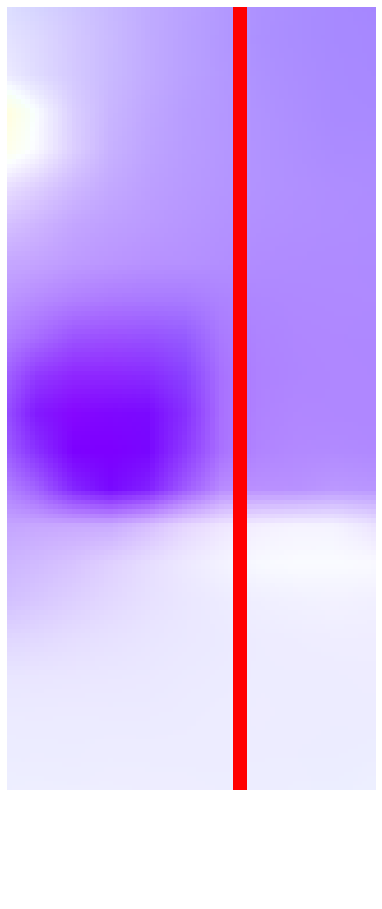}
    \label{fig:forward_backward_analysis_b}
  \end{subfigure}
  \begin{subfigure}[b]{0.340\linewidth}
  \includegraphics[width=1.0\linewidth]{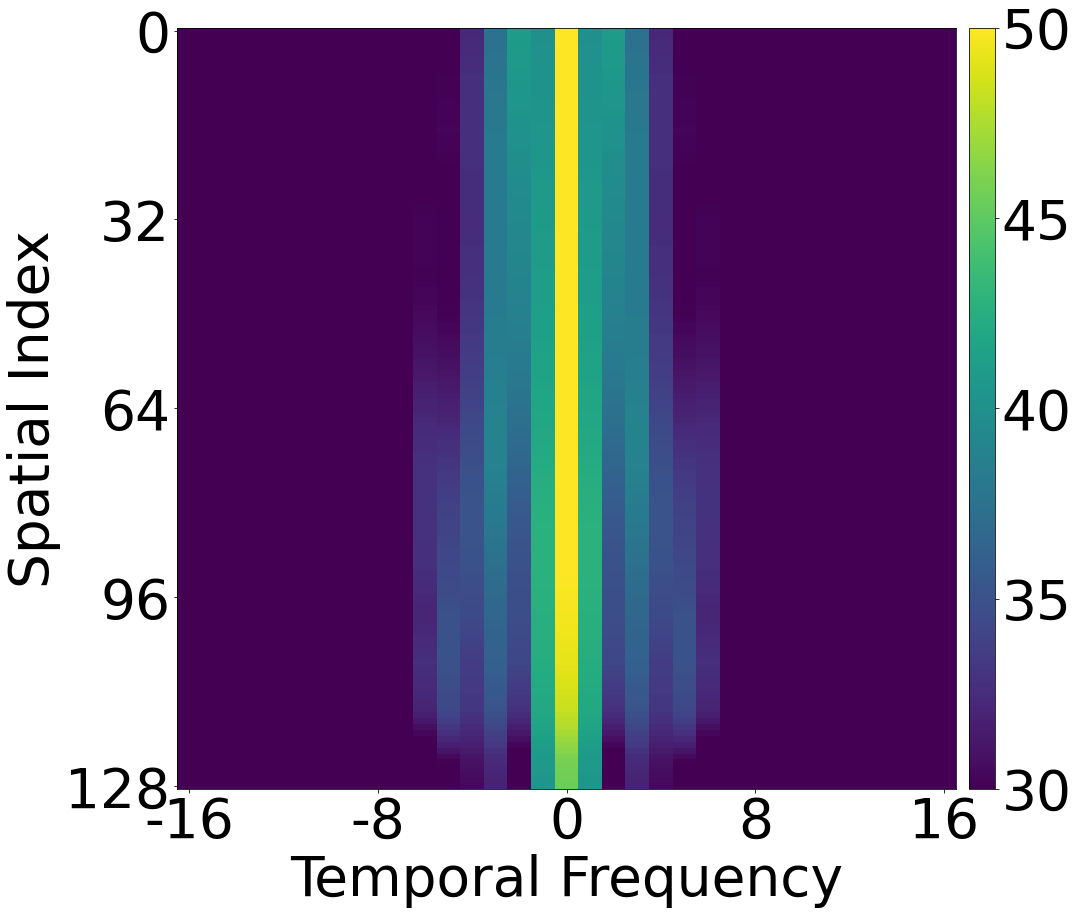}
    \label{fig:forward_backward_analysis_c}
  \end{subfigure}
  \begin{subfigure}[b]{0.340\linewidth}
  \includegraphics[width=1.0\linewidth]{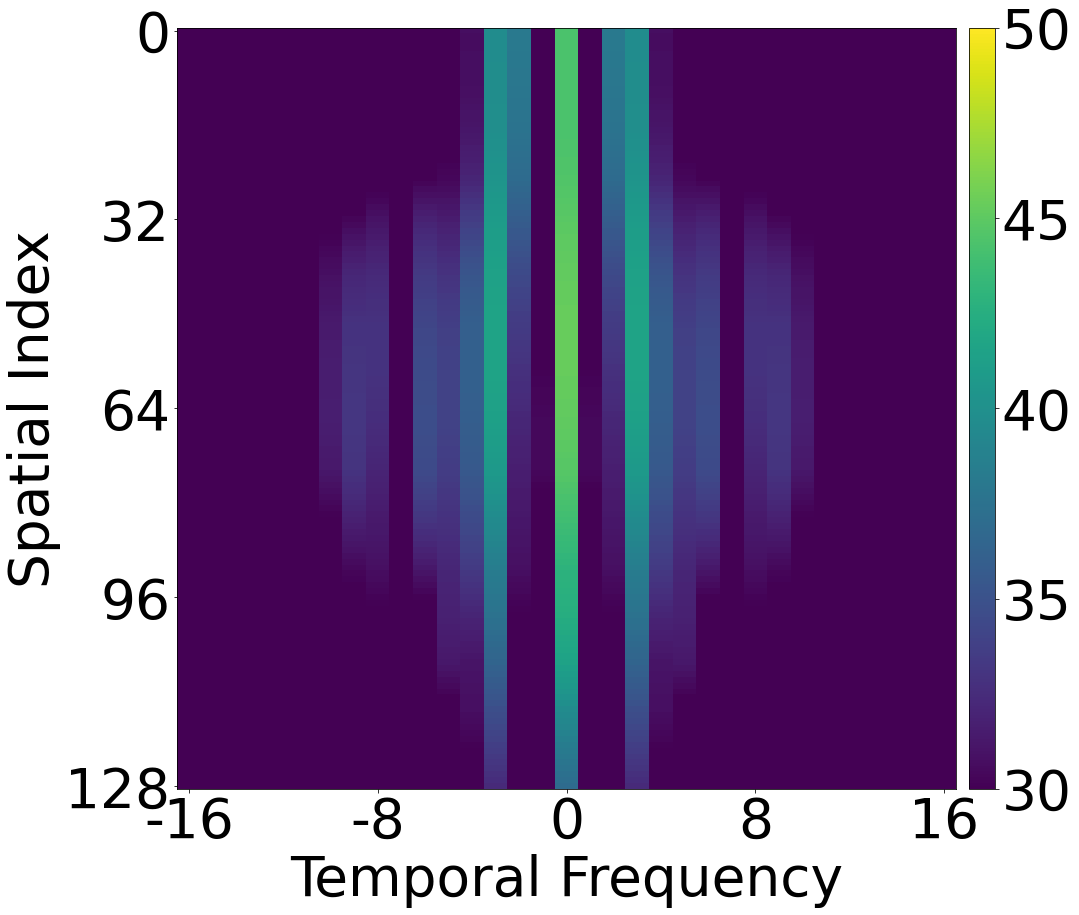}
    \label{fig:forward_backward_analysis_d}
  \end{subfigure}
    \vspace{-1.0em}
   \caption{Fourier analysis of forward and backward motion. The first column shows the slice of pixels whose forward/backward motion are analyzed. The second column is the forward optical flow map. The third column is the temporal signal spectra of the horizontal components of the forward displacement vectors. The forth column is the temporal signal spectra of the horizontal component of the backward displacement vectors. The spectra shown are magnitude responses. Forward motion usually has much stronger responses in the low-frequency bands than backward motion. See Section \ref{sec:fourier}.}
   %\caption{Fourier analysis of forward and backward motion: (a) the slice of pixels whose forward/backward motion are analyzed, (b) the forward optical flow map, (c) the temporal signal spectra of forward motion, (d) the temporal signal spectra of backward motion, (e) the space-time signal spectrum of forward motion, (f) the space-time signal spectrum of backward motion, and (g) the superposition of (e) and (f). The red dots in (c)(d) highlight the cut-off frequencies.}
  \vspace{-1.0em}
   %\caption{We conduct an analysis over 33 consecutive video frames. (a) represents overlayed images we use. (b) visualizes the forward motion. The vertical red line indicates the pixels we sample to analyze. (c) and (d) are the spectrum of forward and backward motion with 1-D Fourier transform applied along the temporal dimension, respectively. The red dots indicates the cut-off frequency for each pixel. (e) and (f) are the 2-D spectrum of forward and backward motion with 2-D Fourier transform applied along both the spatial and temporal dimensions, respectively. (g) is the overlayed (e) and (f).}
  %\caption{We collect 33 temporally consecutive forward and backward optical flow maps, $\{M^H_{0\xrightarrow[]{}i}\}_{i=0}^{32}$ and $\{M^H_{i\xrightarrow[]{}0}\}_{i=0}^{32}$ and apply fast Fourier transform to ground-truth forward and backward motions respectively to analyze the temporal and spatial continuity of forward and backward motions. (a) is the overlayed $I^H_0$ and $I^H_{32}$. The red line indicates the pixels we sample to analyze. (b) is the visualization of $M^H_{0\xrightarrow[]{}32}$. The red line indicates the pixels we sample to analyze. (c) and (d) are the spectrum of forward and backward motions after 1-D fast Fourier transform over time, respectively. The red dots indicates the bandwidth for each pixel (e) is the 3-D spectrum of forward and backward motions after 2-D fast Fourier transform over time and y (the red line in (a) and (b)), respectively. (g) is the overlayed (e) and (f).}
  \label{fig:forward_backward_analysis_xt}
\end{figure*}

\begin{figure*}[t!]
  \centering

  \begin{subfigure}[b]{0.115\linewidth}
  \includegraphics[width=1.0\linewidth]{Figures/forward_backward_analysis/intensity.png}
    \label{fig:forward_backward_analysis_a}
  \end{subfigure}
  \begin{subfigure}[b]{0.115\linewidth}
  \includegraphics[width=1.0\linewidth]{Figures/forward_backward_analysis/motion.png}
    \label{fig:forward_backward_analysis_b}
  \end{subfigure}
  \begin{subfigure}[b]{0.340\linewidth}
  \includegraphics[width=1.0\linewidth]{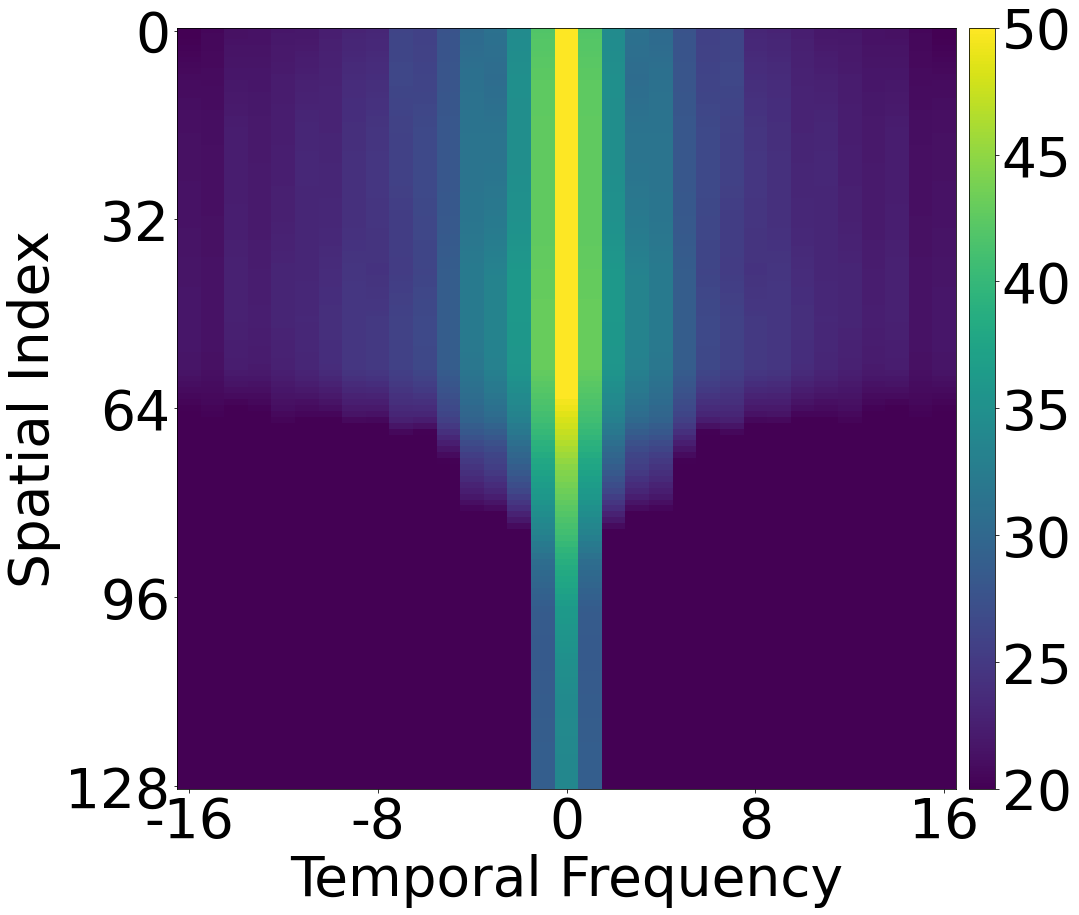}
    \label{fig:forward_backward_analysis_c}
  \end{subfigure}
  \begin{subfigure}[b]{0.340\linewidth}
  \includegraphics[width=1.0\linewidth]{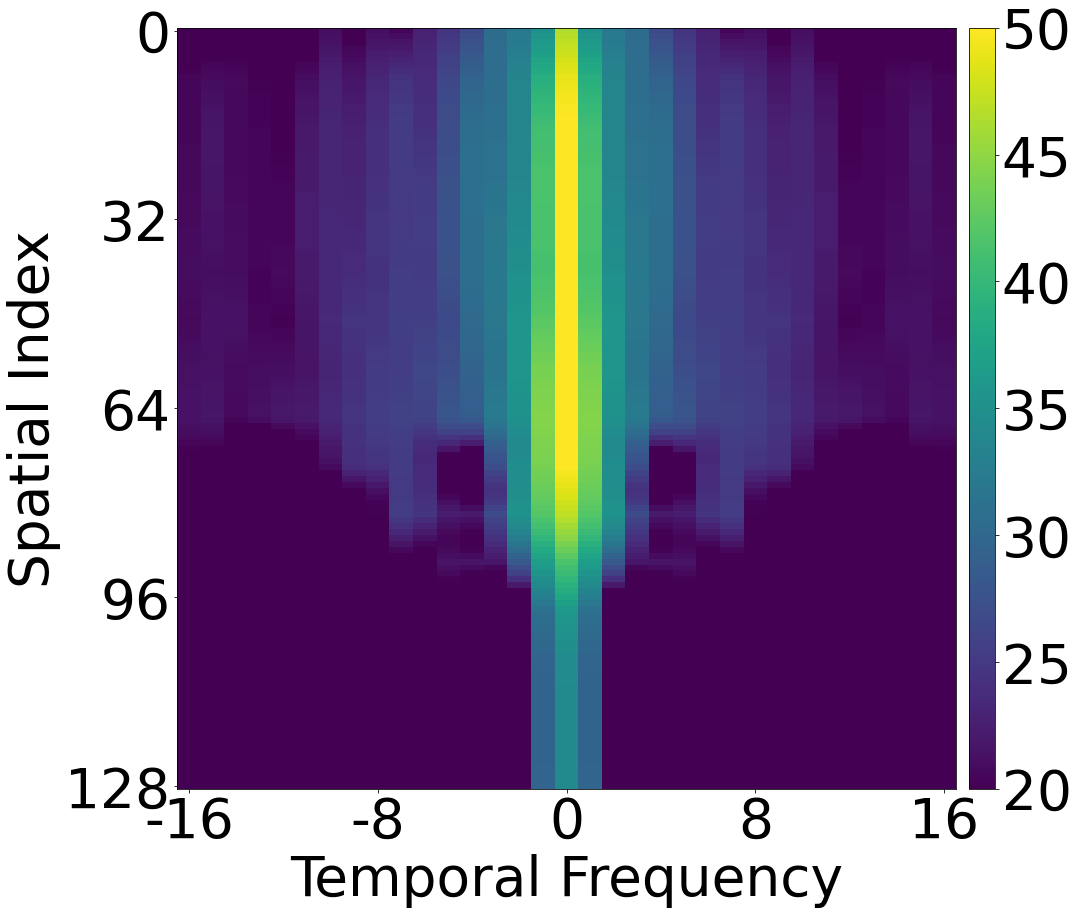}
    \label{fig:forward_backward_analysis_d}
  \end{subfigure}  
  
  \begin{subfigure}[b]{0.115\linewidth}
  \includegraphics[width=1.0\linewidth]{Figures/forward_backward_analysis/2_intensity.png}
    \label{fig:forward_backward_analysis_a}
  \end{subfigure}
  \begin{subfigure}[b]{0.115\linewidth}
  \includegraphics[width=1.0\linewidth]{Figures/forward_backward_analysis/2_motion.png}
    \label{fig:forward_backward_analysis_b}
  \end{subfigure}
  \begin{subfigure}[b]{0.340\linewidth}
  \includegraphics[width=1.0\linewidth]{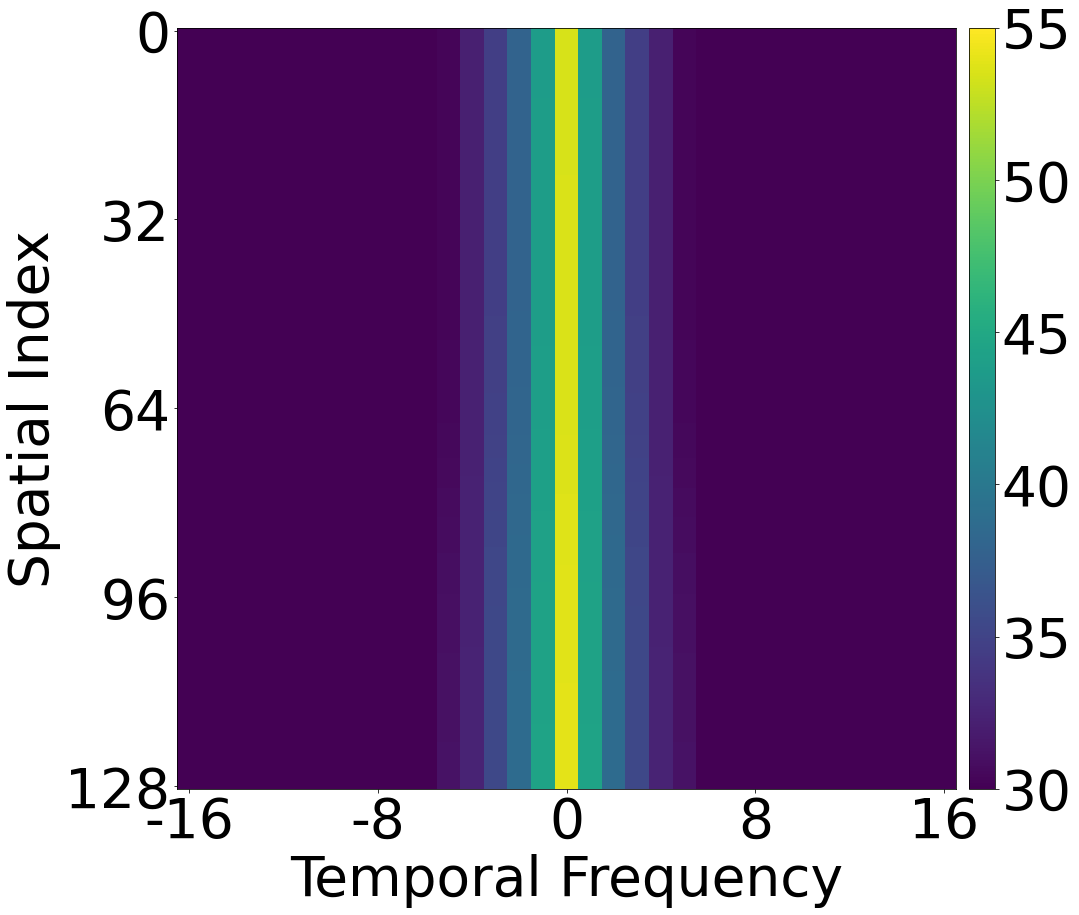}
    \label{fig:forward_backward_analysis_c}
  \end{subfigure}
  \begin{subfigure}[b]{0.340\linewidth}
  \includegraphics[width=1.0\linewidth]{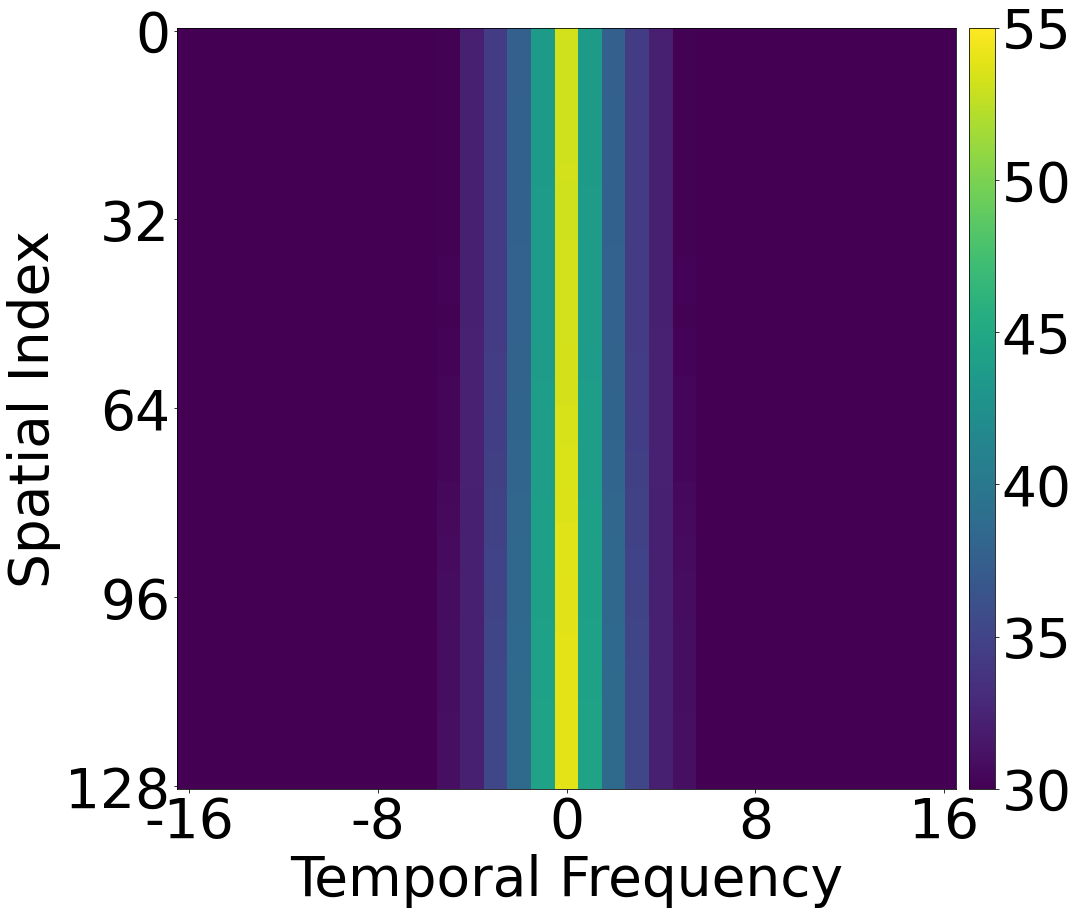}
    \label{fig:forward_backward_analysis_d}
  \end{subfigure}

  \begin{subfigure}[b]{0.115\linewidth}
  \includegraphics[width=1.0\linewidth]{Figures/forward_backward_analysis/3_intensity.png}
    \label{fig:forward_backward_analysis_a}
  \end{subfigure}
  \begin{subfigure}[b]{0.115\linewidth}
  \includegraphics[width=1.0\linewidth]{Figures/forward_backward_analysis/3_motion.png}
    \label{fig:forward_backward_analysis_b}
  \end{subfigure}
  \begin{subfigure}[b]{0.340\linewidth}
  \includegraphics[width=1.0\linewidth]{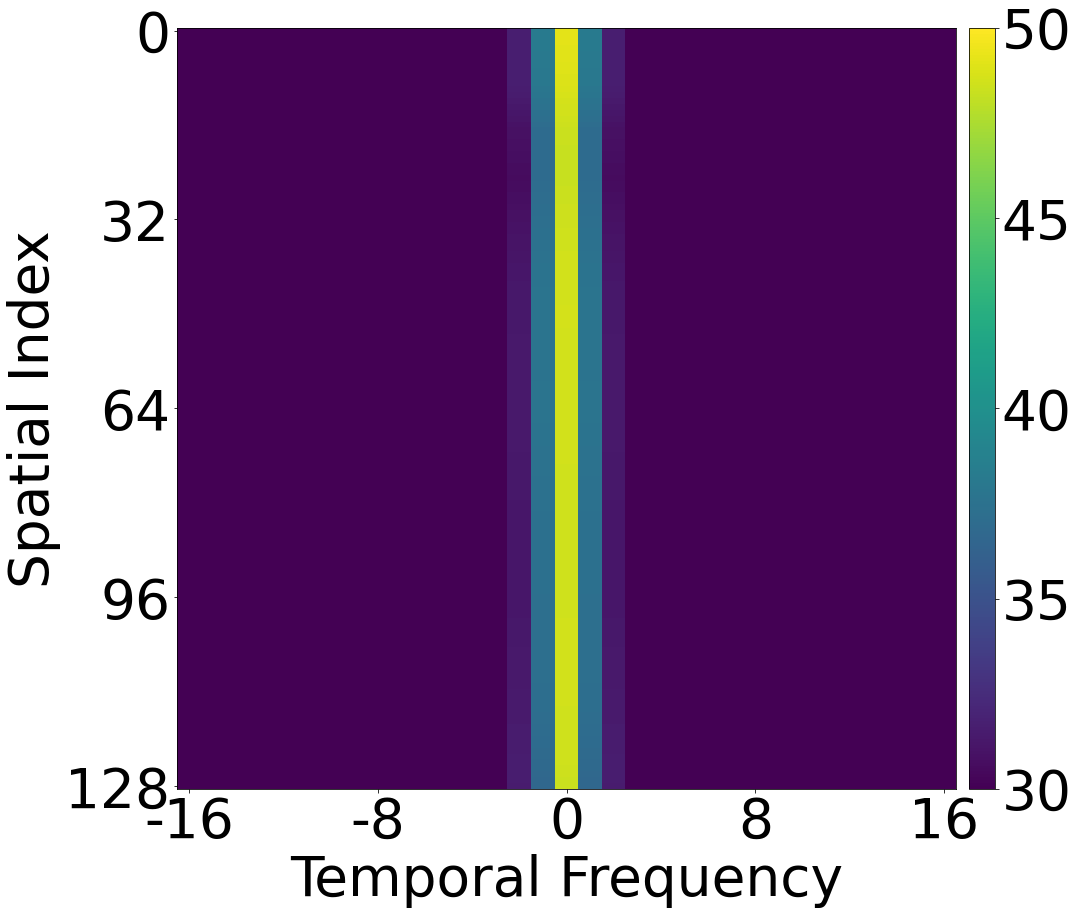}
    \label{fig:forward_backward_analysis_c}
  \end{subfigure}
  \begin{subfigure}[b]{0.340\linewidth}
  \includegraphics[width=1.0\linewidth]{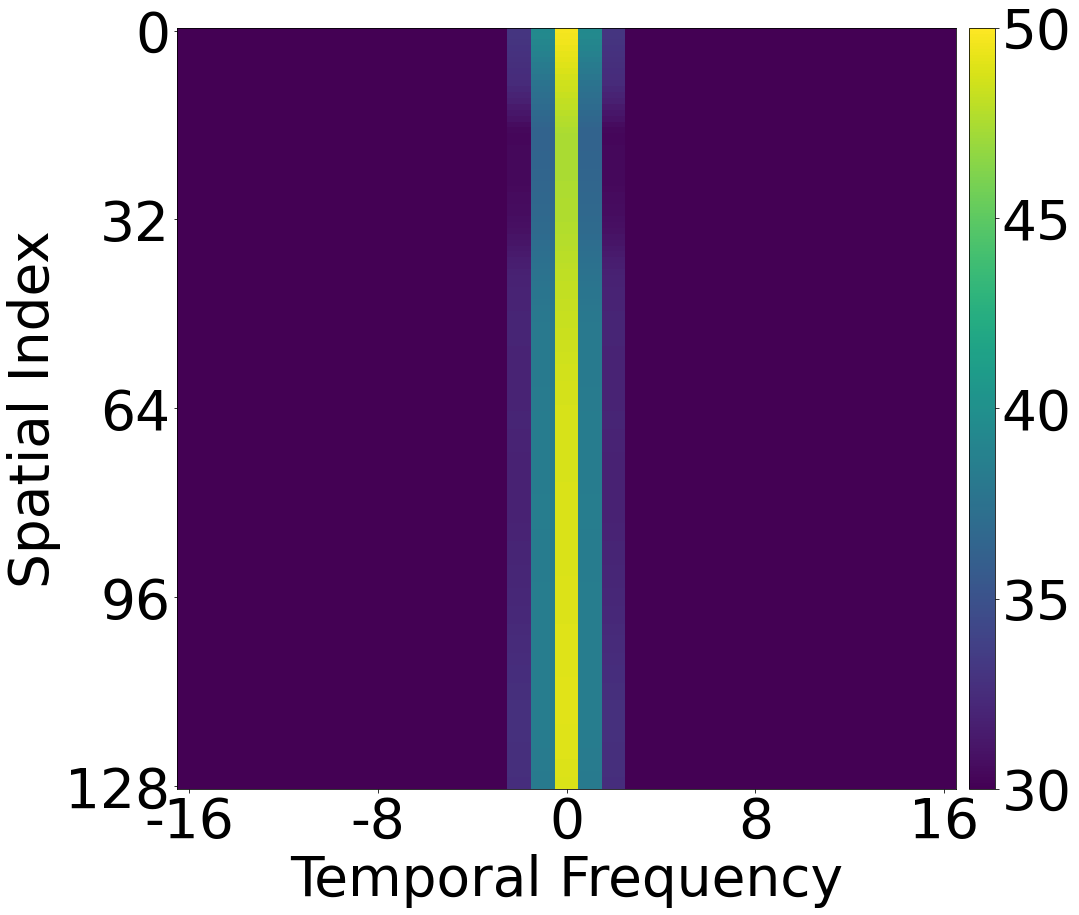}
    \label{fig:forward_backward_analysis_d}
  \end{subfigure}  
  
  \begin{subfigure}[b]{0.115\linewidth}
  \includegraphics[width=1.0\linewidth]{Figures/forward_backward_analysis/4_intensity.png}
    \label{fig:forward_backward_analysis_a}
  \end{subfigure}
  \begin{subfigure}[b]{0.115\linewidth}
  \includegraphics[width=1.0\linewidth]{Figures/forward_backward_analysis/4_motion.png}
    \label{fig:forward_backward_analysis_b}
  \end{subfigure}
  \begin{subfigure}[b]{0.340\linewidth}
  \includegraphics[width=1.0\linewidth]{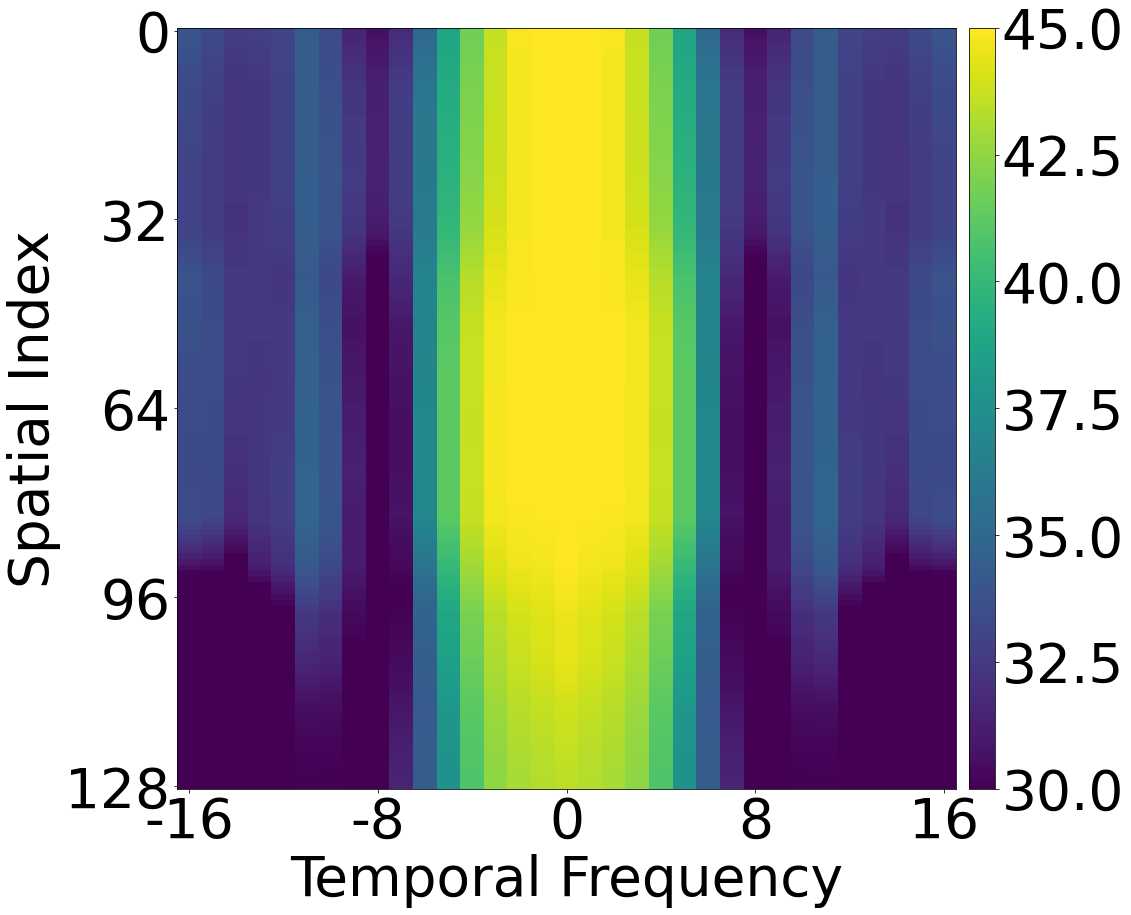}
    \label{fig:forward_backward_analysis_c}
  \end{subfigure}
  \begin{subfigure}[b]{0.340\linewidth}
  \includegraphics[width=1.0\linewidth]{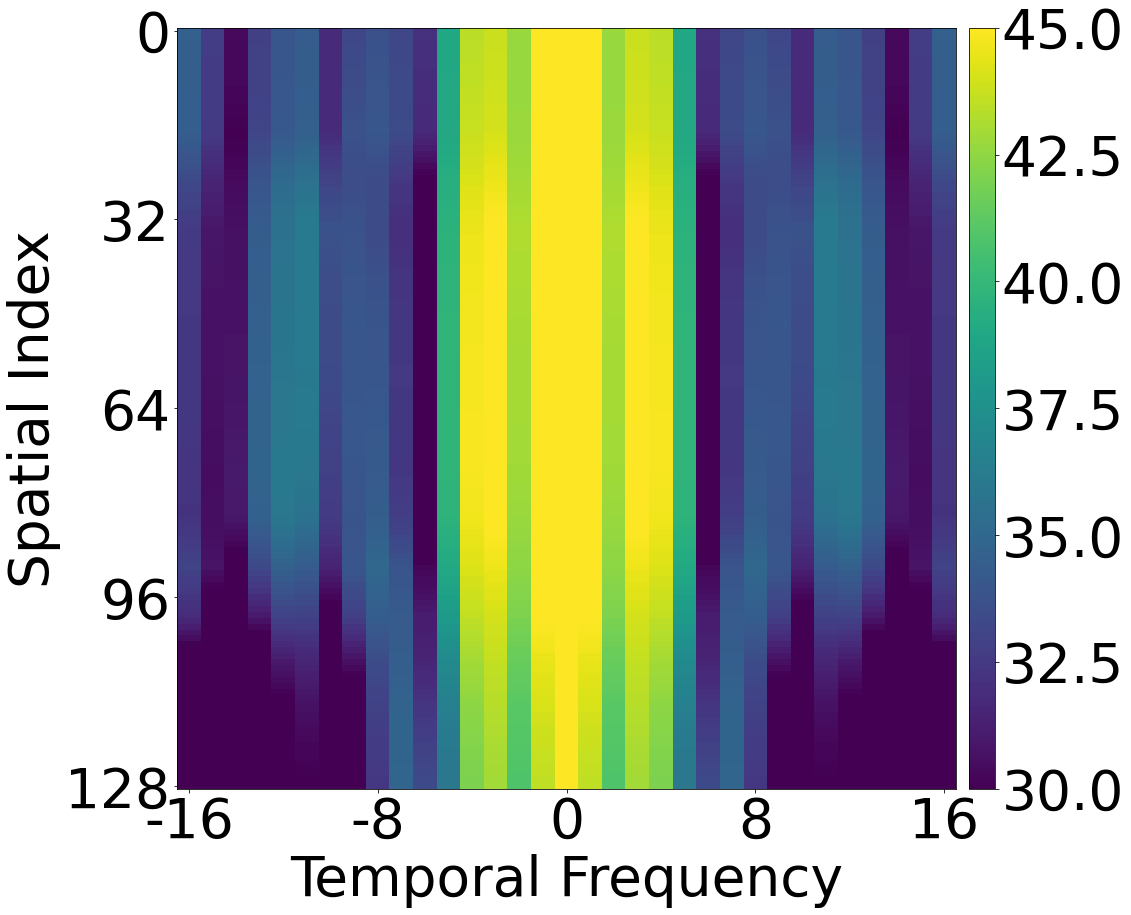}
    \label{fig:forward_backward_analysis_d}
  \end{subfigure}
    \vspace{-2.0em}
    \caption{Fourier analysis of forward and backward motion. The first column shows the slice of pixels whose forward/backward motion are analyzed. The second column is the forward optical flow map. The third column is the temporal signal spectra of the vertical component of the forward displacement vectors. The forth column is the temporal signal spectra of the vertical components of the backward displacement vectors. The spectra shown are magnitude responses. Forward and backward motion have similar frequency responses. This is because most video sequences have less and smaller vertical motion. See Section \ref{sec:fourier}.}

  \label{fig:forward_backward_analysis_yt}
\end{figure*}

%In this section, we provide a Fourier analysis of forward and backward motion, justifying the use of forward motion as a better representation.
Figs.~\ref{fig:forward_backward_analysis_xt} and ~\ref{fig:forward_backward_analysis_yt} analyze the signal spectra of the forward and backward motion representations. %Fourier analysis results of the forward and backward motion representations are provided in . 
%Fourier analysis results of the forward and backward motion representations are provided in Figs.~\ref{fig:forward_backward_analysis_xt} and ~\ref{fig:forward_backward_analysis_yt}. 
We take a vertical slice of pixels in the first columns of Fig.~\ref{fig:forward_backward_analysis_xt} and \ref{fig:forward_backward_analysis_yt} as examples, and represent their forward or backward motion over 33 consecutive video frames as functions of time. At each vertical pixel location, we conduct 1-D Fourier transform of the motion signal along the temporal dimension. In each figure, (1) the first column superimposes the first and the last frames of the test sequence, (2) the second column shows the forward motion from the first frame to the last frame, and (3) the third and the fourth columns visualize the spectra of the forward and backward motion, respectively.

In Fig.~\ref{fig:forward_backward_analysis_xt}, at each spatial location, the 1-D Fourier transform along the temporal dimension is applied to the horizontal component (namely, the x-component) of the displacement vectors. The spectra shown are magnitude responses. We see that forward motion usually has much stronger responses in the low-frequency bands, especially the DC band (temporal frequency=0), than backward motion. On the other hand, backward motion has more high-frequency responses. This implies that the back motion representation is typically less smooth temporally. 

In Fig.~\ref{fig:forward_backward_analysis_yt}, a similar analysis is conducted on the vertical component (namely, the y-component) of the displacement vectors. Interestingly, both the forward and backward motion representations have similar frequency responses. This may be because most video sequences have less and smaller vertical motion. 

\section{More Qualitative Results}
\label{sec:qualitative}
Figs.~\ref{tbl:qualitative_results_2} ,~\ref{tbl:qualitative_results_1}
,~\ref{tbl:qualitative_results_spatial1}
, and~\ref{tbl:qualitative_results_spatial2}
provide more subjective quality comparisons. Our MoTIF preserves more high-frequency details than the other competing methods in tests with both in-distribution and out-of-distribution temporal scaling factors (cf. the buildings in Fig.~\ref{tbl:qualitative_results_2}, the heads of the ducks in Fig.~\ref{tbl:qualitative_results_2}, the edge of the butterfly in Fig.~\ref{tbl:qualitative_results_1}, the paper posted on the door of the train in Fig.~\ref{tbl:qualitative_results_1}, 
the license plate of the taxi in Fig.~\ref{tbl:qualitative_results_spatial1},
and the legs of the race horse in Fig.~\ref{tbl:qualitative_results_spatial2}
).

%with different temporal scales between Our MoTIF, TMNet~\cite{TMNet} and VideoINR~\cite{VideoINR}. 
\section{Implementation Details}
\label{sec:implementation_detials}

\subsection{Reliability Maps}
Following ~\cite{Splatting_based}, we quantify the reliability of a forward optical flow map based on (1) the intensity warping error $Z_{0 \rightarrow 1}^{int}$, (2) the flow warping error $Z_{0 \rightarrow 1}^{flow}$, and (3) the local variances of the flow map. Consider $M_{0 \rightarrow 1}^L$ as an example. These metrics are given, respectively, by  
\begin{equation}
    Z_{0 \rightarrow 1}^{int} = \lVert I_0^L-\omega(I_1^L, M_{0 \rightarrow 1}^L) \rVert, 
\label{int_error}
\end{equation}
\begin{equation}
    Z_{0 \rightarrow 1}^{flow} = \lVert M_{0 \rightarrow 1}^L - (-\omega(M_{1 \rightarrow 0}^L, M_{0 \rightarrow 1}^L)) \rVert,
\label{flow_error}
\end{equation}
\begin{equation}
    Z_{0 \rightarrow 1}^{var} = \sqrt{G((M_{0 \rightarrow 1}^L)^2)-G(M_{0 \rightarrow 1}^L)^2},
\label{flow_var}
\end{equation}
where $\omega(A,B)$ denotes the operation of backward warping $A$ based on $B$,~e.g.~$I_0^L-\omega(I_1^L,M_{0 \rightarrow 1}^L) \equiv I_0^L(p)-I_1^L(p+M_{0 \rightarrow 1}^L(p)),~\forall p$, with $p$ denoting the pixel coordinates in $I_0^L$, and $G(\cdot)$ denotes the $3\times3$ Gaussian kernel. From Eq.~\eqref{int_error}, the intensity warping error evaluates the prediction error of $I_0^L$ by backward warping $I_1^L$ using $M_{0 \rightarrow 1}^L$. The flow warping error in Eq.~\eqref{flow_error} checks the consistency between $M_{0 \rightarrow 1}^L$ and $M_{1 \rightarrow 0}^L$. It is defined as the prediction error of $M_{0 \rightarrow 1}^L$ by backward warping $M_{1 \rightarrow 0}^L$ using $M_{0 \rightarrow 1}^L$. The sign flipping $-\omega(M_{1 \rightarrow 0}^L,M_{0 \rightarrow 1}^L)$ accounts for the difference between $M_{0 \rightarrow 1}^L$ and $M_{1 \rightarrow 0}^L$ in their directions. 
%With these three maps, we then concatenate them channel-wisely to form the reliability map $Z_{0 \rightarrow 1}^L$.

%\subsection{Generalization to More Reference Frames}

%Having access to more reference frames allows us to take more forward motion samples $\{M_{0 \rightarrow i}^L\}_{i=1}^{3}$, which help construct accurate motion trajectories. 

\subsection{Network Architecture}
\begin{figure*}[]
  \centering
  \includegraphics[width=0.4\linewidth]{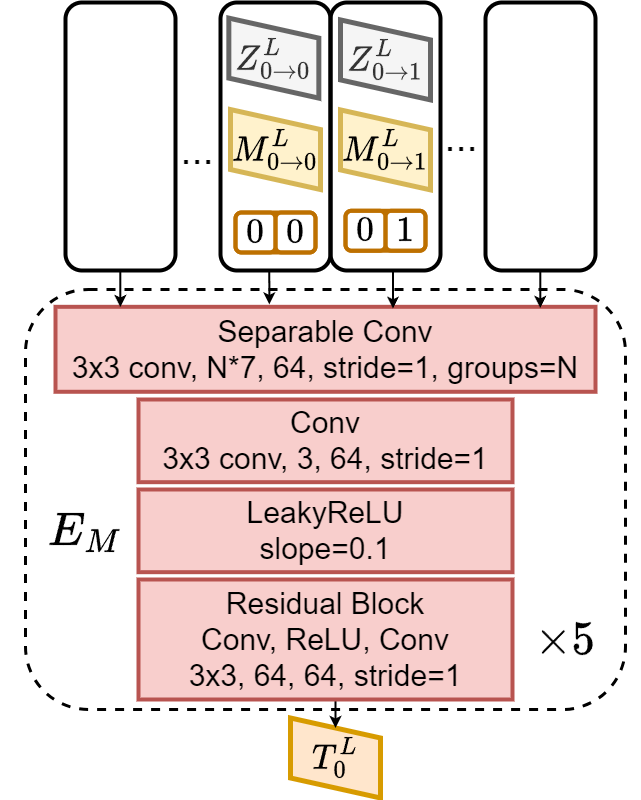}
  \caption{The network architecture of our motion encoder $E_M$. $N$ is the number of motion samples we use. See Section \ref{sec:implementation_detials}.}
  \label{fig:EM}
\end{figure*}

\begin{figure*}[]
  \centering
  \includegraphics[width=0.55\linewidth]{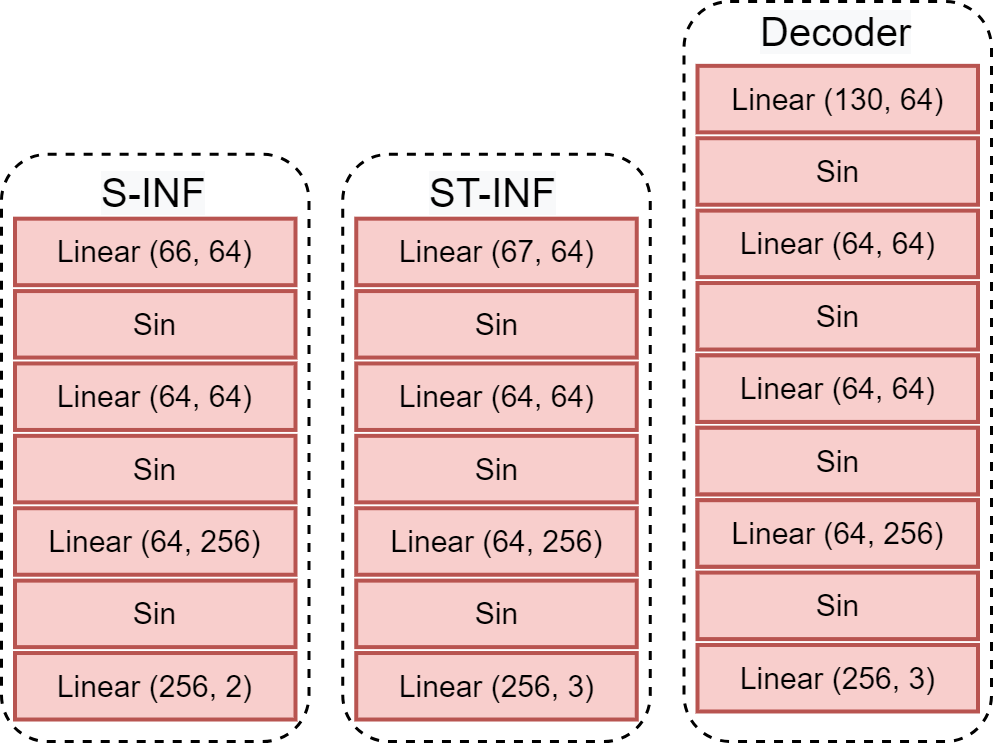}
  \caption{Shown from left to right are the network architectures of our S-INF, ST-INF and decoder, respectively. See Section \ref{sec:implementation_detials}.}
  \label{fig:Siren}
\end{figure*}

We further illustrate details of our network architecture in Fig.~\ref{fig:EM} and Fig.~\ref{fig:Siren}. As shown in Fig.~\ref{fig:EM}, our motion encoder takes $N$ group of motion features as input, where $N$ is the number of motion samples we use. Each motion feature includes the forward motion, the reliability map and two constant maps describing the source time and destination time of the forward motion, respectively.

\input{Figures/quantitative_2_down}
\input{Figures/quantitative_1_down}
\input{Figures/quantitative_spatial_1}
\input{Figures/quantitative_spatial_2}


\end{document}